\documentclass[amssymb,nofootinbib,superscriptaddress]{revtex4-1}
\pdfoutput=1

\usepackage{aas_macros,empheq,esint,fancybox,graphicx,multirow}
\usepackage[bottom]{footmisc}
\usepackage[colorlinks=true]{hyperref}
\let\originalleft\left
\let\originalright\right
\renewcommand{\left}{\mathopen{}\mathclose\bgroup\originalleft}
\renewcommand{\right}{\aftergroup\egroup\originalright}

\newcommand{\ab}[1]{\left|#1\right|}

\newcommand{\br}[1]{\left[#1\right]}
\newcommand{\cu}[1]{\left\{#1\right\}}
\newcommand{\pa}[1]{\left(#1\right)}

\newcommand{\ed}{\mathop{}\!\mathrm{d}}
\newcommand{\pd}{\mathop{}\!\partial}
\renewcommand{\L}{\mathcal{L}}
\renewcommand{\O}[1]{\mathcal{O}\pa{#1}}

\DeclareMathOperator\arctanh{arctanh}
\DeclareMathOperator\arcsinh{arcsinh}
\DeclareMathOperator\dn{dn}
\DeclareMathOperator\sech{sech}
\DeclareMathOperator\sign{sign}
\DeclareMathOperator\sn{sn}

\makeatletter
\def\l@subsubsection#1#2{}
\makeatother

\begin{document}

\title{\vspace*{40pt}\huge{Particle motion near high-spin black holes}\vspace*{40pt}}

\author{Daniel Kapec}
\email{kapec@ias.edu}
\affiliation{School of Natural Sciences, Institute for Advanced Study, Princeton, NJ 08540, USA}
\author{Alexandru Lupsasca}
\email{lupsasca@fas.harvard.edu}
\affiliation{Center for the Fundamental Laws of Nature, Harvard University, Cambridge, MA 02138, USA}
\affiliation{Society of Fellows, Harvard University, Cambridge, MA 02138, USA}

\begin{abstract}
\vspace*{40pt}
General relativity predicts that the Kerr black hole develops qualitatively new and surprising features in the limit of maximal spin. Most strikingly, the region of spacetime near the event horizon stretches into an infinitely long throat and displays an emergent conformal symmetry. Understanding dynamics in this NHEK (Near-Horizon Extreme Kerr) geometry is necessary for connecting theory to upcoming astronomical observations of high-spin black holes. We review essential properties of NHEK and its relationship to the rapidly rotating Kerr black hole. We then completely solve the geodesic equation in the NHEK region and describe how the resulting trajectories transform under the action of its enhanced symmetries. In the process, we derive explicit expressions for the angular integrals appearing in the Kerr geodesic equation and obtain a useful formula, valid at arbitrary spin, for a particle's polar angle in terms of its radial motion. These results will aid in the analytic computation of astrophysical observables relevant to ongoing and future experiments.
\end{abstract}

\maketitle

\clearpage

\tableofcontents

\vspace{30pt}

\section{Introduction}

Extremal black holes have served as a rich source of novel ideas and techniques in quantum gravity and field theory for several decades \cite{Strominger1996,Maldacena1997,Guica2009,Maldacena2016a,Maldacena2016b}. These fundamental advances have led to a mathematical description of numerous interesting quantum-mechanical and gravitational systems, but have yet to connect directly with experiment. However, with the advent of a new generation of powerful astronomical detectors such as LIGO and the Event Horizon Telescope \cite{Abbott2016,EHT2019}, a subclass of astrophysically realistic near-extremal black holes stands poised to bridge this gap between formal theoretical investigation and successful experimental verification. The near-extremal Kerr black hole exhibits a number of striking phenomena showcasing strong-field general relativity, and a confirmation of even the most basic, qualitative prediction derived from the emergent symmetries of its near-horizon region would mark a huge success for both theory and experiment. If high-spin black holes do exist and come within observational reach, they will provide a window into a region of our Universe that is qualitatively similar to the extensively studied Anti-de Sitter (AdS) spacetime, which plays an outsize role in the modern holographic perspective on quantum gravity.

The traditional approach to the modeling of astrophysical black holes is based on extensive numerical simulation across large swaths of parameter space. While this method of analysis is perfectly adequate in the general setting, it must confront new complications that arise in the specific regime of high spin. As a black hole rotates faster, it develops an increasingly deep throat that nonetheless remains confined within a small coordinate distance from the event horizon. As a result, resolving near-horizon physics requires a spacetime mesh of increasingly fine resolution as the spin grows. Meanwhile, the overall size of the grid must remain large in order to accurately capture the asymptotically flat region far from the black hole. Eventually, this large separation of scales can incur a prohibitive computational cost. Fortunately, the same phenomenon that renders the problem numerically intractable also enables the application of a complementary analytic method.

The analytic approach proceeds from the key observation that near extremality, the Kerr spacetime separates into two distinct regions. While the extreme Kerr metric resolves physics away from the horizon, the throat region is instead described by the dramatically simpler Near-Horizon Extreme Kerr (NHEK) geometry, which possesses additional symmetries and can be viewed as a spacetime in its own right. The enhanced symmetry of the NHEK metric fixes the behavior of many near-horizon processes and turns complex dynamical questions into considerably simpler kinematic ones \cite{Lupsasca2014,Zhang2014,Lupsasca2015,Compere2015,Gralla2016a,Chen2017,Gralla2017b,Gralla2018,Hadar2019}. Once a problem has been solved in the NHEK geometry, it can be matched onto the far region in order to produce astrophysical predictions. Indeed, there is a growing body of work that seeks to exploit the symmetries of the NHEK region to constrain physical observables targeted by ongoing and upcoming experiments \cite{Porfyriadis2014,Hadar2014,Hadar2015,Hadar2017,Compere2018,Gralla2015,Gralla2016b,Burko2016,Gralla2016c,Compere2017,Gralla2017a,Porfyriadis2017,Lupsasca2018,Gates2018}. Many, if not all, of these analyses require approximate or exact solutions to the geodesic equation in NHEK and Kerr. While Kerr geodesic motion is an old and well-studied subject, most treatments reduce the problem to first-order form and then simply seek numerical integration of the equations. On the other hand, NHEK geodesic motion has received far less attention. Previous work has focused primarily on equatorial geodesics and those geodesics obtainable from the equatorial case through a symmetry transformation \cite{Porfyriadis2014,Hadar2014,Hadar2015,Hadar2017,Compere2018}. In this paper, we present drastically simplified analytic expressions for the angular geodesic integrals in Kerr, and solve outright the full geodesic equation in NHEK. Our results will directly aid in the calculation of gravitational wave and electromagnetic signals from non-equatorial sources in the NHEK region, which are expected to be relevant for both current and future experiments.

The outline of this paper is as follows. In Sec.~\ref{sec:Kerr}, we revisit the problem of geodesic motion in the Kerr spacetime for arbitrary values of the black hole spin, and derive new, improved expressions for the motion in the poloidal $(r,\theta)$ plane. In Sec.~\ref{sec:NHEK}, we present a pedagogical review of the NHEK geometry and its origin as the near-horizon scaling limit of a (near-)extreme Kerr black hole. We then completely solve the NHEK geodesic equation in Sec.~\ref{sec:GeodesicsInNHEK}, working first in the global strip, then in the Poincar\'e patch, and finally in the near-NHEK patch. We classify all categories of geodesic motion in each coordinate system and obtain explicit formulas for the motion as a function of coordinate time in each case. We conclude with a description of how NHEK geodesics transform under the action of the $\mathsf{SL}(2,\mathbb{R})\times\mathsf{U}(1)$ isometry group. Appendix~\ref{app:EllipticIntegrals} gathers mathematical definitions needed in the main body of the text. Appendix~\ref{app:AdS2} describes the dimensional reduction of the NHEK geometry to AdS$_2$ with a constant electric field, and the projection of NHEK geodesics to the trajectories of charged particles in AdS$_2$ subject to the Lorentz force exerted by the background electromagnetic field.

\section{Geodesics in Kerr}
\label{sec:Kerr}

In this section, we review the standard treatment of geodesic motion in the background of a rotating black hole \cite{Carter1968,Bardeen1973,Chandrasekhar1983,ONeill1995}. We begin in Sec.~\ref{subsec:KerrGeodesics} by re-deriving the Kerr geodesic equation in its first-order formulation. Then, we classify the different possible qualitative behaviors of the polar motion in Sec.~\ref{subsec:QualitativeDescription}, before explicitly evaluating all the angular path integrals appearing in the geodesic equation in Sec.~\ref{subsec:Computation}. We take great care to unpack these integrals into sums of manifestly real and positive elliptic integrals, each of which is represented in Legendre canonical form. This results in compact expressions that are appreciably simpler than those previously given in the literature \cite{Rauch1994,Vazquez2004,Kraniotis2005,Dexter2009,Fujita2009,Kraniotis2011,Hackmann2010,Hackmann2014}, which either did not explicitly unpack the path integrals or did not reduce them to manifestly real and positive expressions. Our formulas then allow us to obtain in Sec.~\ref{subsec:PolarInversion} a simple expression for the polar angle as a function of the radial motion. Readers solely interested in the end results may skip directly to Sec.~\ref{subsec:KerrSummary} for a succinct summary.

\subsection{The Kerr geodesic equation in first-order form}
\label{subsec:KerrGeodesics}

The Kerr metric describes astrophysically realistic rotating black holes of mass $M$ and angular momentum $J=aM$. In Boyer-Lindquist coordinates $\pa{t,r,\theta,\phi}$ and natural units where $G=c=1$, the Kerr line element is
\begin{subequations}
\label{eq:Kerr}
\begin{gather}
	ds^2=-\frac{\Delta}{\Sigma}\pa{\ed t-a\sin^2{\theta}\ed\phi}^2+\frac{\Sigma}{\Delta}\ed r^2+\Sigma\ed\theta^2+\frac{\sin^2{\theta}}{\Sigma}\br{\pa{r^2+a^2}\ed\phi-a\ed t}^2,\\
	\Delta(r)=r^2-2Mr+a^2,\qquad
	\Sigma(r,\theta)=r^2+a^2\cos^2{\theta}.
\end{gather}
\end{subequations}
This metric admits two Killing vectors $\pd_t$ and $\pd_\phi$ generating time-translation symmetry and axisymmetry, respectively. In addition to these isometries, the Kerr metric also admits an irreducible symmetric Killing tensor\footnote{A Killing tensor satisfies $\nabla_{(\lambda}K_{\mu\nu)}=0$. The antisymmetric tensor $J_{\mu\nu}=-J_{\nu\mu}$ satisfies the Killing-Yano equation $\nabla_{(\lambda}J_{\mu)\nu}=0$.}
\begin{align}
	\label{eq:KerrKilling}
	K_{\mu\nu}=-{J_\mu}^\lambda J_{\lambda\nu},\qquad
	J=a\cos{\theta}\ed r\wedge\pa{\ed t-a\sin^2{\theta}\ed\phi}+r\sin{\theta}\ed\theta\wedge\br{\pa{r^2+a^2}\ed\phi-a\ed t}.
\end{align}

The motion of a free particle of mass $\mu$ and four-momentum $p^\mu$ is described by the geodesic equation,
\begin{align}
	p^\mu\nabla_\mu p^\nu=0,\qquad
	g^{\mu\nu}p_\mu p_\nu=-\mu^2.
\end{align}
In the Kerr geometry \eqref{eq:Kerr}, geodesic motion is completely characterized by three conserved quantities,
\begin{align}
	\omega=p_\mu\pd_t^\mu
	=-p_t,\qquad
	\ell=p_\mu\pd_\phi^\mu
	=p_\phi,\qquad
	k=K^{\mu\nu}p_\mu p_\nu
	=p_\theta^2+a^2\mu^2\cos^2{\theta}+\pa{p_\phi\csc{\theta}+p_ta\sin{\theta}}^2,
\end{align}
denoting the total energy, angular momentum parallel to the axis of symmetry, and Carter constant, respectively. The first two quantities are the conserved charges associated with the Killing vectors $\pd_t$ and $\pd_\phi$, respectively, whereas the conservation of the third quantity follows from the existence of the Killing tensor \eqref{eq:KerrKilling}. While the Carter constant has the advantage of being manifestly positive, it is often useful to work instead with the so-called Carter integral
\begin{align}
	Q=k-\pa{\ell-a\omega}^2
	=p_\theta^2+a^2\pa{\mu^2-p_t^2}\cos^2{\theta}+p_\phi^2\cot^2{\theta}.
\end{align}

By inverting the above relations for $\pa{\mu^2,\omega,\ell,k}$, we find that a particle following a geodesic in the Kerr geometry \eqref{eq:Kerr} has an instantaneous four-momentum $p=p_\mu\ed x^\mu$ of the form
\begin{align}
	p(x^\mu,\omega,\ell,k)=-\omega\ed t\pm_r\frac{\sqrt{\mathcal{R}(r)}}{\Delta}\ed r\pm_\theta\sqrt{\Theta(\theta)}\ed\theta+\ell\ed\phi,
\end{align}
where the two choices of sign $\pm_r$ and $\pm_\theta$ depend on the radial and polar directions of travel, respectively. Here, we also introduced radial and polar potentials
\begin{subequations}
\begin{align}
	\mathcal{R}(r)&=\br{\omega\pa{r^2+a^2}-a\ell}^2-\Delta\pa{k+\mu^2r^2},\\
	\Theta(\theta)&=k-a^2\mu^2\cos^2{\theta}-\pa{\ell\csc{\theta}-a\omega\sin{\theta}}^2.
\end{align}
\end{subequations}
One can then raise $p_\mu$ to obtain the equations for the geodesic trajectory,
\begin{subequations}
\begin{align}
	\label{eq:RadialGeodesicEquation}
	\Sigma\frac{dr}{d\sigma}&=\pm_r\sqrt{\mathcal{R}(r)},\\
	\label{eq:AngularGeodesicEquation}
	\Sigma\frac{d\theta}{d\sigma}&=\pm_\theta\sqrt{\Theta(\theta)},\\
	\Sigma\frac{d\phi}{d\sigma}&=\frac{a}{\Delta}\br{\omega\pa{r^2+a^2}-a\ell}+\ell\csc^2{\theta}-a\omega,\\
	\label{eq:TimeGeodesicEquation}
	\Sigma\frac{dt}{d\sigma}&=\frac{\pa{r^2+a^2}}{\Delta}\br{\omega\pa{r^2+a^2}-a\ell}+a\pa{\ell-a\omega\sin^2{\theta}}.
\end{align}
\end{subequations}
The parameter $\sigma$ is the affine parameter for massless particles ($\mu=0$), and is related to the proper time $\delta$ by $\delta=\mu\sigma$ for massive particles. This system is completely integrable because it admits as many constants of motion as momentum variables, and can be integrated by quadratures. To do so, first note that
\begin{align}
	\frac{1}{\pm_r\sqrt{\mathcal{R}(r)}}\frac{dr}{d\sigma}=\frac{1}{\Sigma}=\frac{1}{\pm_\theta\sqrt{\Theta(\theta)}}\frac{d\theta}{d\sigma}.
\end{align}
Integration along the geodesic from $\sigma=\sigma_s$ to $\sigma=\sigma_o$ yields
\begin{align}
	\fint_{\sigma_s}^{\sigma_o}\frac{1}{\pm_r\sqrt{\mathcal{R}(r)}}\frac{dr}{d\sigma}\ed\sigma=\fint_{\sigma_s}^{\sigma_o}\frac{1}{\pm_\theta\sqrt{\Theta(\theta)}}\frac{d\theta}{d\sigma}\ed\sigma,
\end{align}
where the slash notation $\fint$ indicates that these integrals are to be evaluated along the geodesic, with turning points in the radial or polar motion occurring whenever the corresponding potential $\mathcal{R}(r)$ or $\Theta(\theta)$ vanishes. By definition, the signs $\pm_r$ and $\pm_\theta$ in front of $\sqrt{\mathcal{R}(r)}$ and $\sqrt{\Theta(\theta)}$ are always the same as that of $\ed r$ and $\ed\theta$, respectively, so these integrals grow secularly (rather than canceling out) over multiple oscillations.

If the particle is located at $\pa{t_s,r_s,\theta_s,\phi_s}$ when $\sigma=\sigma_s$ and at $\pa{t_o,r_o,\theta_o,\phi_o}$ when $\sigma=\sigma_o$, then this simplifies to
\begin{align}
	\fint_{r_s}^{r_o}\frac{\ed r}{\pm_r\sqrt{\mathcal{R}(r)}}=\fint_{\theta_s}^{\theta_o}\frac{\ed\theta}{\pm_\theta\sqrt{\Theta(\theta)}}.
\end{align}
Likewise,
\begin{subequations}
\begin{align}
	\phi_o-\phi_s&=\fint_{\phi_s}^{\phi_o}\ed\phi
	=\fint_{\sigma_s}^{\sigma_o}\frac{d\phi}{d\sigma}\ed\sigma
	=\fint_{\sigma_s}^{\sigma_o}\cu{\frac{a}{\Delta}\br{\omega\pa{r^2+a^2}-a\ell}+\ell\csc^2{\theta}-a\omega}\frac{\ed\sigma}{\Sigma}\\
	&=\fint_{\sigma_s}^{\sigma_o}\cu{\frac{a}{\Delta}\br{\omega\pa{r^2+a^2}-a\ell}-a\omega}\frac{\ed\sigma}{\Sigma}
	+\fint_{\sigma_s}^{\sigma_o}\ell\csc^2{\theta}\frac{\ed\sigma}{\Sigma}.
\end{align}
\end{subequations}
Aside from $\Sigma$, the first integrand only contains $r$-dependent terms and the second integrand only contains $\theta$-dependent terms. Thus, we naturally replace $\Sigma$ using Eq.~\eqref{eq:RadialGeodesicEquation} in the first integral and using Eq.~\eqref{eq:AngularGeodesicEquation} in the second, resulting in
\begin{align}
	\phi_o-\phi_s=\fint_{r_s}^{r_o}\cu{\frac{a}{\Delta}\br{\omega\pa{r^2+a^2}-a\ell}-a\omega}\frac{\ed r}{\pm_r\sqrt{\mathcal{R}(r)}}+\fint_{\theta_s}^{\theta_o}\frac{\ell\csc^2{\theta}}{\pm_\theta\sqrt{\Theta(\theta)}}\ed\theta.
\end{align}
After repeating the same procedure for $t$ and shuffling constant pieces from one integral into the other, we find that
\begin{subequations}
\begin{align}
	t_o-t_s&=\fint_{t_s}^{t_o}\ed t
	=\fint_{\sigma_s}^{\sigma_o}\frac{dt}{d\sigma}\ed\sigma
	=\fint_{\sigma_s}^{\sigma_o}\cu{\frac{\pa{r^2+a^2}}{\Delta}\br{\omega\pa{r^2+a^2}-a\ell}+a\pa{\ell-a\omega\sin^2{\theta}}}\frac{\ed\sigma}{\Sigma}\\
	&=\fint_{\sigma_s}^{\sigma_o}\cu{\frac{\pa{r^2+a^2}}{\Delta}\br{\omega\pa{r^2+a^2}-a\ell}+a\pa{\ell-a\omega}}\frac{\ed\sigma}{\Sigma}+\fint_{\sigma_s}^{\sigma_o}a^2\omega\cos^2{\theta}\frac{\ed\sigma}{\Sigma}\\
	&=\fint_{r_s}^{r_o}\cu{\frac{\pa{r^2+a^2}}{\Delta}\br{\omega\pa{r^2+a^2}-a\ell}+a\pa{\ell-a\omega}}\frac{\ed r}{\pm_r\sqrt{\mathcal{R}(r)}}+\fint_{\theta_s}^{\theta_o}\frac{a^2\omega\cos^2{\theta}}{\pm_\theta\sqrt{\Theta(\theta)}}\ed\theta.
\end{align}
\end{subequations}
To summarize, a geodesic labeled by $(\omega,\ell,k)$ connects spacetime points $x_s^\mu=\pa{t_s,r_s,\theta_s,\phi_s}$ and $x_o^\mu=\pa{t_o,r_o,\theta_o,\phi_o}$ if
\begin{subequations}
\label{eq:KerrGeodesicEquation}
\begin{align}
	&\fint_{r_s}^{r_o}\frac{\ed r}{\pm_r\sqrt{\mathcal{R}(r)}}=\fint_{\theta_s}^{\theta_o}\frac{\ed\theta}{\pm_\theta\sqrt{\Theta(\theta)}},\\
	\phi_o-\phi_s&=\fint_{r_s}^{r_o}\cu{\frac{a}{\Delta}\br{\omega\pa{r^2+a^2}-a\ell}-a\omega}\frac{\ed r}{\pm_r\sqrt{\mathcal{R}(r)}}+\fint_{\theta_s}^{\theta_o}\frac{\ell\csc^2{\theta}}{\pm_\theta\sqrt{\Theta(\theta)}}\ed\theta,\\
	t_o-t_s&=\fint_{r_s}^{r_o}\cu{\frac{\pa{r^2+a^2}}{\Delta}\br{\omega\pa{r^2+a^2}-a\ell}+a\pa{\ell-a\omega}}\frac{\ed r}{\pm_r\sqrt{\mathcal{R}(r)}}+\fint_{\theta_s}^{\theta_o}\frac{a^2\omega\cos^2{\theta}}{\pm_\theta\sqrt{\Theta(\theta)}}\ed\theta.
\end{align}
\end{subequations}
Generically, Kerr geodesics undergo multiple librations (polar oscillations) and rotations about the axis of symmetry. They may also undergo radial oscillations when they are bound ($\ab{\omega}<\mu$) \cite{Wilkins1972}. Kerr geodesics are therefore characterized by integers $(w,m,n)$ denoting the number of turning points in the radial motion, the number of turning points in the polar motion, and the winding number about the axis of symmetry, respectively.

\subsection{Qualitative description of the polar motion}
\label{subsec:QualitativeDescription}

We now wish to compute the angular integrals that appear in the Kerr geodesic equation \eqref{eq:KerrGeodesicEquation},
\begin{align}
	G_\theta=\fint_{\theta_s}^{\theta_o}\frac{\ed\theta}{\pm_\theta\sqrt{\Theta(\theta)}},\qquad
	G_\phi=\fint_{\theta_s}^{\theta_o}\frac{\csc^2{\theta}}{\pm_\theta\sqrt{\Theta(\theta)}}\ed\theta,\qquad
	G_t=\fint_{\theta_s}^{\theta_o}\frac{\cos^2{\theta}}{\pm_\theta\sqrt{\Theta(\theta)}}\ed\theta,
\end{align}
and then solve for the final angle $\theta_o$ in the $(r,\theta)$ part of the equation, which is of the form $I_r=G_\theta$ with
\begin{align}
	I_r=\fint_{r_s}^{r_o}\frac{\ed r}{\pm_r\sqrt{\mathcal{R}(r)}}.
\end{align}
To do so, it is convenient to rewrite the angular potential as
\begin{align}
	\Theta(\theta)=Q+P\cos^2{\theta}-\ell^2\cot^2{\theta},\qquad
	P=a^2\pa{\omega^2-\mu^2}.
\end{align}
There are three qualitatively different cases that we will consider in turn:
\begin{enumerate}
\item
\underline{$P>0$} corresponds to null geodesics with $\mu=0$, or unbound timelike geodesics with $0<\mu^2<\omega^2$.
\item
\underline{$P=0$} corresponds to marginally bound timelike geodesics with $0<\mu^2=\omega^2$.
\item
\underline{$P<0$} corresponds to bound timelike geodesics with $0\le\omega^2<\mu^2$.
\end{enumerate}
Here, the bound/unbound nomenclature refers to the radial motion---the polar motion is of course always bounded.

The positivity condition $\Theta(\theta)\ge0$ implies that a geodesic can reach a pole at $\theta_N=0$ or $\theta_S=\pi$ if and only if $\ell=0$. We will assume the genericity condition $\ell\neq0$, in which case the polar motion is strictly restricted to oscillations bounded by turning points $\theta_\pm\in\pa{0,\pi}$.\footnote{The special case $\ell=0$ needs to be treated separately, as one must account for the possibility that geodesics may climb over the black hole and pass through the rotation axis.} This oscillatory motion can be of two qualitatively different types:
\begin{itemize}\setlength{\itemindent}{.5in}
\item[\underline{Type A:}]
Oscillatory motion about the equatorial plane between $\theta_-\in\pa{0,\pi/2}$ and $\theta_+\in\pa{\pi/2,\pi}$ with $\theta_+=\pi-\theta_-$.
\item[\underline{Type B:}]
``Vortical" motion between turning points
$0<\theta_-<\theta_+<\pi/2$ or $\pi/2<\theta_+<\theta_-<\pi$, corresponding to geodesics that never cross the equatorial plane and are instead confined to a cone lying either entirely above or entirely below the equatorial plane \cite{deFelice1972}.
\end{itemize}
The Kerr geometry also admits planar geodesics at any fixed polar angle $\theta_0$. These arise in the special limit where the turning points of the angular motion coalesce at $\theta_0=\theta_\pm$. In that case, the geodesic equation \eqref{eq:KerrGeodesicEquation} degenerates and a separate treatment is necessary. In the null case, the planar geodesics are the well-known principal null congruences, which endow the Kerr geometry with many of its special properties.\footnote{The Kerr principal null congruences are shear-free. By the Goldberg-Sachs theorem, this implies that the Kerr spacetime is algebraically special of Petrov Type D. This property guarantees the existence of the Killing-Yano tensor \cite{Stephani1978}, from which many of the special properties of the Kerr geometry---including the separability of the wave equation and integrability of geodesic motion---are derived.} We presently exclude this fine-tuned situation, which has been extensively studied in the literature \cite{Chandrasekhar1983}. To study the two generic types, it is useful to define signs
\begin{subequations}
\begin{alignat}{2}
	\eta_o&=\sign\pa{p_o^\theta}\sign\pa{\cos{\theta}_o}\hspace{10em}
	\eta_s&&=\sign\pa{p_s^\theta}\sign\pa{\cos{\theta}_s}\\
	&=(-1)^m\sign\pa{p_s^\theta}\sign\pa{\cos{\theta}_o},
	&&=(-1)^m\sign\pa{p_o^\theta}\sign\pa{\cos{\theta}_s},
	\end{alignat}
\end{subequations}
where $p_s^\theta$ and $p_o^\theta$ denote the polar momentum evaluated at the endpoints $x_s^\mu$ and $x_o^\mu$ of the geodesic, respectively. By working through all the possible configurations, one can check that the angular path integral unpacks as follows:
\begin{subequations}
\begin{align}
	\text{Type A:}\qquad&
	\fint_{\theta_s}^{\theta_o}=2m\ab{\int_{\pi/2}^{\theta_\pm}}+\eta_s\ab{\int_{\pi/2}^{\theta_s}}-\eta_o\ab{\int_{\pi/2}^{\theta_o}},\\
	\text{Type B:}\qquad&
	\fint_{\theta_s}^{\theta_o}=\pa{m\pm\eta_o\frac{1-(-1)^m}{2}}\ab{\int_{\theta_-}^{\theta_+}}\pm\eta_s\ab{\int_{\theta_\pm}^{\theta_s}}\mp\eta_o\ab{\int_{\theta_\pm}^{\theta_o}},
\end{align}
\end{subequations}
where for both types, we presented two equivalent representations that differ only in the choice of turning point taken as a reference for the integrals. It will turn out that the type of oscillation is picked out by the sign of $Q$:
\begin{enumerate}
\item[i.]
\underline{$Q>0$} corresponds to Type A oscillations. These are allowed for all signs of $P$.
\item[ii.]
\underline{$Q<0$} corresponds to Type B (vortical) oscillations. These are only allowed for $P>0$.\footnote{This is because Eq.~\eqref{eq:AngularGeodesicEquation} implies that $Q=\pa{\Sigma p^\theta}^2-\cos^2{\theta}\pa{P-\ell^2\csc^2{\theta}}$. Thus $Q\le0$ is only possible if $P-\ell^2\csc^2{\theta}\ge0$, which requires that $P\ge\ell^2\csc^2{\theta}\ge\ell^2>0$.}
\item[iii.]
\underline{$Q=0$} corresponds to a singular limit of Type B motion in which the cone of oscillation touches the equatorial plane, where the integrals develop a nonintegrable singularity. Such geodesics are also only allowed for $P>0$.
\end{enumerate}

From now on, we will work with the variable $u=\cos^2{\theta}$, in terms of which
\begin{align}
	\Theta(u)=
	\begin{cases}
		\displaystyle\frac{P}{1-u}\pa{u_+-u}\pa{u-u_-}&P\neq0,\vspace{2pt}\\
		\displaystyle\frac{Q+\ell^2}{1-u}\pa{u_0-u}&P=0.
	\end{cases}
\end{align}
Here we defined
\begin{align}
	\Delta_\theta=\frac{1}{2}\pa{1-\frac{Q+\ell^2}{P}},\qquad
	u_\pm=\Delta_\theta\pm\sqrt{\Delta_\theta^2+\frac{Q}{P}},\qquad
	u_0=\frac{Q}{Q+\ell^2}.
\end{align}
For future convenience, we also introduce the quantities
\begin{align}
	\Psi_j^\pm=\arcsin\sqrt{\frac{\cos^2{\theta_j}}{u_\pm}},\qquad
	\Upsilon_j^\pm=\pm\arcsin\sqrt{\pm\frac{\cos^2{\theta_j}-u_\mp}{u_+-u_-}}.
\end{align}

\subsection{Computation of the angular integrals}
\label{subsec:Computation}

\subsubsection{Case 1: \texorpdfstring{$P=0$}{P=0}}

In this case, we must necessarily have $Q>0$ and $0\le\cos^2{\theta}\le u_0<1$. Hence, the oscillation is of Type A with turning points at $\theta_\mp=\arccos{\pm\sqrt{u_0}}$. The only integrals we need are
\begin{subequations}
\begin{align}
	\ab{\int_{\pi/2}^{\theta_j}\frac{\ed\theta}{\sqrt{\Theta(\theta)}}}&=\frac{1}{2}\sqrt{\frac{u_0}{Q}}\int_0^{u_j}\frac{\ed u}{\sqrt{u\pa{u_0-u}}}
	=\sqrt{\frac{u_0}{Q}}\arcsin\sqrt{\frac{u_j}{u_0}},\\
	\ab{\int_{\pi/2}^{\theta_j}\frac{\csc^2{\theta}}{\sqrt{\Theta(\theta)}}\ed\theta}&=\frac{1}{2}\sqrt{\frac{u_0}{Q}}\int_0^{u_j}\frac{\ed u}{\pa{1-u}\sqrt{u\pa{u_0-u}}}
	=\frac{1}{\sqrt{Q}}\sqrt{\frac{u_0}{1-u_0}}\arcsin\sqrt{\frac{u_j}{u_0}\pa{\frac{1-u_0}{1-u_j}}},\\
	\ab{\int_{\pi/2}^{\theta_j}\frac{\cos^2{\theta}}{\sqrt{\Theta(\theta)}}\ed\theta}&=\frac{1}{2}\sqrt{\frac{u_0}{Q}}\int_0^{u_j}\frac{u\ed u}{\sqrt{u\pa{u_0-u}}}
	=\frac{1}{2}\sqrt{\frac{u_0}{Q}}\br{u_0\arcsin\sqrt{\frac{u_j}{u_0}}-\sqrt{u_j\pa{u_0-u_j}}},
\end{align}
\end{subequations}
where, in order to ensure that each integral is real and positive, we used the substitution
\begin{align}
	u=u_0t^2.
\end{align}
Thus, in the $P=0$ case (where necessarily $Q>0$), we obtain
\begin{subequations}
\begin{empheq}[box=\ovalbox]{align}
	\shortintertext{\centering $P=0$ ($Q>0$ required):}
	\label{eq:P=0,Q>0}
	G_\theta&=\sqrt{\frac{u_0}{Q}}\br{\pi m+\eta_s\arcsin\sqrt{\frac{u_s}{u_0}}-\eta_o\arcsin\sqrt{\frac{u_o}{u_0}}},\\
	G_\phi&=\frac{1}{\sqrt{Q}}\sqrt{\frac{u_0}{1-u_0}}\br{\pi m+\eta_s\arcsin\sqrt{\frac{u_s}{u_0}\pa{\frac{1-u_0}{1-u_s}}}-\eta_o\arcsin\sqrt{\frac{u_o}{u_0}\pa{\frac{1-u_0}{1-u_o}}}},\\
	G_t&=\frac{1}{2}\cu{u_0G_\theta-\sqrt{\frac{u_0}{Q}}\br{\eta_s\sqrt{u_s\pa{u_0-u_s}}-\eta_o\sqrt{u_o\pa{u_0-u_o}}}}.
\end{empheq}
\end{subequations}

\subsubsection{Case 2: \texorpdfstring{$P<0$}{P<0}}

In this case, we must necessarily have $Q>0$ and $0\le\cos^2{\theta}\le u_-<1$. Hence, the oscillation is of Type A with turning points at $\theta_\mp=\arccos{\pm\sqrt{u_-}}$. The only integrals we need are
\begin{subequations}
\begin{align}
	\ab{\int_{\pi/2}^{\theta_j}\frac{\ed\theta}{\sqrt{\Theta(\theta)}}}&=\frac{1}{2\sqrt{-P}}\int_0^{u_j}\frac{\ed u}{\sqrt{u\pa{u_+-u}\pa{u_--u}}}
	=\frac{1}{\sqrt{-u_+P}}F\pa{\Psi_j^-\left|\frac{u_-}{u_+}\right.},\\
	\ab{\int_{\pi/2}^{\theta_j}\frac{\csc^2{\theta}}{\sqrt{\Theta(\theta)}}\ed\theta}&=\frac{1}{2\sqrt{-P}}\int_0^{u_j}\frac{\ed u}{\pa{1-u}\sqrt{u\pa{u_+-u}\pa{u_--u}}}
	=\frac{1}{\sqrt{-u_+P}}\Pi\pa{u_-;\Psi_j^-\left|\frac{u_-}{u_+}\right.},\\
	\ab{\int_{\pi/2}^{\theta_j}\frac{\cos^2{\theta}}{\sqrt{\Theta(\theta)}}\ed\theta}&=\frac{1}{2\sqrt{-P}}\int_0^{u_j}\frac{u\ed u}{\sqrt{u\pa{u_+-u}\pa{u_--u}}}
	=-\frac{2u_-}{\sqrt{-u_+P}}E'\pa{\Psi_j^-\left|\frac{u_-}{u_+}\right.},
\end{align}
\end{subequations}
where we defined $E'(x|k)\equiv\pd_kE(x|k)=\br{E(x|k)-F(x|k)}/(2k)$ and, in order to ensure that each integral is real and positive, we used the substitution
\begin{align}
	u=u_-t^2.
\end{align}
Thus, in the $P<0$ case (where necessarily $Q>0$), we obtain
\begin{subequations}
\begin{empheq}[box=\ovalbox]{align}
	\shortintertext{\centering $P<0$ ($Q>0$ required):}
	\label{eq:P<0,Q>0}
	G_\theta&=\frac{1}{\sqrt{-u_+P}}\br{2mK\pa{\frac{u_-}{u_+}}+\eta_sF\pa{\Psi_s^-\left|\frac{u_-}{u_+}\right.}-\eta_oF\pa{\Psi_o^-\left|\frac{u_-}{u_+}\right.}},\\
	G_\phi&=\frac{1}{\sqrt{-u_+P}}\br{2m\Pi\pa{u_-\left|\frac{u_-}{u_+}\right.}+\eta_s\Pi\pa{u_-;\Psi_s^-\left|\frac{u_-}{u_+}\right.}-\eta_o\Pi\pa{u_-;\Psi_o^-\left|\frac{u_-}{u_+}\right.}},\\
	G_t&=-\frac{2u_-}{\sqrt{-u_+P}}\br{2mE'\pa{\frac{u_-}{u_+}}+\eta_sE'\pa{\Psi_s^-\left|\frac{u_-}{u_+}\right.}-\eta_oE'\pa{\Psi_o^-\left|\frac{u_-}{u_+}\right.}}.
\end{empheq}
\end{subequations}

\subsubsection{Case 3: \texorpdfstring{$P>0$}{P>0}}

If $Q>0$, then $u_-<0\le\cos^2{\theta}\le u_+<1$ and the oscillation is of Type A with turning points at $\theta_\mp=\arccos{\pm\sqrt{u_+}}$. The only integrals we need are
\begin{subequations}
\begin{align}
	\ab{\int_{\pi/2}^{\theta_j}\frac{\ed\theta}{\sqrt{\Theta(\theta)}}}&=\frac{1}{2\sqrt{P}}\int_0^{u_j}\frac{\ed u}{\sqrt{u\pa{u_+-u}\pa{u-u_-}}}
	=\frac{1}{\sqrt{-u_-P}}F\pa{\Psi_j^+\left|\frac{u_+}{u_-}\right.},\\
	\ab{\int_{\pi/2}^{\theta_j}\frac{\csc^2{\theta}}{\sqrt{\Theta(\theta)}}\ed\theta}&=\frac{1}{2\sqrt{P}}\int_0^{u_j}\frac{\ed u}{\pa{1-u}\sqrt{u\pa{u_+-u}\pa{u-u_-}}}
	=\frac{1}{\sqrt{-u_-P}}\Pi\pa{u_+;\Psi_j^+\left|\frac{u_+}{u_-}\right.},\\
	\ab{\int_{\pi/2}^{\theta_j}\frac{\cos^2{\theta}}{\sqrt{\Theta(\theta)}}\ed\theta}&=\frac{1}{2\sqrt{P}}\int_0^{u_j}\frac{u\ed u}{\sqrt{u\pa{u_+-u}\pa{u-u_-}}}
	=-\frac{2u_+}{\sqrt{-u_-P}}E'\pa{\Psi_j^+\left|\frac{u_+}{u_-}\right.},
\end{align}
\end{subequations}
where, in order to ensure that each integral is real and positive, we used the substitution
\begin{align}
	u=u_+t^2.
\end{align}
Thus, in the $P>0$, $Q>0$ case, we obtain
\begin{subequations}
\begin{empheq}[box=\ovalbox]{align}
	\shortintertext{\centering $P>0$, $Q>0$:}
	\label{eq:P>0,Q>0}
	G_\theta&=\frac{1}{\sqrt{-u_-P}}\br{2mK\pa{\frac{u_+}{u_-}}+\eta_sF\pa{\Psi_s^+\left|\frac{u_+}{u_-}\right.}-\eta_oF\pa{\Psi_o^+\left|\frac{u_+}{u_-}\right.}},\\
	G_\phi&=\frac{1}{\sqrt{-u_-P}}\br{2m\Pi\pa{u_+\left|\frac{u_+}{u_-}\right.}+\eta_s\Pi\pa{u_+;\Psi_s^+\left|\frac{u_+}{u_-}\right.}-\eta_o\Pi\pa{u_+;\Psi_o^+\left|\frac{u_+}{u_-}\right.}},\\
	G_t&=-\frac{2u_+}{\sqrt{-u_-P}}\br{2mE'\pa{\frac{u_+}{u_-}}+\eta_sE'\pa{\Psi_s^+\left|\frac{u_+}{u_-}\right.}-\eta_oE'\pa{\Psi_o^+\left|\frac{u_+}{u_-}\right.}}.
\end{empheq}
\end{subequations}

If $Q<0$, then $0<u_-\le\cos^2{\theta}\le u_+<1$ and the oscillation is of Type B with turning points at $\theta_-=\arccos{\pm\sqrt{u_+}}$ and $\theta_+=\arccos{\pm\sqrt{u_-}}$, where the upper/lower sign corresponds to vortical oscillation within a cone lying entirely above/below the equatorial plane. If we use $\theta_\pm$ as the reference turning point ($i.e.$, in both hemispheres, we integrate from the turning point closest to/farthest from the equator), then the only integrals we need are
\begin{subequations}
\begin{align}
	\ab{\int_{\theta_\pm}^{\theta_j}\frac{\ed\theta}{\sqrt{\Theta(\theta)}}}&=\pm\frac{1}{2\sqrt{P}}\int_{u_\mp}^{u_j}\frac{\ed u}{\sqrt{u\pa{u_+-u}\pa{u-u_-}}}
	=\pm\frac{1}{\sqrt{u_\mp P}}F\pa{\Upsilon_j^\pm\left|1-\frac{u_\pm}{u_\mp}\right.},\\
	\ab{\int_{\theta_\pm}^{\theta_j}\frac{\csc^2{\theta}}{\sqrt{\Theta(\theta)}}\ed\theta}&=\pm\frac{1}{2\sqrt{P}}\int_{u_\mp}^{u_j}\frac{\ed u}{\pa{1-u}\sqrt{u\pa{u_+-u}\pa{u-u_-}}}
	=\pm\frac{1}{\pa{1-u_\mp}\sqrt{u_\mp P}}\Pi\pa{\frac{u_\pm-u_\mp}{1-u_\mp};\Upsilon_j^\pm\left|1-\frac{u_\pm}{u_\mp}\right.},\\
	\ab{\int_{\theta_\pm}^{\theta_j}\frac{\cos^2{\theta}}{\sqrt{\Theta(\theta)}}\ed\theta}&=\pm\frac{1}{2\sqrt{P}}\int_{u_\mp}^{u_j}\frac{u\ed u}{\sqrt{u\pa{u_+-u}\pa{u-u_-}}}
	=\pm\sqrt{\frac{u_\mp}{P}}E\pa{\Upsilon_j^\pm\left|1-\frac{u_\pm}{u_\mp}\right.},
\end{align}
\end{subequations}
where we used the substitution
\begin{align}
	u=u_\mp\pm\pa{u_+-u_-}t^2
\end{align}
in order to ensure that each integral is real and positive. Thus, in the $P>0$, $Q<0$ case, we obtain
\begin{subequations}
\begin{empheq}[box=\ovalbox]{align}
	\shortintertext{\centering $P>0$, $Q<0$:}
	\label{eq:P>0,Q<0}
	G_\theta&=\frac{1}{\sqrt{u_\mp P}}\br{\pa{m\pm\eta_o\frac{1-(-1)^m}{2}}K\pa{1-\frac{u_\pm}{u_\mp}}+\eta_sF\pa{\Upsilon_s^\pm\left|1-\frac{u_\pm}{u_\mp}\right.}-\eta_oF\pa{\Upsilon_o^\pm\left|1-\frac{u_\pm}{u_\mp}\right.}},\\
	G_\phi&=\frac{1}{\pa{1-u_\mp}\sqrt{u_\mp P}}\bigg[\pa{m\pm\eta_o\frac{1-(-1)^m}{2}}\Pi\pa{\frac{u_\pm-u_\mp}{1-u_\mp}\left|1-\frac{u_\pm}{u_\mp}\right.}\\
	&\qquad\qquad\qquad\qquad\qquad+\eta_s\Pi\pa{\frac{u_\pm-u_\mp}{1-u_\mp};\Upsilon_s^\pm\left|1-\frac{u_\pm}{u_\mp}\right.}-\eta_o\Pi\pa{\frac{u_\pm-u_\mp}{1-u_\mp};\Upsilon_o^\pm\left|1-\frac{u_\pm}{u_\mp}\right.}\bigg],\notag\\
	G_t&=\sqrt{\frac{u_\mp}{P}}\br{\pa{m\pm\eta_o\frac{1-(-1)^m}{2}}E\pa{1-\frac{u_\pm}{u_\mp}}+\eta_sE\pa{\Upsilon_s^\pm\left|1-\frac{u_\pm}{u_\mp}\right.}-\eta_oE\pa{\Upsilon_o^\pm\left|1-\frac{u_\pm}{u_\mp}\right.}}.
\end{empheq}
\end{subequations}
To be clear, in these equations, the choice of upper/lower sign leads to equivalent representations of the integrals corresponding to different choices of reference turning point.

If $Q=0$, then $0=u_-\le\cos^2{\theta}\le u_+<1$ and the oscillation appears to be of Type B with turning points at the equator and $\theta_-=\arccos{\pm\sqrt{u_+}}$, where the upper/lower sign corresponds to vortical oscillation within a cone lying entirely above/below the equatorial plane. However, in practice, the geodesic motion can only turn at $\theta_-$, as it is barred from reaching the equator at $u=0$, which corresponds to a nonintegrable singularity of the angular integrals. Hence, the complete motion can undergo at most one libration, $i.e.$, it can only have $m=0$ or $m=1$. Therefore, in this special situation,
\begin{align}
	\fint_{\theta_s}^{\theta_o}=\eta_o\ab{\int_{\theta_-}^{\theta_o}}-\eta_s\ab{\int_{\theta_-}^{\theta_s}}.
\end{align}
Thus, using $\theta_-$ as the reference turning point (which we must, in order to avoid the singularity at the equator), the only integrals we need are
\begin{subequations}
\begin{align}
	\ab{\int_{\theta_-}^{\theta_j}\frac{\ed\theta}{\sqrt{\Theta(\theta)}}}&=\frac{1}{2\sqrt{P}}\int_{u_j}^{u_+}\frac{\ed u}{u\sqrt{u_+-u}}
	=\frac{1}{\sqrt{u_+P}}\arctanh\sqrt{1-\frac{u_j}{u_+}},\\
	\ab{\int_{\theta_-}^{\theta_j}\frac{\csc^2{\theta}}{\sqrt{\Theta(\theta)}}\ed\theta}&=\frac{1}{2\sqrt{P}}\int_{u_j}^{u_+}\frac{\ed u}{\pa{1-u}u\sqrt{u_+-u}}
	=\frac{1}{\sqrt{u_+P}}\br{\arctanh\sqrt{1-\frac{u_j}{u_+}}+\sqrt{\frac{u_+}{1-u_+}}\arctan\sqrt{\frac{u_+-u_j}{1-u_+}}},\\
	\ab{\int_{\theta_-}^{\theta_j}\frac{\cos^2{\theta}}{\sqrt{\Theta(\theta)}}\ed\theta}&=\frac{1}{2\sqrt{P}}\int_{u_j}^{u_+}\frac{\ed u}{\sqrt{u_+-u}}
	=\sqrt{\frac{u_+-u_j}{P}},
\end{align}
\end{subequations}
where we did not need any substitution to obtain simpler trigonometric representations of the integrals, which are all real and positive. In conclusion, in the $P>0$, $Q=0$ case, we obtain
\begin{subequations}
\begin{empheq}[box=\ovalbox]{align}
	\shortintertext{\centering $P>0$, $Q=0$:}
	\label{eq:P>0,Q=0}
	G_\theta&=\frac{1}{\sqrt{u_+P}}\br{\eta_o\arctanh\sqrt{1-\frac{u_o}{u_+}}-\eta_s\arctanh\sqrt{1-\frac{u_s}{u_+}}},\\
	G_\phi&=G_\theta+\frac{1}{\sqrt{\pa{1-u_+}P}}\br{\eta_o\arctan\sqrt{\frac{u_+-u_o}{1-u_+}}-\eta_s\arctan\sqrt{\frac{u_+-u_s}{1-u_+}}},\\
	G_t&=\frac{1}{\sqrt{P}}\br{\eta_o\sqrt{u_+-u_o}-\eta_s\sqrt{u_+-u_s}}.
\end{empheq}
\end{subequations}

\subsection{Solution to the \texorpdfstring{$(r,\theta)$}{(r,theta)} equation}
\label{subsec:PolarInversion}

The $(r,\theta)$ part of the Kerr geodesic equation \eqref{eq:KerrGeodesicEquation} is of the form $I_r=G_\theta$, with
\begin{align}
	I_r=\fint_{r_s}^{r_o}\frac{\ed r}{\pm\sqrt{\mathcal{R}(r)}}.
\end{align}
We want to solve this equation for $\theta_o$. We will proceed by considering each case in turn, starting with the simplest.

In the $P=0$ case (where necessarily $Q>0$), Eq.~\eqref{eq:P=0,Q>0} tells us that
\begin{align}
	\sqrt{\frac{Q}{u_0}}I_r=\pi m+\eta_s\arcsin\sqrt{\frac{u_s}{u_0}}-\eta_o\arcsin\sqrt{\frac{u_o}{u_0}}.
\end{align}
Using the fact that $\arcsin\pa{-x}=-\arcsin{x}$ is an odd function, this can be rewritten
\begin{align}
	\arcsin\pa{\frac{\cos{\theta_o}}{\sqrt{u_0}}}=(-1)^m\br{\arcsin\pa{\frac{\cos{\theta_s}}{\sqrt{u_0}}}+\sign\pa{p_s^\theta}\pa{\pi m-\sqrt{\frac{Q}{u_0}}I_r}},
\end{align}
from which it follows that
\begin{align}
	\cos{\theta_o}=\sqrt{u_0}\sin\br{(-1)^mW_m},\qquad
	W_m=\arcsin\pa{\frac{\cos{\theta_s}}{\sqrt{u_0}}}+\sign\pa{p_s^\theta}\pa{\pi m-\sqrt{\frac{Q}{u_0}}I_r}.
\end{align}
This expression can be further simplified by noting that
\begin{align}
	W_m=W_{m-1}+\sign\pa{p_s^\theta}\pi.
\end{align}
Since the function $\sin{x}$ satisfies the periodicity condition $\sin\pa{x\pm\pi}=-\sin{x}$, it follows that
\begin{align}
	\sin\br{(-1)^mW_m}=\sin\br{(-1)^{m-1}W_{m-1}}
	=\sin{W_0},
\end{align}
from which we conclude that $\cos{\theta_o}$ is in fact independent of the number of turning points along the trajectory:
\begin{empheq}[box=\ovalbox]{align}
	\shortintertext{\centering $P=0$ ($Q>0$ required):}
	\cos{\theta_o}=\sqrt{u_0}\sin{W_0},\qquad
	W_0=\arcsin\pa{\frac{\cos{\theta_s}}{\sqrt{u_0}}}-\sign\pa{p_s^\theta}\sqrt{\frac{Q}{u_0}}I_r.
\end{empheq}

We now turn to the remaining cases (with $P\neq0$), which are slightly more complicated but can nonetheless be treated using a similar approach. As a preliminary, note that because $\arcsin\pa{-x}=-\arcsin{x}$ is an odd function,
\begin{align}
	\label{eq:SignAbsorption}
	\sign\pa{\cos{\theta_j}}\Psi_j^\pm=\arcsin\pa{\frac{\cos{\theta_j}}{\sqrt{u_\pm}}}.
\end{align}

In the $P\neq0$ case with $Q>0$, Eqs.~\eqref{eq:P<0,Q>0} and \eqref{eq:P>0,Q>0} tell us that
\begin{align}
	\sqrt{-u_\mp P}I_r=2mK\pa{\frac{u_\pm}{u_\mp}}+\eta_sF\pa{\Psi_s^\pm\left|\frac{u_\pm}{u_\mp}\right.}-\eta_oF\pa{\Psi_o^\pm\left|\frac{u_\pm}{u_\mp}\right.},
\end{align}
where $\pm=\sign\pa{P}$. Using Eq.~\eqref{eq:SignAbsorption}, this can be rewritten
\begin{align}
	\label{eq:IntermediateStep+}
	F\pa{\arcsin\pa{\frac{\cos{\theta_o}}{\sqrt{u_\pm}}}\left|\frac{u_\pm}{u_\mp}\right.}=(-1)^mX_m^\pm,
\end{align}
where we defined
\begin{align}
	X_m^\pm=F\pa{\arcsin\pa{\frac{\cos{\theta_s}}{\sqrt{u_\pm}}}\left|\frac{u_\pm}{u_\mp}\right.}+\sign\pa{p_s^\theta}\br{2mK\pa{\frac{u_\pm}{u_\mp}}-\sqrt{-u_\mp P}I_r},
\end{align}
and used the fact that $F(-x|k)=-F(x|k)$ is odd in its first argument. The inverse function of the elliptic integral of the first kind is the Jacobi elliptic function $\sn(x|k)$, which satisfies $\sn(F(\arcsin{x}|k)|k)=x$. Using this identity, it immediately follows from Eq.~\eqref{eq:IntermediateStep+} that
\begin{align}
	\frac{\cos{\theta_o}}{\sqrt{u_\pm}}=\sn\pa{(-1)^mX_m^\pm\left|\frac{u_\pm}{u_\mp}\right.}.
\end{align}
This expression can be further simplified by noting that
\begin{align}
	X_m^\pm=X_{m-1}^\pm+2\sign\pa{p_s^\theta}K\pa{\frac{u_\pm}{u_\mp}}.
\end{align}
Since the function $\sn(x|k)$ satisfies the periodicity condition $\sn(x\pm2K(k)|k)=-\sn(x|k)$, it follows that
\begin{align}
	\sn\pa{(-1)^mX_m^\pm\left|\frac{u_\pm}{u_\mp}\right.}=\sn\pa{(-1)^{m-1}X_{m-1}^\pm\left|\frac{u_\pm}{u_\mp}\right.}
	=\sn\pa{X_0^\pm\left|\frac{u_\pm}{u_\mp}\right.},
\end{align}
from which we again conclude that $\cos{\theta_o}$ is independent of the number of turning points along the trajectory:
\begin{empheq}[box=\ovalbox]{align}
	\shortintertext{\centering $P\neq0$, $Q>0$ with $\pm=\sign\pa{P}$:}
	\label{eq:ObserverAngleQ>0}
	\cos{\theta_o}=\sqrt{u_\pm}\sn\pa{X_0^\pm\left|\frac{u_\pm}{u_\mp}\right.},\qquad
	X_0^\pm=F\pa{\arcsin\pa{\frac{\cos{\theta_s}}{\sqrt{u_\pm}}}\left|\frac{u_\pm}{u_\mp}\right.}-\sign\pa{p_s^\theta}\sqrt{-u_\mp P}I_r.
\end{empheq}

In the $Q<0$ case (where necessarily $P>0$), Eq.~\eqref{eq:P>0,Q<0} tells us that 
\begin{align}
	\sqrt{u_\mp P}I_r=\pa{m\pm\eta_o\frac{1-(-1)^m}{2}}K\pa{1-\frac{u_\pm}{u_\mp}}+\eta_sF\pa{\Upsilon_s^\pm\left|1-\frac{u_\pm}{u_\mp}\right.}-\eta_oF\pa{\Upsilon_o^\pm\left|1-\frac{u_\pm}{u_\mp}\right.},
\end{align}
where either choice of sign $\pm$ is equally valid. 
This can be rewritten as
\begin{align}
	\label{eq:IntermediateStep-}
	F\pa{\Upsilon_o^\pm\left|1-\frac{u_\pm}{u_\mp}\right.}=(-1)^m\sign\pa{\cos{\theta_o}}Y_m^\pm,
\end{align}
where we introduced
\begin{align}
	Y_m^\pm=\sign\pa{\cos{\theta_s}}F\pa{\Upsilon_s^\pm\left|1-\frac{u_\pm}{u_\mp}\right.}+\sign\pa{p_s^\theta}\br{\pa{m\pm\eta_o\frac{1-(-1)^m}{2}}K\pa{1-\frac{u_\pm}{u_\mp}}-\sqrt{u_\mp P}I_r}.
\end{align}
Next, we use the fact that the Jacobi elliptic function $\dn(x,k)$ satisfies $\dn(F(\arcsin{x}|k)|k)=\sqrt{1-kx^2}$, and therefore
\begin{align}
	\ab{y}=\dn\pa{\left.F\pa{\left.\arcsin{\sqrt{\frac{1-y^2}{k}}}\right|k}\right|k}.
\end{align}
Applying this identity to $y=\sin{\Psi_o^\mp}$ yields
\begin{align}
	\ab{\sin{\Psi_o^\mp}}=\dn\pa{\left.F\pa{\left.\arcsin{\sqrt{\frac{1-\sin^2{\Psi_o^\mp}}{1-\frac{u_\pm}{u_\mp}}}}\right|1-\frac{u_\pm}{u_\mp}}\right|1-\frac{u_\pm}{u_\mp}}
	=\dn\pa{\pm(-1)^m\sign\pa{\cos{\theta_o}}Y_m^\pm\left|1-\frac{u_\pm}{u_\mp}\right.}.
\end{align}
The last step follows from Eq.~\eqref{eq:IntermediateStep-}, together with the observation that
\begin{align}
	\label{eq:UpsilonPsiRelation}
	\arcsin\sqrt{\frac{1-\sin^2{\Psi_j^\mp}}{1-\frac{u_\pm}{u_\mp}}}=\pm\Upsilon_j^\pm,
\end{align}
which follows from
\begin{align}
	1-\sin^2{\Psi_j^\mp}=\pa{1-\frac{u_\pm}{u_\mp}}\sin^2{\Upsilon_j^\pm}.
\end{align}
Using Eq.~\eqref{eq:SignAbsorption} and the fact that $\dn(-x|k)=\dn(x|k)$ is even in its first argument, we find that
\begin{align}
	\frac{\cos{\theta_o}}{\sqrt{u_\mp}}=\sign\pa{\cos{\theta_o}}\ab{\sin{\Psi_o^\mp}}
	=\sign\pa{\cos{\theta_o}}\dn\pa{Y_m^\pm\left|1-\frac{u_\pm}{u_\mp}\right.}
	=\sign\pa{\cos{\theta_s}}\dn\pa{Y_m^\pm\left|1-\frac{u_\pm}{u_\mp}\right.},
\end{align}
where in the last step, we used the fact that $\sign\pa{\cos{\theta_s}}=\sign\pa{\cos{\theta_o}}$ for Type B vortical geodesics. This expression can be further simplified by noting that
\begin{align}
	Y_m^\pm=Y_{m-1}^\pm+\br{1\pm\eta_o(-1)^{m-1}}\sign\pa{p_s^\theta}K\pa{1-\frac{u_\pm}{u_\mp}}.
\end{align}
Since the function $\dn(x|k)$ satisfies the periodicity condition $\dn(x\pm2K(k)|k)=\dn(x|k)$, it follows that
\begin{align}
	\dn\pa{Y_m^\pm\left|1-\frac{u_\pm}{u_\mp}\right.}=\dn\pa{Y_{m-1}^\pm\left|1-\frac{u_\pm}{u_\mp}\right.}
	=\dn\pa{Y_0^\pm\left|1-\frac{u_\pm}{u_\mp}\right.},
\end{align}
from which we again conclude that $\cos{\theta_o}$ is independent of the number of turning points along the trajectory:
\begin{empheq}[box=\ovalbox]{align}
	\shortintertext{\centering $P>0$, $Q<0$ with $\pm$ arbitrary:}
	\label{eq:ObserverAngleQ<0}
	\cos{\theta_o}=\sign\pa{\cos{\theta_s}}\sqrt{u_\mp}\dn\pa{Y_0^\pm\left|1-\frac{u_\pm}{u_\mp}\right.},\qquad
	Y_0^\pm=\sign\pa{\cos{\theta_s}}F\pa{\Upsilon_s^\pm\left|1-\frac{u_\pm}{u_\mp}\right.}-\sign\pa{p_s^\theta}\sqrt{u_\mp P}I_r.
\end{empheq}

While this expression for $\cos{\theta_o}$ (valid for $Q<0$) superficially differs from that given in Eq.~\eqref{eq:ObserverAngleQ>0} (valid for $Q>0$), the two expressions can be brought into the same form. For $k\not\in \mathbb{R}$, the reciprocal-modulus theorem \cite{Fettis1970}
\begin{align}
	\frac{1}{\sqrt{k}}K\pa{\frac{1}{k}}=K(k)\mp iK(1-k),\qquad
	\pm=\sign\pa{\operatorname{Im}{k}},
\end{align}
determines the real and imaginary parts of $K(1/k)$ in terms of $K(k)$ and $K(1-k).$ The difference in sign results from the branch cut in $K(k)$ extending from $k=1$ to $+\infty$ along the positive real axis. For real $k$, choosing the primary branch for $k<1$ fixes
\begin{align}
	\label{eq:EllipticIdentity}
	\frac{1}{\sqrt{k}}K\pa{\frac{1}{k}}=K(k)\mp iK(1-k),\qquad
	\pm=\sign\pa{1-k}.
\end{align}
With this choice, identical manipulations demonstrate that
\begin{subequations}
\begin{empheq}[left={\displaystyle\frac{1}{\sqrt{k}}K\pa{\frac{1}{k}}=\empheqlbrace}]{alignat=2}
	\label{eq:EllipticIdentity+}
	&F\pa{\left.\arcsin\sqrt{\frac{y^2}{k}}\right|k}+iF\pa{\left.\arcsin\sqrt{\frac{1-y^2}{1-k}}\right|1-k},\qquad
	1<y^2<k,\\
	\label{eq:EllipticIdentity-}
	&F\pa{\left.\arcsin\sqrt{\frac{y^2}{k}}\right|k}-iF\pa{\left.\arcsin\sqrt{\frac{1-y^2}{1-k}}\right|1-k},\qquad
	0<k<y^2<1.
\end{empheq}
\end{subequations}
We now apply these identities with $y=\sin{\Psi_s^\mp}=\sqrt{u_s/u_\mp}$ and $k=u_\pm/u_\mp$. Since $0<u_-<u_s<u_+<1$, we apply \eqref{eq:EllipticIdentity+} for the upper choice of sign  and \eqref{eq:EllipticIdentity-} for the lower choice of sign. Combined with Eq.~\eqref{eq:UpsilonPsiRelation}, we find that
\begin{align}
	\sqrt{\frac{u_\mp}{u_\pm}}K\pa{\frac{u_\mp}{u_\pm}}=F\pa{\Psi_s^\pm\left|\frac{u_\pm}{u_\mp}\right.}+iF\pa{\Upsilon_s^\pm\left|1-\frac{u_\pm}{u_\mp}\right.}.
\end{align}
After multiplying by $\sign\pa{\cos{\theta_s}}$, this becomes
\begin{align}
	\label{eq:IntermediateStep}
	\sign\pa{\cos{\theta_s}}\sqrt{\frac{u_\mp}{u_\pm}}K\pa{\frac{u_\mp}{u_\pm}}=F\pa{\arcsin\pa{\frac{\cos{\theta_s}}{\sqrt{u_\pm}}}\left|\frac{u_\pm}{u_\mp}\right.}+iY_0^\pm+\sign\pa{p_s^\theta}\sqrt{-u_\mp P}I_r.
\end{align}
In order to proceed, we also need the identity\footnote{This can be derived, for instance, from the two standard identities $\mathrm{sc}(x\pm iK(1-k)+ K(k)|k)=\pm\frac{i}{\sqrt{1-k}}\dn(x|k)$ and $\sn(ix|k)=i\,\mathrm{sc}(x|1-k)$ (Eqs.~(16.8.9) and (16.20.1) of Ref.~\cite{Abramowitz1972}) which imply that $\dn(x|1-k)=\pm\sqrt{k}\sn(ix\pm K(k)+iK(1-k)|k)$. The result is then obtained by using Eq.~\eqref{eq:EllipticIdentity} together with the imaginary periodicity condition $\sn(x\pm2iK(1-k)|k)=\sn(x|k)$ whenever it is needed.}
\begin{align}
	\dn(x|1-k)=\pm\sqrt{k}\sn\pa{\left.ix\pm\frac{1}{\sqrt{k}}K\pa{\frac{1}{k}}\right|k}.
\end{align}
This relation holds for either choice of sign, which we take to be $\pm=-\sign\pa{\cos{\theta_s}}$. Substituting $k=u_\pm/u_\mp$ and combining with Eq.~\eqref{eq:ObserverAngleQ<0} yields
\begin{align}
	\cos{\theta_o}=\sign\pa{\cos{\theta_s}}\sqrt{u_\mp}\dn\pa{Y_0^\pm\left|1-\frac{u_\pm}{u_\mp}\right.}
	=\sqrt{u_\pm}\sn\pa{-iY_0^\pm+\sign\pa{\cos{\theta_s}}\sqrt{\frac{u_\mp}{u_\pm}}K\pa{\frac{u_\mp}{u_\pm}}\left|\frac{u_\pm}{u_\mp}\right.}.
\end{align}
Finally, using Eq.~\eqref{eq:IntermediateStep}, this reduces to
\begin{align}
	\cos{\theta_o}=\sqrt{u_\pm}\sn\pa{F\pa{\arcsin\pa{\frac{\cos{\theta_s}}{\sqrt{u_\pm}}}\left|\frac{u_\pm}{u_\mp}\right.}+\sign\pa{p_s^\theta}\sqrt{-u_\mp P}I_r\left|\frac{u_\pm}{u_\mp}\right.}.
\end{align}
Therefore  Eqs.~\eqref{eq:ObserverAngleQ>0} and \eqref{eq:ObserverAngleQ<0} can be combined into the single expression
\begin{empheq}[box=\ovalbox]{align}
	\shortintertext{\centering $P\neq0$, $Q\neq0$  with $\pm=\sign\pa{P}$:}
	\cos{\theta_o}=\sqrt{u_\pm}\sn\pa{X_0^\pm\left|\frac{u_\pm}{u_\mp}\right.},\qquad
	X_0^\pm=F\pa{\arcsin\pa{\frac{\cos{\theta_s}}{\sqrt{u_\pm}}}\left|\frac{u_\pm}{u_\mp}\right.}-\sign\pa{p_s^\theta Q}\sqrt{-u_\mp P}I_r.
\end{empheq}

In the $Q=0$ case (where necessarily $P>0$), Eq.~\eqref{eq:P>0,Q=0} tells us that
\begin{align}
	\sqrt{u_+P}I_r&=\eta_o\arctanh\sqrt{1-\frac{u_o}{u_+}}-\eta_s\arctanh\sqrt{1-\frac{u_s}{u_+}},
\end{align}
which can be rewritten
\begin{align}
	\arctanh\sqrt{1-\frac{u_o}{u_+}}=(-1)^m\sign\pa{\cos{\theta_o}}Z,\qquad
	Z=\sign\pa{\cos{\theta_s}}\arctanh\sqrt{1-\frac{u_s}{u_+}}+\sign\pa{p_s^\theta}\sqrt{u_+P}I_r.
\end{align}
Since $\tanh\pa{-x}=-\tanh{x}$ is an odd function, it follows that
\begin{align}
	\sqrt{1-\frac{u_o}{u_+}}=(-1)^m\sign\pa{\cos{\theta_o}}\tanh{Z}.
\end{align}
Therefore, the overall signs disappear upon squaring, leaving
\begin{align}
	u_o=u_+\sech^2{Z}.
\end{align}
Using the fact that $\sech\pa{-x}=\sech{x}$ is an even function, and that $\sign\pa{\cos{\theta_s}}=\sign\pa{\cos{\theta_o}}$ for vortical geodesics, we finally conclude:
\begin{empheq}[box=\ovalbox]{align}
	\shortintertext{\centering $P>0$, $Q=0$:}
	\cos{\theta_o}=\sign\pa{\cos{\theta_s}}\sqrt{u_+}\sech{Z},\qquad
	Z=\arctanh\sqrt{1-\frac{u_s}{u_+}}+\eta_s\sqrt{u_+P}I_r.
\end{empheq}

\subsection{Summary of results}
\label{subsec:KerrSummary}

Here, we collect the simplified expressions for the angular integrals $G_\theta$, $G_\phi$ and $G_t$, as well as expressions for the final angle $\theta_o$ obtained by solving the $(r,\theta)$ part of the geodesic equation \eqref{eq:KerrGeodesicEquation}. Our conventions are such that all of the terms appearing in these expressions are real and positive.

\begin{subequations}
\label{eq:Q>0,P=0}
\begin{empheq}[box=\ovalbox]{align}
	\shortintertext{\centering\underline{$Q>0$ with $P=0$:}
	$\quad0\le\cos^2{\theta}\le u_0<1$}
	G_\theta&=\sqrt{\frac{u_0}{Q}}\br{\pi m+\eta_s\arcsin\sqrt{\frac{u_s}{u_0}}-\eta_o\arcsin\sqrt{\frac{u_o}{u_0}}},\\
	G_\phi&=\frac{1}{\sqrt{Q}}\sqrt{\frac{u_0}{1-u_0}}\br{\pi m+\eta_s\arcsin\sqrt{\frac{u_s}{u_0}\pa{\frac{1-u_0}{1-u_s}}}-\eta_o\arcsin\sqrt{\frac{u_o}{u_0}\pa{\frac{1-u_0}{1-u_o}}}},\\
	G_t&=\frac{1}{2}\cu{u_0G_\theta-\sqrt{\frac{u_0}{Q}}\br{\eta_s\sqrt{u_s\pa{u_0-u_s}}-\eta_o\sqrt{u_o\pa{u_0-u_o}}}},\\
	&\cos{\theta_o}=\sqrt{u_0}\sin{W_0},\qquad
	W_0=\arcsin\pa{\frac{\cos{\theta_s}}{\sqrt{u_0}}}-\sign\pa{p_s^\theta}\sqrt{\frac{Q}{u_0}}I_r.
\end{empheq}
\end{subequations}

\begin{subequations}
\label{eq:Q>0,P!=0}
\begin{empheq}[box=\ovalbox]{align}
	\shortintertext{\centering\underline{$Q>0$ with $P\neq0$:}
	$\quad0\le\cos^2{\theta}\le u_\pm<1$ with $\pm=\sign\pa{P}$}
	G_\theta&=\frac{1}{\sqrt{-u_\mp P}}\br{2mK\pa{\frac{u_\pm}{u_\mp}}+\eta_sF\pa{\Psi_s^\pm\left|\frac{u_\pm}{u_\mp}\right.}-\eta_oF\pa{\Psi_o^\pm\left|\frac{u_\pm}{u_\mp}\right.}},\\
	G_\phi&=\frac{1}{\sqrt{-u_\mp P}}\br{2m\Pi\pa{u_\pm\left|\frac{u_\pm}{u_\mp}\right.}+\eta_s\Pi\pa{u_\pm;\Psi_s^\pm\left|\frac{u_\pm}{u_\mp}\right.}-\eta_o\Pi\pa{u_\pm;\Psi_o^\pm\left|\frac{u_\pm}{u_\mp}\right.}},\\
	G_t&=-\frac{2u_\pm}{\sqrt{-u_\mp P}}\br{2mE'\pa{\frac{u_\pm}{u_\mp}}+\eta_sE'\pa{\Psi_s^\pm\left|\frac{u_\pm}{u_\mp}\right.}-\eta_oE'\pa{\Psi_o^\pm\left|\frac{u_\pm}{u_\mp}\right.}},\\
	\label{eq:ObserverAngleP!=0,Q>0}
	&\cos{\theta_o}=\sqrt{u_\pm}\sn\pa{X_0^\pm\left|\frac{u_\pm}{u_\mp}\right.},\qquad
	X_0^\pm=F\pa{\arcsin\pa{\frac{\cos{\theta_s}}{\sqrt{u_\pm}}}\left|\frac{u_\pm}{u_\mp}\right.}-\sign\pa{p_s^\theta}\sqrt{-u_\mp P}I_r.
\end{empheq}
\end{subequations}

\begin{subequations}
\begin{empheq}[box=\ovalbox]{align}
	\shortintertext{\centering\underline{$Q<0$ (with $P>0$ required):}
	$\quad0<u_-\le\cos^2{\theta}\le u_+<1$ with $\pm$ arbitrary}
	G_\theta&=\frac{1}{\sqrt{u_\mp P}}\br{\pa{m\pm\eta_o\frac{1-(-1)^m}{2}}K\pa{1-\frac{u_\pm}{u_\mp}}+\eta_sF\pa{\Upsilon_s^\pm\left|1-\frac{u_\pm}{u_\mp}\right.}-\eta_oF\pa{\Upsilon_o^\pm\left|1-\frac{u_\pm}{u_\mp}\right.}},\\
	G_\phi&=\frac{1}{\pa{1-u_\mp}\sqrt{u_\mp P}}\bigg[\pa{m\pm\eta_o\frac{1-(-1)^m}{2}}\Pi\pa{\frac{u_\pm-u_\mp}{1-u_\mp}\left|1-\frac{u_\pm}{u_\mp}\right.}\\
	&\qquad\qquad\qquad\qquad\qquad+\eta_s\Pi\pa{\frac{u_\pm-u_\mp}{1-u_\mp};\Upsilon_s^\pm\left|1-\frac{u_\pm}{u_\mp}\right.}-\eta_o\Pi\pa{\frac{u_\pm-u_\mp}{1-u_\mp};\Upsilon_o^\pm\left|1-\frac{u_\pm}{u_\mp}\right.}\bigg],\notag\\
	G_t&=\sqrt{\frac{u_\mp}{P}}\br{\pa{m\pm\eta_o\frac{1-(-1)^m}{2}}E\pa{1-\frac{u_\pm}{u_\mp}}+\eta_sE\pa{\Upsilon_s^\pm\left|1-\frac{u_\pm}{u_\mp}\right.}-\eta_oE\pa{\Upsilon_o^\pm\left|1-\frac{u_\pm}{u_\mp}\right.}},\\
	\label{eq:ObserverAngleP>0,Q<0}
	&\cos{\theta_o}=\sign\pa{\cos{\theta_s}}\sqrt{u_\mp}\dn\pa{Y_0^\pm\left|1-\frac{u_\pm}{u_\mp}\right.},\quad
	Y_0^\pm=\sign\pa{\cos{\theta_s}}F\pa{\Upsilon_s^\pm\left|1-\frac{u_\pm}{u_\mp}\right.}-\sign\pa{p_s^\theta}\sqrt{u_\mp P}I_r.
\end{empheq}
\end{subequations}

\begin{subequations}
\begin{empheq}[box=\ovalbox]{align}
	\shortintertext{\centering\underline{$Q=0$ (with $P>0$ required):}
	$\quad0<\cos^2{\theta}<u_+<1$}
	G_\theta&=\frac{1}{\sqrt{u_+P}}\br{\eta_o\arctanh\sqrt{1-\frac{u_o}{u_+}}-\eta_s\arctanh\sqrt{1-\frac{u_s}{u_+}}},\\
	G_\phi&=G_\theta+\frac{1}{\sqrt{\pa{1-u_+}P}}\br{\eta_o\arctan\sqrt{\frac{u_+-u_o}{1-u_+}}-\eta_s\arctan\sqrt{\frac{u_+-u_s}{1-u_+}}},\\
	G_t&=\frac{1}{\sqrt{P}}\br{\eta_o\sqrt{u_+-u_o}-\eta_s\sqrt{u_+-u_s}},\\
	&\cos{\theta_o}=\sign\pa{\cos{\theta_s}}\sqrt{u_+}\sech{Z},\qquad
	Z=\arctanh\sqrt{1-\frac{u_s}{u_+}}+\eta_s\sqrt{u_+P}I_r.
\end{empheq}
\end{subequations}

Finally, we note that Eqs.~\eqref{eq:ObserverAngleP!=0,Q>0} and \eqref{eq:ObserverAngleP>0,Q<0} can be conveniently combined into a single expression:
\begin{empheq}[box=\ovalbox]{align}
	\shortintertext{\centering\underline{$Q\neq0$ and $P\neq0$ with $\pm=\sign\pa{P}$:}}
	\label{eq:ObserverAngle}
	&\cos{\theta_o}=\sqrt{u_\pm}\sn\pa{X_0^\pm\left|\frac{u_\pm}{u_\mp}\right.},\qquad
	X_0^\pm=F\pa{\arcsin\pa{\frac{\cos{\theta_s}}{\sqrt{u_\pm}}}\left|\frac{u_\pm}{u_\mp}\right.}-\sign\pa{p_s^\theta Q}\sqrt{-u_\mp P}I_r.
\end{empheq}

\section{Near-horizon geometry of (near-)extreme Kerr}
\label{sec:NHEK}

The Kerr family of metrics has two adjustable parameters corresponding to the mass $M$ and angular momentum $J$ of the black hole. Geometries satisfying the Kerr bound $\ab{J}\le M^2$ have smooth event horizons concealing a ring singularity, while solutions that violate this bound exhibit naked singularities visible from infinity. Black holes that (nearly) saturate the Kerr bound are termed (near-)extremal, and there is strong evidence to suggest that no physical process can drive a (sub-)extremal black hole over the Kerr bound \cite{Sorce2017} (such super-extremal black holes would behave very differently---see, $e.g.$, Refs.~\cite{Stuchlik2010,Stuchlik2011}). However, one would expect accretion of matter onto an astrophysical black hole to push it towards extremality, and indeed the vast majority of measured supermassive black holes spins are close to maximal \cite{Brenneman2013,Reynolds2019}. The limiting behavior of the Kerr metric in the extremal limit $\ab{J}\to M^2$ is therefore of both theoretical and astronomical interest.

This section is dedicated to a pedagogical review of the qualitatively new and surprising features that the Kerr black hole develops in the high-spin regime. We begin in Sec.~\ref{subsec:Puzzle} by presenting a timelike equatorial orbit that seemingly lies on a null hypersurface. This apparent paradox is resolved in Sec.~\ref{subsec:EmergentThroat} by the presence of an infinitely deep throat-like region bunched up near the event horizon: the Near-Horizon Extreme Kerr (NHEK) geometry. This motivates the more systematic investigation of near-horizon scaling limits that we conduct in Sec.~\ref{subsec:ScalingLimits}. These different limits are then related by an emergent conformal symmetry of the throat in Sec.~\ref{subsec:ConformalSymmetryInTheSky}. Finally, in Sec.~\ref{subsec:GlobalNHEK}, we introduce global coordinates and describe the causal structure of the throat geometry using its Carter-Penrose diagram.

\subsection{Peculiar features of the extremal limit}
\label{subsec:Puzzle}

The $\ab{J}\to M^2$ limit of the Kerr geometry poses a series of puzzles whose resolution requires a careful analysis of the near-horizon geometry of the extreme black hole. A particle orbiting a Kerr black hole on a prograde, circular, equatorial geodesic at radius $r=r_s$ has four-velocity \cite{Bardeen1972}
\begin{align}
	\label{eq:EquatorialGeodesics}
	u_s=u_s^t\pa{\pd_t+\Omega_s\pd_\phi},\qquad
	u_s^t=\frac{r_s^{3/2}+aM^{1/2}}{\sqrt{r_s^3-3Mr_s^2+2aM^{1/2}r_s^{3/2}}},\qquad
	\Omega_s=\frac{M^{1/2}}{r_s^{3/2}+aM^{1/2}}.
\end{align}
The energy and angular momentum of this geodesic is given by
\begin{align}
	\label{eq:EquatorialGeodesicsParameters}
	\frac{\omega_s}{\mu}=\frac{r_s^{3/2}-2Mr_s^{1/2}+aM^{1/2}}{r_s^{3/4}\sqrt{r_s^{3/2}-3Mr_s^{1/2}+2aM^{1/2}}},\qquad
	\frac{\ell_s}{\mu}=\frac{M^{1/2}\big(r_s^2-2aM^{1/2}r_s^{1/2}+a^2\big)}{r_s^{3/4}\sqrt{r_s^{3/2}-3Mr_s^{1/2}+2aM^{1/2}}}.
\end{align}

This orbit is stable provided that the orbital radius $r_s$ exceeds the marginally stable radius $r_\mathrm{ms}$ of the Innermost Stable Circular Orbit (ISCO), given by
\begin{subequations}
\label{eq:ISCO}
\begin{gather}
	r_\mathrm{ms}=M\pa{3+Z_2-\sqrt{\pa{3-Z_1}\pa{3+Z_1+2Z_2}}},\\
	Z_1=1+\pa{1-a_\star^2}^{1/3}\br{\pa{1+a_\star}^{1/3}+\pa{1-a_\star}^{1/3}},\qquad
	Z_2=\pa{3a_\star^2+Z_1^2}^{1/2},\qquad
	a_\star=\frac{a}{M}.
\end{gather}
\end{subequations}
In the context of black hole astrophysics, these orbits provide a simple model for accretion onto a black hole: to a very good approximation, a thin disk of slowly accreting matter consists of particles following the geodesics \eqref{eq:EquatorialGeodesics} \cite{Novikov1973,Page1974}. In reality, their trajectories also have a small inward radial component, but it can be neglected down to the ISCO radius, which delineates the innermost edge of the disk where accretion terminates. Beyond this edge, the particles quickly plunge into the black hole and their radial momentum can no longer be ignored. Instead, their motion is described by infalling geodesics with the conserved quantities of the marginally stable orbit \cite{Cunningham1975,Penna2012}.

Consider a Kerr black hole with spin parameter $a=M\sqrt{1-\kappa^2}$. When the deviation from extremality is small, $0<\kappa\ll1$, the black hole has a small Hawking temperature of order $\kappa$,
\begin{align}
	T_H=\frac{1}{4\pi r_+}\pa{\frac{r_+-r_-}{r_++r_-}}
	=\frac{\kappa}{4\pi M}+\O{\kappa^2},
\end{align}
where $r_\pm$ denotes the radius of the (outer/inner) event horizon,
\begin{align}
	\label{eq:EventHorizon}
	r_\pm=M\pm\sqrt{M^2-a^2}
	=M\pa{1\pm\kappa}.
\end{align}
Thus, the extremal limit $\ab{J}\to M^2$ is equivalent to a low-temperature limit $\kappa\to0$. A detailed investigation of this limit raises several puzzles:
\begin{enumerate}
\item
The first puzzle pertains to the fate of the ISCO in the extremal limit. According to Eq.~\eqref{eq:ISCO},
\begin{align}
	r_\mathrm{ms}\stackrel{\kappa\to0}{=}M\br{1+2^{1/3}\kappa^{2/3}+\O{\kappa^{4/3}}}.
\end{align}
Comparing with Eq.~\eqref{eq:EventHorizon}, it would appear that the ISCO moves onto the event horizon in the extremal limit:
\begin{align}
	\lim_{a\to M}r_\mathrm{ms}=M
	=\lim_{a\to M}r_+.
\end{align}
However, for any sub-extremal black hole, the ISCO is a timelike geodesic, while the event horizon is ruled by null geodesics. Clearly, the extreme Kerr metric fails to accurately portray the spacetime geometry in the ISCO region correctly.
\item
Indeed, although the ISCO and extremal horizon appear to coincide, the proper radial distance (as measured on a Boyer-Lindquist time-slice) between the two actually diverges logarithmically in this limit:
\begin{align}
	\int_{r_+}^{r_\mathrm{ms}}ds\stackrel{\kappa\to0}{\sim}M\ab{\log{\kappa}}.
\end{align}
Thus, even though generic timelike (null) geodesics fall into the horizon in finite proper (affine) time, the near-horizon region acquires an infinite proper three-volume in the extremal limit. 
\end{enumerate}
These observations were noted early on by Bardeen, Press and Teukolsky \cite{Bardeen1972} and later revisited in Refs.~\cite{Jacobson2011,Gralla2016a}. Taken together, these peculiarities indicate that the extremal Kerr metric grossly misrepresents the spacetime geometry near the event horizon of the extremal black hole. While it is true that the Boyer-Lindquist coordinates become singular at the horizon, we stress that these problems are not a coordinate artifact: they still arise even in coordinates that are smooth across the horizon. The existence of the infinite throat region is a coordinate-invariant statement, and describing it requires a careful resolution of the near-horizon geometry. This was accomplished by Bardeen and Horowitz \cite{Bardeen1999} by introducing a horizon-scaling limit tailored to this task, to which we now turn.

\subsection{The extreme Kerr throat}
\label{subsec:EmergentThroat}

In Sec.~\ref{subsec:Puzzle}, we saw that as a black hole spins up and approaches the limiting extremal geometry with $\ab{J}\to M^2$, a deep throat of divergent proper depth develops outside of its event horizon. Moreover, from the perspective of a distant observer, particles on the ISCO co-rotate with the black hole horizon in this limit. This motivates a coordinate transformation from Boyer-Lindquist coordinates $(t,r,\phi)$ to Bardeen-Horowitz coordinates $(T,R,\Phi)$ given by
\begin{align}
	\label{eq:BardeenHorowitzCoordinates}
	t=\frac{T}{\Omega_H},\qquad
	r=r_+\pa{1+R},\qquad
	\phi=\Phi+T,
\end{align}
where $\Omega_H$ denotes the angular velocity of the extremal black hole horizon,
\begin{align}
	\Omega_H=\frac{a}{2Mr_+}
	\stackrel{a\to M}{=}\frac{1}{2M}.
\end{align}
These coordinates are adapted to a local near-horizon observer co-rotating with the black hole, since
\begin{align}
	\Omega_H\pd_T=\pd_t+\Omega_H\pd_\phi.
\end{align}
Local finite-energy excitations near the horizon of a black hole have large gravitational redshift relative to an observer at infinity. For black holes far from extremality, this region of spacetime is small and contains no stable orbits. However, for extremal black holes, the stable orbits extend down the throat, and the high-redshift emissions from sources in this region are phenomenologically interesting. In order to resolve the degeneracy arising from the infinite redshift while zooming into the horizon, we perform an infinite dilation onto the horizon, implemented by the rescaling
\begin{align}
	\label{eq:HorizonScaling}
	\pa{T,R}\to\pa{\frac{T}{\epsilon},\epsilon R},\qquad
	\epsilon\to0.
\end{align}
If the black hole is precisely extremal ($a=M$), this scaling procedure has a finite limit and yields the NHEK geometry, with non-degenerate line element \cite{Bardeen1999}
\begin{subequations}
\label{eq:NHEK}
\begin{gather}
	d\hat{s}^2=2M^2\Gamma\br{-R^2\ed T^2+\frac{\ed R^2}{R^2}+\ed\theta^2+\Lambda^2\pa{\ed\Phi+R\ed T}^2},\\
	\Gamma(\theta)=\frac{1+\cos^2{\theta}}{2},\qquad
	\Lambda(\theta)=\frac{2\sin{\theta}}{1+\cos^2{\theta}}.
\end{gather}
\end{subequations}
Since the NHEK geometry arises as a non-singular scaling limit of the extreme Kerr solution, it manifestly solves the vacuum Einstein equations and can be studied as a spacetime in its own right. Moreover, since in the limit $\epsilon\to0$, the resulting metric is $\epsilon$-independent, further coordinate rescalings $(T,R)\to(T/\epsilon,\epsilon R)$ leave the NHEK line element \eqref{eq:NHEK} invariant: physically, the throat-like region is sufficiently deep that it becomes self-similar in the extremal limit.

Therefore, the region of spacetime in the throat displays an emergent scaling symmetry, which is generated at the infinitesimal level by the dilation Killing vector $H_0=T\pd_T-R\pd_R$. Surprisingly, yet another, no-less constraining symmetry---invariance under special conformal transformations generated by $H_-$---also emerges in this limit. Together, these symmetries generate the global conformal group $\mathsf{SL}(2,\mathbb{R})$, with commutation relations
\begin{align}
	\br{H_0,H_\pm}=\mp H_\pm,\qquad
	\br{H_+,H_-}=2H_0.
\end{align}
Hence, in the high-spin regime, the $\mathbb{R}\times\mathsf{U}(1)$ symmetry of the Kerr metric \eqref{eq:Kerr} generated by the Killing vectors $\pd_t$ and $\pd_\phi$ (associated with stationarity and axisymmetry, respectively) is enlarged within the near-horizon region to an $\mathsf{SL}(2,\mathbb{R})\times\mathsf{U}(1)$ isometry group generated by
\begin{align}
	\label{eq:KillingFieldsNHEK}
	H_0=T\pd_T-R\pd_R,\qquad
	H_+=\pd_T,\qquad
	H_-=\pa{T^2+\frac{1}{R^2}}\pd_T-2TR\pd_R-\frac{2}{R}\pd_\Phi,\qquad
	W_0=\pd_\Phi.
\end{align}
In fact, although the Killing tensor \eqref{eq:KerrKilling} in extreme Kerr is associated to a non-geometrically-realized symmetry, in the near-horizon limit, this irreducible Killing tensor descends to a reducible Killing tensor in NHEK \cite{Galajinsky2010,AlZahrani2011}. More precisely, it is given (up to a mass term) by the Casimir of $\mathsf{SL}(2,\mathbb{R})\times\mathsf{U}(1)$:
\begin{align}
	\label{eq:KillingTensor}
	\hat{K}^{\mu\nu}=M^2\hat{g}^{\mu\nu}-H_0^\mu H_0^\nu+\frac{1}{2}\pa{H_+^\mu H_-^\nu+H_-^\mu H_+^\nu}+W_0^\mu W_0^\nu.
\end{align}
Thus, the Kerr metric's hidden symmetries become explicit in the emergent throat region, where it decomposes into
\begin{align}
	\hat{g}^{\mu\nu}=-\frac{1}{2M^2\Gamma}\br{-H_0^\mu H_0^\nu+\frac{1}{2}\pa{H_+^\mu H_-^\nu+H_-^\mu H_+^\nu}-\pd_\theta^\mu\pd_\theta^\nu+\pa{1-\frac{1}{\Lambda^2}}W_0^\mu W_0^\nu}.
\end{align}
Finally, we note that the induced metric on a hypersurface of fixed $\theta$ is that of warped three-dimensional Anti de-Sitter space (WAdS$_3$) with warp factor $\Lambda(\theta)$ \cite{Song2009}. This warp factor goes to unity, $\Lambda(\theta_c)=1$, at the critical angle $\theta_c=\arctan\pa{\sqrt{2}/3^{1/4}}$, which corresponds to the so-called ``light surface" of the extreme Kerr black hole \cite{Aman2012}. The NHEK metric becomes precisely that of AdS$_3$ on this surface, which seems to play a priviledged role in the propagation of light out of the throat \cite{Gralla2017a}.

\subsection{Near-horizon scaling limits for near-extreme Kerr}
\label{subsec:ScalingLimits}

In Sec.~\ref{subsec:Puzzle}, we saw that the extreme Kerr metric fails to resolve near-horizon physics. Then, we argued in Sec.~\ref{subsec:EmergentThroat} that this failure is caused by the emergence, in a certain scaling limit, of a near-horizon region of the Kerr spacetime that resembles an infinite gravitational potential well. This throat-like region is sufficiently deep that in the extremal limit, it becomes self-similiar and enjoys an enhanced isometry group: the global conformal group $\mathsf{SL}(2,\mathbb{R})\times\mathsf{U}(1)$.

The scaling limit to the NHEK region is unique for a precisely extremal black hole. But, according to the classical laws of black hole thermodynamics, such a black hole is unphysical since it has zero temperature, and no adiabatic process can turn a non-extremal black hole extremal \cite{Israel1986}. Thus, it is more realistic to consider black holes with a small deviation from extremality and a correspondingly small temperature. However, for such a near-extremal black hole, there exist infinitely many bands of near-horizon radii that become infinitely separated from each other in the extremal limit \cite{Gralla2015}. More precisely, if one considers two Boyer-Lindquist radii $r_1$ and $r_2$ that scale to the horizon at different rates as extremality is approached, so that
\begin{align}
	a=M\sqrt{1-\pa{\epsilon\kappa}^2},\qquad
	r_1=M\br{1+\epsilon^qR_1},\qquad
	r_2=M\br{1+\epsilon^sR_2},\qquad
	0<s\leq q\leq1,
\end{align}
then the proper radial separations along a Boyer-Lindquist time-slice have the limiting form
\begin{align}
	\label{eq:GeodesicDistance}
	\lim_{\epsilon\to0}\int_{r_1}^{r_2}ds=
	\begin{cases}
		\displaystyle M\log\pa{\frac{R_2}{R_1}}+\pa{q-s}M\ab{\log{\epsilon}},
		&q<1,\ s<1,\vspace{2pt}\\
		\displaystyle M\log\pa{\frac{2R_2}{R_1+\sqrt{R_1^2-\kappa^2}}}+\pa{1-s}M\ab{\log{\epsilon}},
		&q=1,\ s<1,\vspace{2pt}\\
		\displaystyle M\log\pa{\frac{R_2+\sqrt{R_2^2-\kappa^2}}{R_1+\sqrt{R_1^2-\kappa^2}}},
		&q=1,\ s=1.
	\end{cases}
\end{align}
Only radii that scale to the horizon at the same rate have finite radial separation in the extremal limit. This indicates that there are in fact infinitely many physically distinct near-horizon limits, each of which resolves the throat physics at different scales. The relevant scaling limits straightforwardly generalize Eqs.~\eqref{eq:BardeenHorowitzCoordinates} and \eqref{eq:HorizonScaling} to
\begin{align}
	\label{eq:Parameters}
	a=M\sqrt{1-\pa{\epsilon\kappa}^2},\qquad
	T=\epsilon^p\frac{t}{2M},\qquad
	R=\frac{r-M}{\epsilon^pM},\qquad
	\Phi=\phi-\frac{t}{2M},\qquad
	0<p\le1,\qquad
	\epsilon\to0.
\end{align}
The $\epsilon\to0$ limit with $p=1$ (which physically amounts to zooming into the near-horizon region at the same rate that the black hole is dialed into extremality) and $(T,R,\Phi,\kappa)$ held fixed yields the so-called near-NHEK geometry \cite{Bredberg2010}
\begin{align}
	\label{eq:NearNHEK}
	d\bar{s}^2=2M^2\Gamma\br{-\pa{R^2-\kappa^2}\ed T^2+\frac{\ed R^2}{R^2-\kappa^2}+\ed\theta^2+\Lambda^2\pa{\ed\Phi+R\ed T}^2},
\end{align}
which is the finite-temperature analogue of the NHEK metric \eqref{eq:NHEK}. It also has an $\mathsf{SL}(2,\mathbb{R})\times\mathsf{U}(1)$ isometry group generated by
\begin{align}
	\label{eq:KillingFieldsNearNHEK}
	H_0=\frac{1}{\kappa}\pd_T,\qquad
	H_\pm=\frac{e^{\mp\kappa T}}{\sqrt{R^2-\kappa^2}}\br{\frac{R}{\kappa}\pd_T\pm\pa{R^2-\kappa^2}\pd_R-\kappa\pd_\Phi},\qquad
	W_0=\pd_\Phi.
\end{align}
This region lies deepest in the throat and resolves the horizon at $r=M\pa{1+\epsilon\kappa}$, along with all other radii that also scale like $r\sim M\pa{1+\epsilon R}$ in the $\epsilon\to0$ limit. Examples of physically interesting radii that scale into near-NHEK include the photon orbit at $r_\mathrm{ph}$ and the (prograde) marginally bound orbit at $r_\mathrm{mb}$, also known as the Innermost Bound Circular Orbit (IBCO) radius \cite{Bardeen1972},
\begin{subequations}
\begin{align}
	r_\mathrm{ph}&=4M\cos^2\br{\frac{1}{3}\arccos\pa{\frac{a}{M}}-\frac{\pi}{3}}
	=M\br{1+\frac{2}{\sqrt{3}}\epsilon\kappa+\O{\epsilon^2}},\\
	r_\mathrm{mb}&=2M-a+2\sqrt{M\pa{M-a}}
	=M\br{1+\sqrt{2}\epsilon\kappa+\O{\epsilon^2}}.
\end{align}
\end{subequations}
The proper radial separation between two prograde, equatorial, circular geodesics in near-NHEK is given by
\begin{align}
	d\bar{s}\pa{R_1,R_2}=M\int^{ R_2}_{ R_1}\frac{\ed R}{\sqrt{R^2-\kappa^2}}
	=M\log\pa{\frac{R_2+\sqrt{R_2^2-\kappa^2}}{R_1+\sqrt{R_1^2-\kappa^2}}}.
\end{align}
This expression matches the limiting radial separation of equatorial geodesics calculated in the Kerr geometry \eqref{eq:GeodesicDistance}, provided that one identifies the near-NHEK radius $R$ with the radius in Kerr scaling as $r=M\pa{1+\epsilon R}$. Note that this distance is not scale-invariant due to the presence of the horizon: physically, the presence of a small temperature $\kappa$ breaks the scaling symmetry exhibited by the NHEK distance. Mathematically, this can also be seen from Eq.~\eqref{eq:Parameters}, where the presence of the temperature $\kappa>0$ precludes the scaling limit $\epsilon\to0$ from being a coordinate limit: unlike the NHEK scaling \eqref{eq:HorizonScaling}, the dilation into near-NHEK also acts on the parameter $a$, which is why it is not forced to become an isometry in the limit.

The $\epsilon\to0$ limit with any $0<p<1$ physically corresponds to spinning up the black hole faster than one zooms into the horizon, and always produces the same NHEK metric \eqref{eq:NHEK}. However, each geometry thus obtained corresponds to a physically distinct region of the throat: a given choice of $p$ resolves a band of Boyer-Lindquist radii that scale like $r=M\pa{1+\epsilon^pR}$.\footnote{For a precisely extremal black hole, this distinction is irrelevant since all the scales we discuss lie precisely at $r=M$, so there is a single NHEK limit that resolves them.} The proper radial separation in NHEK,
\begin{align}
	d\hat{s}\pa{R_1,R_2}=M\int^{R_2}_{R_1}\frac{\ed R}{R}
	=M\log\pa{\frac{R_2}{R_1}},
\end{align}
matches the corresponding limiting radial separation of equatorial geodesics calculated in the Kerr geometry \eqref{eq:GeodesicDistance}. However, because the NHEK expression is scale-invariant, the identification of NHEK radii with Kerr radii is ambiguous: one identifies the NHEK radius $R$ with the radius in Kerr scaling as $r=M\pa{1+\epsilon^pR}$, up to an overall $R$-independent factor. It is only after the throat is reattached to the asymptotically flat region, and the dilation symmetry is broken, that NHEK radii can be unambiguously identified with Kerr radii. 

A physically interesting Kerr radius that scales into NHEK is the ISCO radius, which according to Eq.~\eqref{eq:ISCO} has a near-horizon limit $r_\mathrm{ms}=M\br{1+2^{1/3}\pa{\epsilon\kappa}^{2/3}}$. The band of radii in Kerr with a finite radial separation from the ISCO in the extremal limit all scale like $r=M\pa{1+\epsilon^{2/3}R}$, and the limit \eqref{eq:Parameters} with $p=2/3$ produces precisely the NHEK metric \eqref{eq:NHEK}. In this limit, the four-velocity of the ISCO becomes \cite{Gralla2016a}
\begin{align}
	\label{eq:TangentISCO}
	U=\frac{1}{2M}\frac{4}{\sqrt{3}R}\pa{\pd_T-\frac{3}{4}R\pd_\Phi},
\end{align}
which is both timelike and finite.\footnote{The ISCO has finite Kerr energy given by Eq.~\eqref{eq:EquatorialGeodesicsParameters}, and retains finite energy in the NHEK limit. However, plunging trajectories that were resolved by the far region will be null in this limit, and their NHEK energy will typically diverge; see, $e.g.$, Ref.~\cite{Gralla2016a} and Fig.~1 therein.} Therefore, the ``$p=2/3$ NHEK" resolves the part of the throat in which ISCO-scale physics occurs. Again, note that while the near-horizon limit of Eq.~\eqref{eq:ISCO} appears to identify the ISCO radius with the NHEK radius $R=2^{1/3}\kappa^{2/3}$, the dilation invariance of the NHEK metric \eqref{eq:NHEK} indicates that there is in fact no meaningful way to assign a definite radius to the ISCO within NHEK. In fact, in contrast to near-NHEK, all circular geodesics in NHEK are marginally stable: from the near-horizon viewpoint, the ISCO is in some sense \textit{everywhere} within the $p=2/3$ NHEK \cite{Gralla2015}. 

\begin{figure}[!ht]
	\centering
	\begin{tabular}{ c c c c }
		\multirow{12}{*}{\includegraphics[width=.55\textwidth]{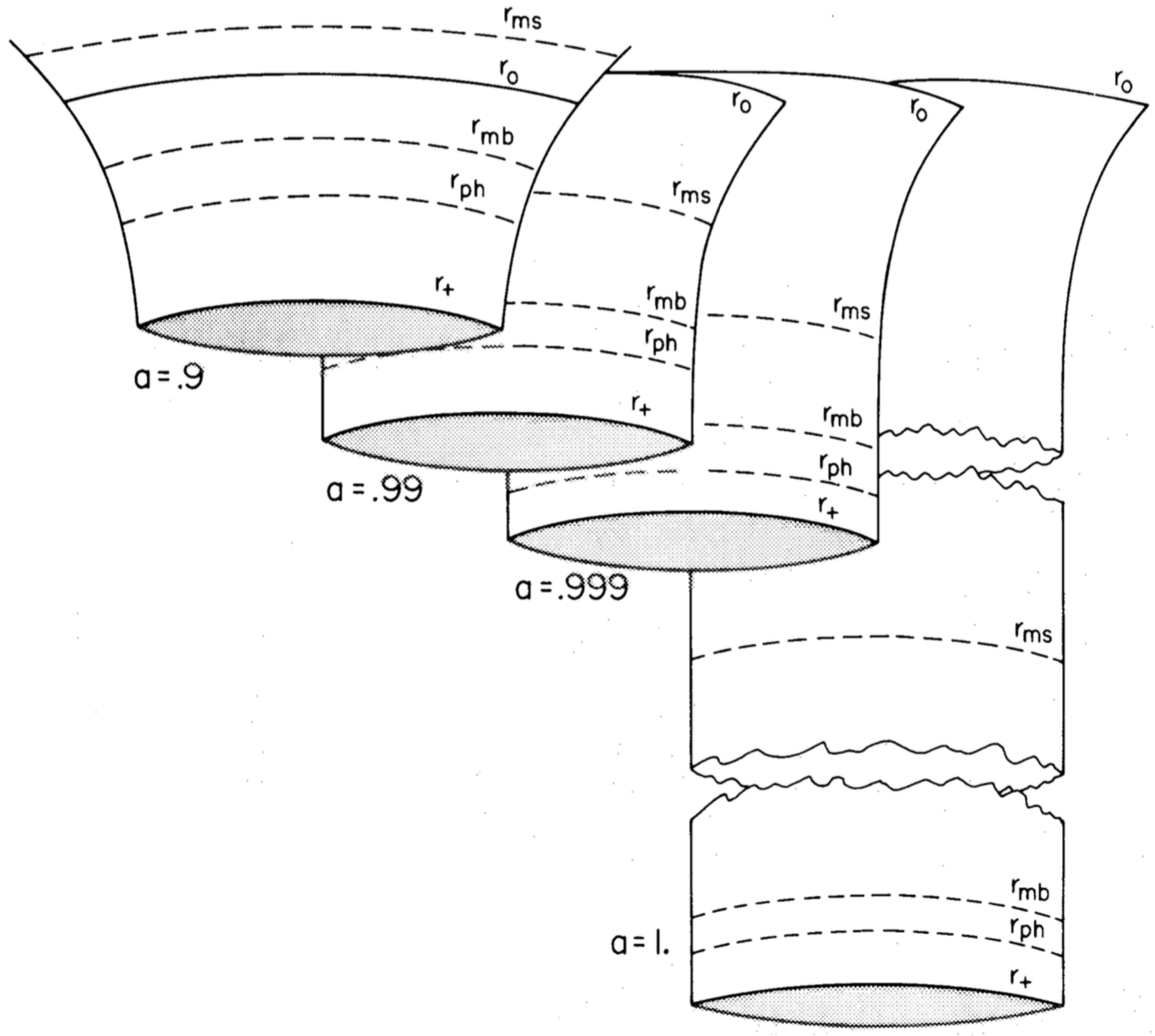}} & \vspace{20pt} & \\
		& \multicolumn{2}{c}{$a=M\sqrt{1-\pa{\epsilon\kappa}^2}$\qquad\qquad\qquad} & \\
		& & \vspace{10pt} & \\
		& \multicolumn{2}{c}{$r\sim M\pa{1+\epsilon^0R}$} & Far region \\
		& & \vspace{10pt} & \\
		& $\Big\}\ ds\sim M\ab{\log{\epsilon}}$ & & \\
		& & \vspace{10pt} & \\
		& \multicolumn{2}{c}{$r\sim M\pa{1+\epsilon^pR}$} & \hfill NHEK $\pa{0<p<1}$ \\
		& & \vspace{10pt} & \\
		& $\Big\}\ ds\sim M\ab{\log{\epsilon}}$ & & \\
		& & \vspace{10pt} & \\
		& \multicolumn{2}{c}{$r\sim M\pa{1+\epsilon R}$} & near-NHEK $\pa{p=1}$ \\
		& & \vspace{5pt} &
	\end{tabular}
	\caption{Throat geometry of a high-spin black hole. (Figure reproduced from Ref.~\cite{Bardeen1972} with additional labels.) ``High spin" means $a=M\sqrt{1-\pa{\epsilon\kappa}^2}$ with $\epsilon\ll1$. As the black hole approaches extremality ($\epsilon\to0$), its near-horizon region develops an infinitely deep throat, which becomes a spacetime in its own right. Its geometry is described by the Near-Horizon Extreme Kerr (NHEK) metric \eqref{eq:NHEK} and exhibits an emergent scaling symmetry consisting of dilations of the throat (or equivalently, translations along the infinite length of the cylinder).}
	\label{fig:Throat}
\end{figure}

To summarize, any two radii that scale to the horizon at the same rate (``lie in the same band'') end up in the same near-horizon geometry at a finite proper radial distance from each other. For instance, the photon orbit and IBCO radii both lie in the horizon band, and hence scale to the same near-NHEK region. Accordingly, the proper radial distance between these scales tends to a finite limit:
\begin{subequations}
\begin{align}
	ds\pa{r_+,r_\mathrm{ph}}&=M\log\sqrt{3}+\O{\epsilon},\\
	ds\pa{r_+,r_\mathrm{mb}}&=M\log\pa{1+\sqrt{2}}+\O{\epsilon},\\
	ds\pa{r_\mathrm{ph},r_\mathrm{mb}}&
	=M\log\pa{\frac{1+\sqrt{2}}{\sqrt{3}}}+\O{\epsilon}.
\end{align}
\end{subequations}
On the other hand, radii that lie in different bands, $i.e.$, that scale to the horizon at different rates, end up in different NHEKs that are infinitely far apart (separated by a divergent proper radial distance). For instance, the ISCO band is infinitely far from both the horizon band, as well as from the mouth of the throat, which we may for instance define as the spin-independent equatorial radius of the ergosphere, $r_0=2M$:
\begin{subequations}
\begin{align}
	ds\pa{r_+,r_\mathrm{ms}}&=\frac{M}{3}\log\pa{\frac{2^4}{\epsilon\kappa}}+\O{\epsilon^{2/3}},\\
	ds\pa{r_\mathrm{ms},r_0}&=M\br{1+\frac{2}{3}\log\pa{\frac{1}{\sqrt{2}\epsilon\kappa}}}+\O{\epsilon^{2/3}},\\
	ds\pa{r_+,r_0}&
	=M\br{1+\log\pa{\frac{2}{\epsilon\kappa}}}+\O{\epsilon^2}.
\end{align}
\end{subequations}
These facts are summarized in Fig.~\ref{fig:Throat}, where the different bands appear stacked on top of one another, with cracks denoting the logarithmically divergent proper radial distance separating them. From this point of view, the precisely extremal, zero-temperature black hole is a degenerate limit in which all the throat geometries merge: near-NHEK disappears and the different NHEKs coalesce into one.

Finally, note that the expansion about extremality defined by Eq.~\eqref{eq:Parameters} can be viewed both as a small-temperature expansion and an expansion in the divergent proper depth $D=ds\pa{r_+,r_0}=M\ab{\log{\epsilon}}+\O{\epsilon^0}$ of the throat: indeed, at leading order,
\begin{align}
	\epsilon=e^{-D/M}.
\end{align}
Thus, subleading corrections due to deviations from extremality are exponentially suppressed in the characteristic length scale of the system, which diverges in the extremal limit. Similar behavior is of course observed near critical points in condensed matter systems---this analogy was further developed in Ref.~\cite{Gralla2016a}.

When studying the extremal Kerr black hole, it is important to note that neither the far metric (extreme Kerr) nor the near metric (NHEK) is more fundamental than the other: away from extremality, the Kerr metric resolves physics in the entire spacetime, but near extremality, the spacetime decouples into two regions. Each of these two regions is described by its own metric, which fails in the other region: while NHEK resolves the near-horizon region, it fails to resolve the far region (for instance, it is not asymptotically flat), and the far metric does not resolve the throat region. As is usual for smooth extremal solutions in general relativity, the extreme Kerr geometry serves to interpolate between two separate vacuum solutions: flat space in the far region and NHEK in the near region. The two regions of spacetime are on equal footing.

\subsection{Emergent conformal symmetry}
\label{subsec:ConformalSymmetryInTheSky}

In many situations (including those of astrophysical interest), it is appropriate to treat the Kerr geometry as a fixed background while neglecting gravitational backreaction of the matter system (as well as gravitational excitations). When this approximation is valid, it suffices to work strictly with the NHEK metric and its exact isometries. In other applications, one is interested not only in the vacuum NHEK geometry, but in all spacetimes that approach NHEK asymptotically in some appropriate sense.\footnote{The precise choice of boundary conditions is delicate \cite{Guica2009,Dias2009,Amsel2009,Compere2012}.} Although these geometries all possess a long throat and approximate scale-invariance, generic members of this class of spacetimes have no exact isometries. It is the symmetries of the class of spacetimes, rather than the symmetries of a specific spacetime, that control gravitational dynamics in the throat.

In attempting to compute this generalized symmetry group, one often finds an enhancement of the global conformal isometry group $\mathsf{SL}(2,\mathbb{R})$ to an infinite-dimensional local conformal symmetry \cite{Compere2012}. The details of calculations of this type depend delicately on the choice of boundary conditions. We will focus on a particular class of symmetry transformations that have been repeatedly utilized \cite{Porfyriadis2014,Hadar2014,Hadar2015,Hadar2017,Compere2018} in calculating geodesics in NHEK and near-NHEK, and defer a complete asymptotic symmetry group analysis to future work. These large diffeomorphisms are the NHEK analogue of boundary reparameterizations of the AdS$_2$ throat discussed in Ref.~\cite{Maldacena2016b} and should be related to inequivalent ways of reattaching the Kerr throat region to the exterior geometry.

Starting with the NHEK line element \eqref{eq:NHEK},
\begin{align}
	d\hat{s}^2=2M^2\Gamma\br{-R^2\ed T^2+\frac{\ed R^2}{R^2}+\ed\theta^2+\Lambda^2\pa{\ed\Phi+R\ed T}^2},
\end{align}
we consider a coordinate transformation of the form\footnote{We thank Abhishek Pathak for help in deriving this transformation from its AdS$_3$ analogue \cite{Roberts2012}.}
\begin{align}
	\label{eq:ConformalTransformation}
	T=f(t)+\frac{2f''(t)\br{f'(t)}^2}{4r^2\br{f'(t)}^2-\br{f''(t)}^2},\qquad
	R=\frac{4r^2\br{f'(t)}^2-\br{f''(t)}^2}{4r\br{f'(t)}^3},\qquad
	\Phi=\phi+\log\br{\frac{2rf'(t)-f''(t)}{2rf'(t)+f''(t)}}.
\end{align}
The resulting line element is given by
\begin{align}
	\label{eq:VirasoroNHEK}
	d\hat{s}^2=2M^2\Gamma\cu{-r^2\pa{1+\frac{\cu{f(t);t}}{2r^2}}^2\ed t^2+\frac{\ed r^2}{r^2}+\ed\theta^2+\Lambda^2\br{\ed\phi+r\pa{1-\frac{\cu{f(t);t}}{2r^2}}\ed t}^2},
\end{align}
where we introduced the Schwarzian derivative
\begin{align}
	\cu{f(\cdot);\cdot}=\pa{\frac{f''}{f'}}'-\frac{1}{2}\pa{\frac{f''}{f'}}^2.
\end{align}
These metrics are the NHEK analogues of the AdS$_3$ Ba\~nados metrics \cite{Banados1999,Compere2016}. Note that at the boundary $r\to\infty$, this coordinate change implements a time reparameterization $T\to f(t)$, and that as a result, subleading components of the metric transform like the expectation value of a stress-tensor component in CFT$_2$.

Infinitesimally, the conformal transformation \eqref{eq:ConformalTransformation} is implemented by the action of the vector field
\begin{align}
	\xi\br{\epsilon(t)}=\pa{\epsilon(t)+\frac{\epsilon''(t)}{2r^2}}\pd_t-r\epsilon'(t)\pd_r-\frac{\epsilon''(t)}{r}\pd_\phi,\qquad
	f(t)=t+\epsilon(t)+\O{\epsilon^2}.
\end{align}
This can be decomposed into modes
\begin{align}
	\xi_n=\xi\br{t^{1-n}},\qquad
	n\in\mathbb{Z},
\end{align}
which obey the Witt algebra at the boundary,
\begin{align}
	\br{\xi_m,\xi_n}=(m-n)\xi_{m+n}+\O{\frac{1}{r^3}}.
\end{align}
The $\mathsf{SL}(2,\mathbb{R})$ isometry group of NHEK is generated by the vector fields
\begin{align}
	\xi_0=H_0,\qquad
	\xi_{\pm1}=H_\pm,
\end{align}
whose corresponding finite diffeomorphisms are given by the M\"obius transformations with vanishing Schwarzian:\footnote{More precisely: dilations by $a^2$ are obtained by setting $b=c=0$ and $d=a^{-1}$; time-translations by $b$ are obtained by setting $a=d=1$ and $c=0$; special conformal transformations by a parameter $c$ are obtained by setting $a=d=1$ and $b=0$.}
\begin{align}
	f(t)=\frac{at+b}{ct+d},\qquad
	ad-bc=1.
\end{align}
The rest of the symmetry transformations with nonvanishing $\cu{f(t);t}$ are spontaneously broken. Of particular interest here is the exponential map
\begin{align}
	f(t)=e^{\kappa t}
	\qquad\Longrightarrow\qquad
	\cu{f(t);t}=-\frac{\kappa^2}{2},
\end{align}
for which the metric becomes
\begin{align}
	\label{eq:ExponentialNHEK}
	d\hat{s}^2=2M^2\Gamma\cu{-r^2\pa{1-\pa{\frac{\kappa}{2r}}^2}^2\ed t^2+\frac{\ed r^2}{r^2}+\ed\theta^2+\Lambda^2\br{\ed\phi+r\pa{1+\pa{\frac{\kappa}{2r}}^2}\ed t}^2}.
\end{align}
Near the boundary, this diffeomorphism acts as the exponential map on the boundary time. It is the analogue of the usual conformal transformation from the plane to the cylinder, which puts a CFT$_2$ at finite temperature. In fact, the metric \eqref{eq:ExponentialNHEK} is actually near-NHEK, as can be seen by performing a further (small) diffeomorphism
\begin{align}
	r=\frac{1}{2}\pa{\bar{r}+\sqrt{\bar{r}^2-\kappa^2}},
\end{align}
which puts it in the form of Eq.~\eqref{eq:NearNHEK}. By composing these transformations, one can directly map near-NHEK,
\begin{align}
	d\bar{s}^2=2M^2\Gamma\br{-\pa{r^2-\kappa^2}\ed t^2+\frac{\ed r^2}{r^2-\kappa^2}+\ed\theta^2+\Lambda^2\pa{\ed\phi+r\ed t}^2},
\end{align}
into NHEK via the coordinate change
\begin{align}
	\label{eq:NHEK2NearNHEK}
	T=e^{\kappa t}\frac{r}{\sqrt{r^2-\kappa^2}},\qquad
	R=e^{-\kappa t}\frac{\sqrt{r^2-\kappa^2}}{\kappa},\qquad
	\Phi=\phi+\frac{1}{2}\log\pa{\frac{r-\kappa}{r+\kappa}}.
\end{align}
This transformation also maps the NHEK Killing vectors \eqref{eq:KillingFieldsNHEK} and near-NHEK Killing vectors \eqref{eq:KillingFieldsNearNHEK} into each other. It is important to note that this map is not surjective: its range covers only a subset of the NHEK Poincar\'e patch. Within that image, the inverse transformation is
\begin{align}
	t=\frac{1}{2\kappa}\log\pa{T^2-\frac{1}{R^2}},\qquad
	r=\kappa TR,\qquad
	\phi=\Phi-\frac{1}{2}\log\pa{\frac{TR-1}{TR+1}}.
\end{align}
Since the map \eqref{eq:NHEK2NearNHEK} is a diffeomorphism between near-NHEK and a subset of the Poincar\'e patch in NHEK (rather than its entirety), the near-NHEK and NHEK patches are locally (but not globally) diffeomorphic. Of course, since both near-NHEK and NHEK have horizons, they are not geodesically complete spacetimes. As we will discuss in the next section, they have the same maximal extension: global NHEK.

\subsection{The global strip and the Poincar\'e patch}
\label{subsec:GlobalNHEK}

To obtain the maximal extension of the NHEK spacetime, we pass from the Poincar\'e coordinates $(T,R,\Phi)$ with a coordinate singularity at $TR=1$ to global coordinates $(\tau,y,\varphi)$ that can be smoothly continued past this surface. The transformation from Poincar\'e NHEK to a patch $\tau\in(-\pi,\pi)$ of global NHEK is given by
\begin{align}
	\label{eq:PoincareToGlobal}
	\tau=\arctan\br{\frac{2TR^2}{\pa{1-T^2}R^2+1}},\qquad
	y=\frac{\pa{1+T^2}R^2-1}{2R},\qquad
	\varphi=\Phi-\log\pa{\frac{\pa{1-TR}^2+R^2}{\sqrt{\br{\pa{1+T^2}R^2-1}^2+4R^2}}}.
\end{align}
Different branches of the arctangent map Poincar\'e NHEK to diffeomorphic patches of global NHEK which differ by $\tau\to\tau+2\pi n$ translations in global time. For this reason, all geodesic motion in global NHEK is oscillatory with period $2\pi$. The inverse map from the patch $\tau\in(-\pi,\pi)$ of global NHEK to Poincar\'e NHEK is given by
\begin{align}
	\label{eq:GlobalToPoincare}
	T=\frac{\sin{\tau}\sqrt{1+y^2}}{\cos{\tau}\sqrt{1+y^2}+y},\qquad
	R=\cos{\tau}\sqrt{1+y^2}+y,\qquad
	\Phi=\varphi+\log\ab{\frac{\cos{\tau}+y\sin{\tau}}{1+\sin{\tau}\sqrt{1+y^2}}}.
\end{align}
Under this coordinate transformation, the NHEK metric \eqref{eq:NHEK} becomes
\begin{align}
	\label{eq:GlobalNHEK}
	d\tilde{s}^2=2M^2\Gamma\br{-\pa{1+y^2}\ed\tau^2+\frac{\ed y^2}{1+y^2}+\ed\theta^2+\Lambda^2\pa{\ed\varphi+y\ed\tau}^2},
\end{align}
and the NHEK Killing vector fields \eqref{eq:KillingFieldsNHEK} are mapped into
\begin{subequations}
\label{eq:KillingFieldsGlobalNHEK}
\begin{align}
	H_\pm&=\pa{1\pm\frac{y\cos{\tau}}{\sqrt{1+y^2}}}\pd_\tau\pm\sin{\tau}\sqrt{1+y^2}\pd_y\pm\frac{\cos{\tau}}{\sqrt{1+y^2}}\pd_\varphi,\\
	H_0&=\frac{y\sin{\tau}}{\sqrt{1+y^2}}\pd_\tau-\cos{\tau}\sqrt{1+y^2}\pd_y+\frac{\sin{\tau}}{\sqrt{1+y^2}}\pd_\varphi,\\
	W_0&=\pd_\varphi.
\end{align}
\end{subequations}
It will be convenient for us to complexify this algebra by introducing new (complex) generators
\begin{align}
	L_0=i\pd_\tau,\qquad
	L_\pm=e^{\pm i\tau}\br{\pm\frac{y}{\sqrt{1+y^2}}\pd_\tau-i\sqrt{1+y^2}\pd_y\pm\frac{1}{\sqrt{1+y^2}}\pd_\varphi},
\end{align}
which obey the same commutation relations:
\begin{align}
	\br{L_0,L_\pm}=\mp L_\pm,\qquad
	\br{L_+,L_-}=2L_0,\qquad
	\br{W_0,L_\pm}=\br{W_0,L_0}=0.
\end{align}
These two sets of generators are related by
\begin{align}
    \label{eq:Automorphism}
	H_\pm=-iL_0\pm\frac{1}{2}\pa{L_+-L_-},\qquad
	H_0=-\frac{i}{2}\pa{L_++L_-}.
\end{align}
The generator $L_0$ of global time translations is the analogue of the Hamiltonian of a CFT$_2$ defined on the cylinder.

\begin{figure}[!ht]
	\centering
	\includegraphics[scale=.7]{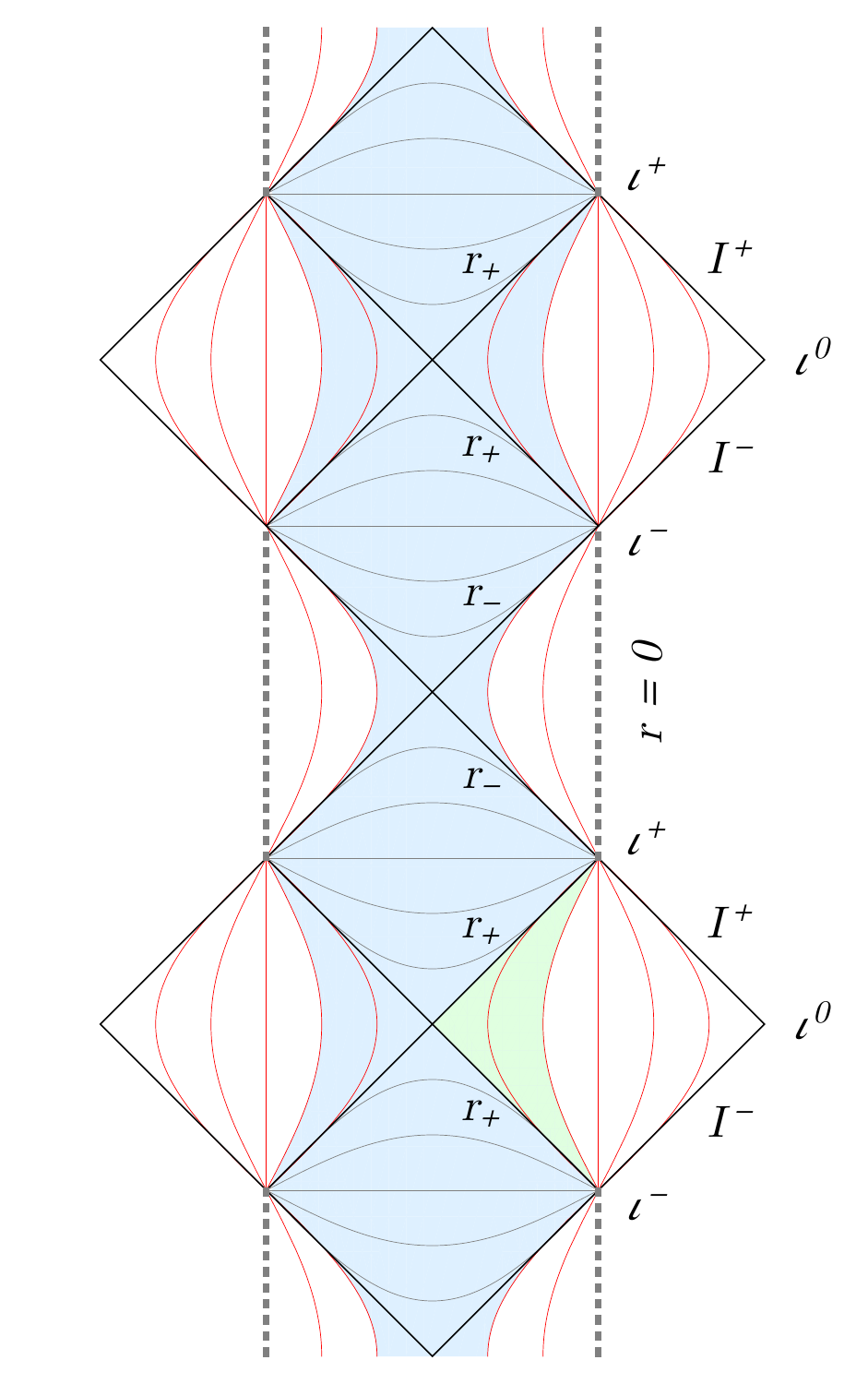}
	\includegraphics[scale=.7]{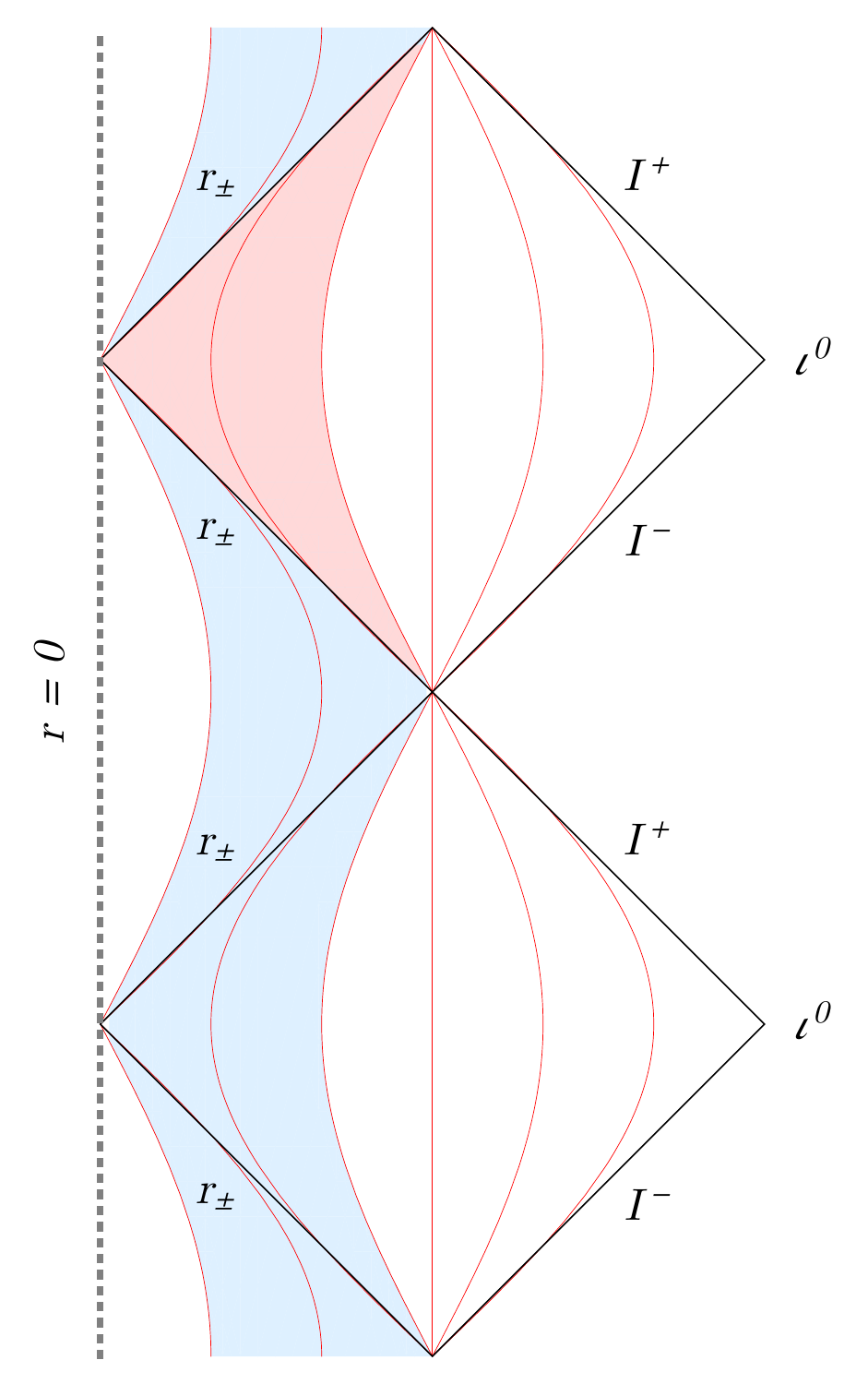}
	\includegraphics[scale=.7]{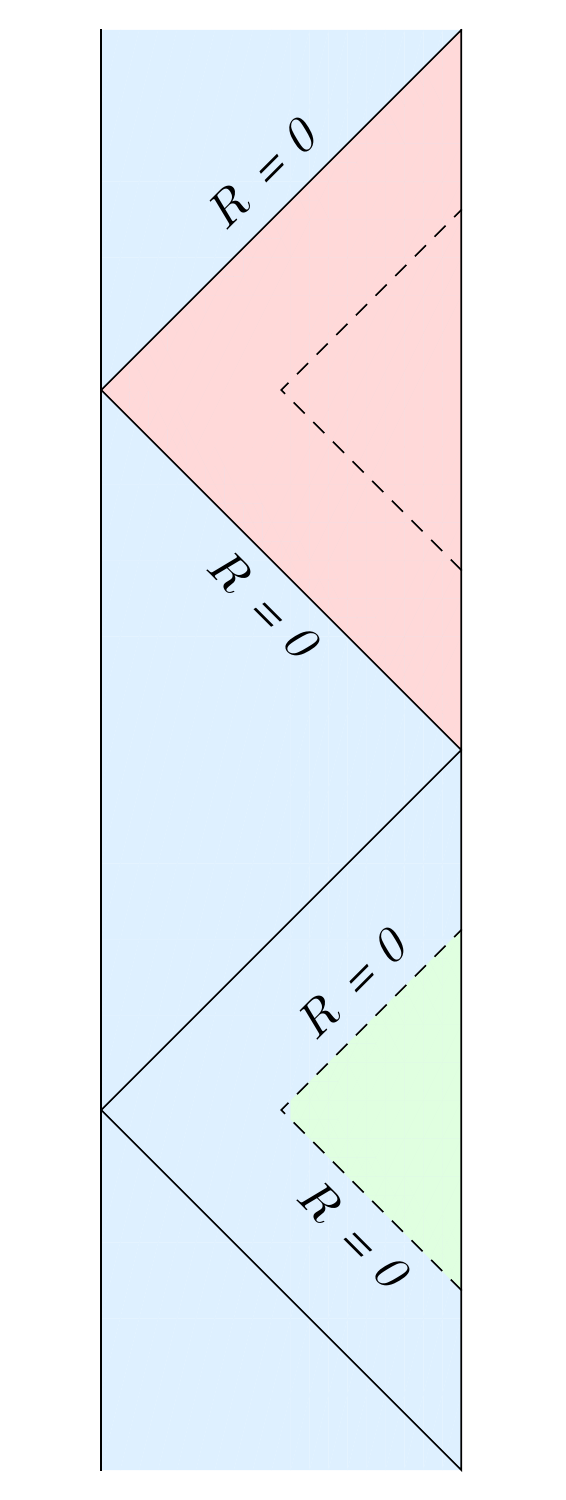}
	\caption{Carter-Penrose diagrams for the equatorial plane of near-extreme Kerr [left], extreme Kerr [middle] and NHEK [right]. (The diagrams depend on $\theta$ because these geometries are not spherically symmetric and, away from the equatorial plane, it is possible to continue past the singularity.) In the diagrams for (near-)extreme Kerr, $\mathcal{I}^\pm$ denotes future/past null infinity ($r=+\infty$ with $t=\pm\infty$), $\iota^0$ is spacelike infinity ($r=+\infty$), $r_\pm$ is the outer/inner horizon, and the singularity lies at $r=0$. The thin lines correspond to hypersurfaces of constant $r$, which are timelike outside the black hole [red] and spacelike inside [gray].}
	\label{fig:PenroseDiagrams}
\end{figure}

The causal structure of the global NHEK geometry is best understood by introducing a compactified radius\footnote{The inverse transformation on the domain $\psi\in\br{0,\pi}$ is given by $\psi=\pi/2+\arctan{y}$.}
\begin{align}
	y=-\cot\psi,\qquad
	\psi\in\br{0,\pi},
\end{align}
in terms of which the global NHEK line element \eqref{eq:GlobalNHEK} becomes
\begin{align}
	d\tilde{s}^2=2M^2\Gamma\br{\frac{-\ed\tau^2+\ed\psi^2}{\sin^2{\psi}}+\ed\theta^2+\Lambda^2\pa{\ed\varphi-\cot{\psi}\ed\tau}^2}.
\end{align}
As explained in detail in App.~\ref{app:AdS2}, the $(\tau,\psi)$ part of the metric describes the global strip of AdS$_2$ parameterized by $\tau\in\pa{-\infty,+\infty}$ and $\psi\in\br{0,\pi}$. Since this two-dimensional metric is conformally flat, with lines of $\tau=\pm\psi+\tau_0$ manifestly null, it is straightforward to obtain its Carter-Penrose diagram, depicted in Fig.~\ref{fig:PenroseDiagrams}.

\begin{figure}[!ht]
	\begin{minipage}[c]{0.45\textwidth}
	\includegraphics[width=.8\textwidth]{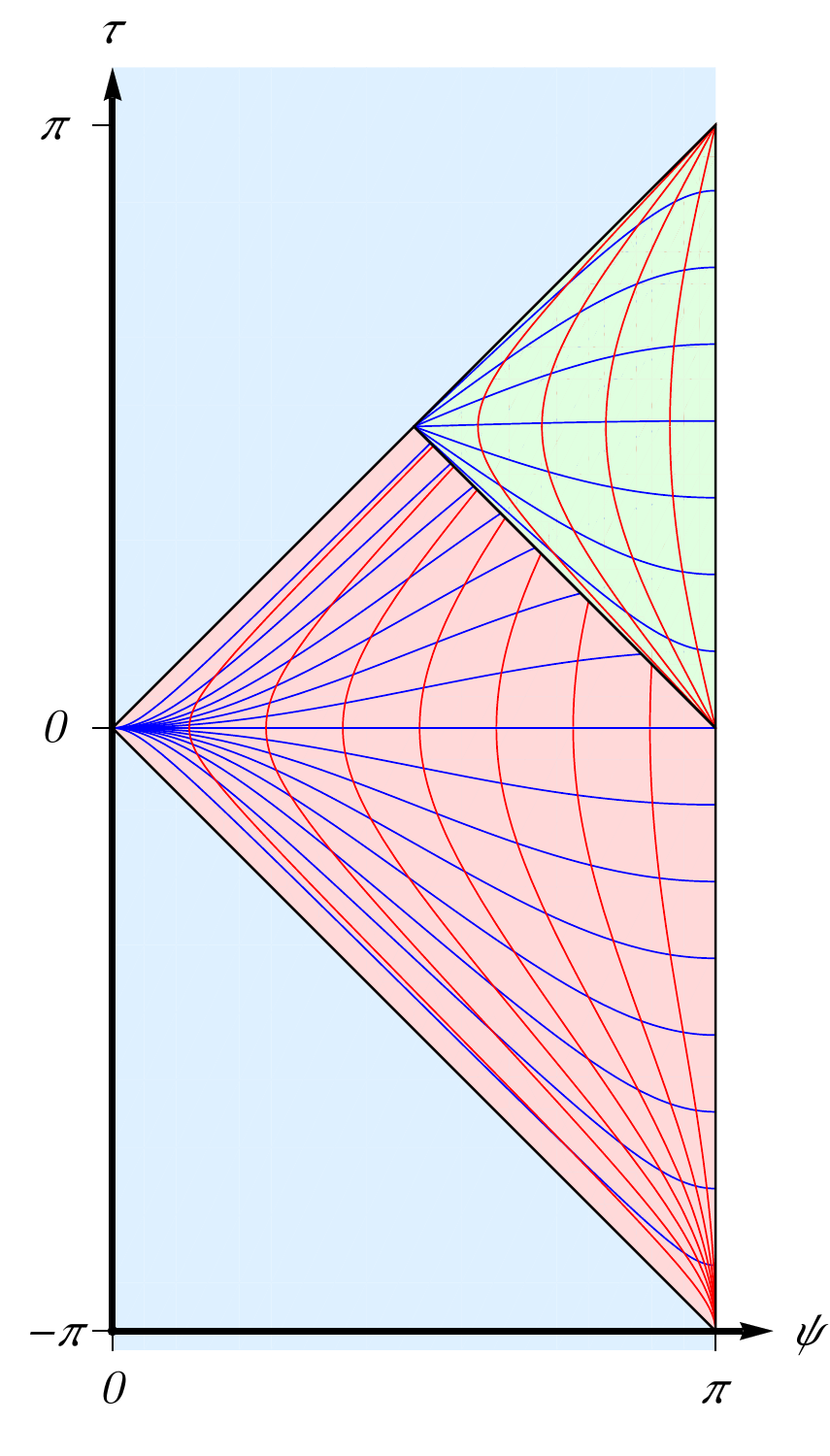}
	\end{minipage}
	\hfill
	\begin{minipage}[c]{0.5\textwidth}
	\caption{Carter-Penrose diagrams for the equatorial plane of global NHEK [blue] and the Poincar\'e patch for NHEK [red] and near-NHEK [green]. The global NHEK strip is covered by the global coordinates $(\psi,\tau)$ with $\psi\in\br{0,\pi}$ and $\tau\in\pa{-\infty,+\infty}$. The Poincar\'e patch for NHEK [red] can be embedded at any vertical position along the strip; here, it is depicted as the red patch at $\tau\in\br{-\pi,+\pi}$, corresponding to the mapping \eqref{eq:PoincareToGlobal}. Likewise, near-NHEK can also be placed anywhere on the strip; here, it is depicted as the green patch embedded at $\tau\in\br{0,+\pi}$, $\psi\in\br{\pi/2,\pi}$, corresponding to the composition of the mappings \eqref{eq:NHEK2NearNHEK} and \eqref{eq:PoincareToGlobal}. In both the NHEK and near-NHEK patches, slices of constant Poincar\'e time $T$ and radius $R$ are depicted by blue and red lines, respectively. For this choice of NHEK patch, the future/past horizon is located at $\tau=\pm\psi$ (or, equivalently, $R=0$ with $TR=\pm1$).}
	\label{fig:GlobalStrip}
	\end{minipage}
\end{figure}

Under the embedding \eqref{eq:GlobalToPoincare} of Poincar\'e NHEK into global NHEK, we see that the Poincar\'e coordinates $(T,R)$ cover the patch $\tau\in\pa{-\pi,+\pi}$ with $\ab{\tau}\le\psi\le\pi$ (in particular, the future/past horizon is located at $TR=\pm1$):
\begin{align}
	T=\frac{\sin{\tau}}{\cos{\tau}-\cos{\psi}},\qquad
	R=\frac{\cos{\tau}-\cos{\psi}}{\sin{\psi}}.
\end{align}
Hence, the lines of constant $T$ and constant $R$ are respectively given by
\begin{align}
	\cos{\tau}=\frac{T^2\cos{\psi}+\sqrt{1+T^2\sin^2{\psi}}}{1+T^2},\qquad
	\cos{\tau}=\cos{\psi}+R\sin{\psi},
\end{align}
and in Fig.~\ref{fig:GlobalStrip}, they are plotted at constant intervals of $x$, where $T=\tan{x/2}$ and $R=\tan{x/2}$.

Under the embedding of near-NHEK into global NHEK obtained by composing Eqs.~\eqref{eq:NHEK2NearNHEK} and \eqref{eq:GlobalToPoincare}, we see that the Poincar\'e coordinates $(T,R)$ cover the patch $\tau\in\pa{0,+\pi}$ with $\pi-\psi\le\tau\le\psi$:
\begin{align}
	T=\frac{1}{\kappa}\log\sqrt{1-\frac{2\cos{\tau}}{\cos{\tau}-\cos{\psi}}},\qquad
	R=\kappa\frac{\sin{\tau}}{\sin{\psi}}.
\end{align}
Hence, the lines of constant $T$ and constant $R$ are respectively given by
\begin{align}
	\cos{\tau}=\frac{e^{2\kappa T}-1}{e^{2\kappa T}+1}\cos{\psi},\qquad
	\sin{\tau}=\frac{R}{\kappa}\sin{\psi},
\end{align}
and in Fig.~\ref{fig:GlobalStrip} they are plotted at constant intervals of $x$, where $T=\log\pa{\tan{x/2}}^{1/\kappa}$ and $R=\kappa/\sin{x}$.

Note that the finite $\mathsf{SL}(2,\mathbb{R})$ transformations---Eq.~\eqref{eq:ConformalTransformation} with $f(t)$ a M\"obius transformation---leave the NHEK metric \eqref{eq:NHEK} invariant, but not its geodesics: they are mapped into each other under the action of the global conformal group. Equivalently, we can study geodesics in global coordinates and obtain different geodesics in the Poincar\'e patch by using the embedding \eqref{eq:PoincareToGlobal} composed with the $\mathsf{SL}(2,\mathbb{R})$ transformations. This strategy can be used to map circular orbits to (slow or fast) plunges and was used to great effect in Refs.~\cite{Hadar2014,Hadar2015,Hadar2017,Compere2018}. In fact, most preexisting analyses of geodesics in NHEK have focused on equatorial circular geodesics or plunging geodesics obtainable through the above mapping from the ISCO geodesic \eqref{eq:TangentISCO}.

\section{Geodesics in NHEK}
\label{sec:GeodesicsInNHEK}

In this section, we analyze geodesic motion in global NHEK, the Poincar\'e patch, and near-NHEK. For each of these spacetimes, we follow the procedure outlined for Kerr in Sec.~\ref{sec:Kerr} and recast the geodesic equation in first-order form. We then obtain explicit expressions for all the path integrals appearing in the equation. Finally, inverting the expression for the time-lapse allows us to solve for the radial motion as a function of coordinate time and thereby derive a complete and explicit parameterization of all NHEK geodesics.

We begin by analyzing the geodesic equation in global NHEK (Sec.~\ref{subsec:GlobalGeodesics}), as it is a geodesically complete spacetime, unlike Poincar\'e NHEK (Sec.~\ref{subsec:PoincareGeodesics}) and near-NHEK (Sec.~\ref{subsec:NearGeodesics}). As a byproduct of our analysis, we elucidate the action of the NHEK isometry group $\mathsf{SL}(2,\mathbb{R})\times\mathsf{U}(1)$ on the space of geodesics. We find that the group orbits are classified by the $\mathsf{SL}(2,\mathbb{R})$ Casimir $\mathcal{C}$ and angular momentum $L$, which completely determine the polar motion. Any two NHEK geodesics with the same polar motion can be mapped into each other by the action of $\mathsf{SL}(2,\mathbb{R})\times\mathsf{U}(1)$ using the explicit transformations presented at the end of App.~\ref{app:AdS2}.

\subsection{Geodesics in global NHEK}
\label{subsec:GlobalGeodesics}

Recall from Sec.~\ref{subsec:GlobalNHEK} that in global coordinates, the NHEK line element is
\begin{align}
	\label{eq:GlobalMetric}
	d\tilde{s}^2=2M^2\Gamma\br{-\pa{1+y^2}\ed\tau^2+\frac{\ed y^2}{1+y^2}+\ed\theta^2+\Lambda^2\pa{\ed\varphi+y\ed\tau}^2},
\end{align}
and the generators of $\mathsf{SL}(2,\mathbb{R})\times\mathsf{U}(1)$ are
\begin{align}
	L_0=i\pd_\tau,\qquad
	L_\pm=e^{\pm i\tau}\br{\pm\frac{y}{\sqrt{1+y^2}}\pd_\tau-i\sqrt{1+y^2}\pd_y\pm\frac{1}{\sqrt{1+y^2}}\pd_\varphi},\qquad W_0=\partial_\varphi.
\end{align}
The Casimir of $\mathsf{SL}(2,\mathbb{R})$ is the (manifestly reducible) symmetric Killing tensor
\begin{align}
	\mathcal{C}^{\mu\nu}=-L_0^\mu L_0^\nu+\frac{1}{2}\pa{L_+^\mu L_-^\nu+L_-^\mu L_+^\nu}.
\end{align}
It is related to $\tilde{K}^{\mu\nu}$, the NHEK limit \eqref{eq:KillingTensor} of the irreducible Killing tensor on Kerr \eqref{eq:KerrKilling}, by
\begin{align}
	\tilde{K}^{\mu\nu}=\mathcal{C}^{\mu\nu}+W_0^\mu W_0^\nu+M^2\tilde{g}^{\mu\nu}.
\end{align}

The motion of a free particle of mass $\mu$ and four-momentum $P^\mu$ is described by the geodesic equation,
\begin{align}
	P^\mu\tilde{\nabla}_\mu P^\nu=0,\qquad
	\tilde{g}^{\mu\nu}P_\mu P_\nu=-\mu^2.
\end{align} 
Geodesic motion in global NHEK is completely characterized by the three conserved quantities\footnote{Observe that $\mathcal{C}^{\mu\nu}P_\mu P_\nu=-\mathcal{H}_0^2+\frac{1}{2}\pa{\mathcal{H}_+\mathcal{H}_-+\mathcal{H}_-\mathcal{H}_+}$, where $\mathcal{H}_i=H_i^\mu P_\mu$ or $\mathcal{H}_i=L_i^\mu P_\mu$ are the conserved quantities associated with the generators of $\mathsf{SL}(2,\mathbb{R})$. Since these are not independent of each other, we use the $\mathsf{SL}(2,\mathbb{R})$ Casimir, which is in involution with all the $\mathcal{H}_i$ \cite{AlZahrani2011}. This is exactly analogous to exploiting the conservation of $J_z$ and $J^2$, rather than $(J_x,J_y,J_z)$, in a problem with $\mathsf{SO}(3)$ symmetry.}
\begin{subequations}
\label{eq:GlobalConservedQuantities}
\begin{gather}
	\triangle=iL_0^\mu P_\mu
	=-P_\tau,\qquad
	L=W_0^\mu P_\mu
	=P_\varphi,\\
	\mathcal{C}=\mathcal{C}^{\mu\nu}P_\mu P_\nu
	=P_\theta^2-\pa{1-\frac{1}{\Lambda^2}}P_\varphi^2+\pa{2M^2\Gamma}\mu^2
	=\frac{\pa{P_\tau-P_\varphi y}^2}{1+y^2}-\pa{1+y^2}P_y^2-P_\varphi^2,
\end{gather}
\end{subequations}
denoting the global energy, angular momentum parallel to the axis of symmetry, and ``Casimir" constant, respectively. When connecting NHEK geodesics to the far region in Kerr, it is more convenient to work with the Carter constant
\begin{align}
	K=\tilde{K}^{\mu\nu}P_\mu P_\nu
	=\mathcal{C}+L^2-\mu^2M^2,
\end{align}
which is directly related to its Kerr analogue $k$. In this paper, however, we restrict our attention to motion within NHEK, which is more easily characterized by the $\mathsf{SL}(2,\mathbb{R})$ Casimir $\mathcal{C}$.

By inverting the above relations for $\pa{\mu^2,\triangle,L,\mathcal{C}}$, we find that a particle following a geodesic in the global NHEK geometry \eqref{eq:GlobalMetric} has an instantaneous four-momentum $P=P_\mu\ed X^\mu$ of the form
\begin{align}
	\label{eq:KerrGeodesic}
	P\pa{X^\mu,\triangle,L,\mathcal{C}}=-\triangle\ed\tau\pm_y\frac{\sqrt{\mathcal{Y}(y)}}{1+y^2}\ed y\pm_\theta\sqrt{\Theta_n(\theta)}\ed\theta+L\ed\varphi,
\end{align}
where the two choices of sign $\pm_y$ and $\pm_\theta$ depend on the radial and polar directions of travel, respectively. Here, we also introduced radial and polar potentials
\begin{subequations}
\begin{align}
	\label{eq:GlobalRadialPotential}
	\mathcal{Y}(y)&=\pa{\triangle+Ly}^2-\pa{\mathcal{C}+L^2}\pa{1+y^2},\\
	\label{eq:PolarPotentialNHEK}
	\Theta_n(\theta)&=\mathcal{C}+\pa{1-\frac{1}{\Lambda^2}}L^2-\pa{2M^2\Gamma}\mu^2.
\end{align}
\end{subequations}
One can then raise $P_\mu$ to obtain the equations for the geodesic trajectory,
\begin{subequations}
\label{eq:GlobalMomentum}
\begin{align}
	\label{eq:GlobalRadialEquation}
	2M^2\Gamma\frac{dy}{d\sigma}&=\pm_y\sqrt{\mathcal{Y}(y)},\\
	2M^2\Gamma\frac{d\theta}{d\sigma}&=\pm_\theta\sqrt{\Theta_n(\theta)},\\
	2M^2\Gamma\frac{d\varphi}{d\sigma}&=-y\pa{\frac{\triangle+Ly}{1+y^2}}+\frac{L}{\Lambda^2},\\
	\label{eq:GlobalTimeEquation}
	2M^2\Gamma\frac{d\tau}{d\sigma}&=\frac{\triangle+Ly}{1+y^2}.
\end{align}
\end{subequations}
The parameter $\sigma$ is the affine parameter for massless particles ($\mu=0$), and is related to the proper time $\delta$ by $\delta=\mu\sigma$ for massive particles.

Following the same procedure as in Kerr, we find from Eqs.~\eqref{eq:GlobalMomentum} that a geodesic labeled by $(\triangle,L,\mathcal{C})$ connects spacetime points $X_s^\mu=\pa{\tau_s,y_s,\theta_s,\varphi_s}$ and $X_o^\mu=\pa{\tau_o,y_o,\theta_o,\varphi_o}$ if
\begin{subequations}
\begin{align}
	&\fint_{y_s}^{y_o}\frac{\ed y}{\pm_y\sqrt{\mathcal{Y}(y)}}=\fint_{\theta_s}^{\theta_o}\frac{\ed\theta}{\pm_\theta\sqrt{\Theta_n(\theta)}},\\
	\varphi_o-\varphi_s&=\fint_{y_s}^{y_o}\br{-y\pa{\frac{\triangle+Ly}{1+y^2}}}\frac{\ed y}{\pm_y\sqrt{\mathcal{Y}(y)}}+\fint_{\theta_s}^{\theta_o}\frac{L\Lambda^{-2}(\theta)}{\pm_\theta\sqrt{\Theta_n(\theta)}}\ed\theta,\\
	\tau_o-\tau_s&=\fint_{y_s}^{y_o}\pa{\frac{\triangle+Ly}{1+y^2}}\frac{\ed y}{\pm_y\sqrt{\mathcal{Y}(y)}}.
\end{align}
\end{subequations}
We may rewrite these conditions as
\begin{align}
	\label{eq:GlobalGeodesics}
	\tilde{I}_y=\tilde{G}_\theta,\qquad
	\varphi_o-\varphi_s=\tilde{G}_\varphi-\tilde{I}_\varphi,\qquad
	\tau_o-\tau_s=\tilde{I}_\tau,
\end{align}
where we have defined the integrals
\begin{subequations}
\begin{gather}
	\tilde{I}_y=\fint_{y_s}^{y_o}\frac{\ed y}{\pm_y\sqrt{\mathcal{Y}(y)}},\qquad
	\tilde{I}_\varphi=\fint_{y_s}^{y_o}y\pa{\frac{\triangle+Ly}{1+y^2}}\frac{\ed y}{\pm_y\sqrt{\mathcal{Y}(y)}},\qquad
	\tilde{I}_\tau=\fint_{y_s}^{y_o}\pa{\frac{\triangle+Ly}{1+y^2}}\frac{\ed y}{\pm_y\sqrt{\mathcal{Y}(y)}},\\
	\tilde{G}_\theta=\fint_{\theta_s}^{\theta_o}\frac{\ed\theta}{\pm_\theta\sqrt{\Theta_n(\theta)}},\qquad
	\tilde{G}_\varphi=\fint_{\theta_s}^{\theta_o}\frac{L\Lambda^{-2}(\theta)}{\pm_\theta\sqrt{\Theta_n(\theta)}}\ed\theta.
\end{gather}
\end{subequations}

\subsubsection{Qualitative description of geodesic motion}
\label{subsec:GlobalQualitative}

Before solving this geodesic equation outright, it is useful to determine the qualitative behavior of the geodesics projected onto the poloidal $(y,\theta)$ plane. The analysis of the polar motion follows directly from our earlier discussion in Sec.~\ref{sec:Kerr} for Kerr: indeed, it suffices to note that $\Theta_n(\theta)$ takes the form $Q+P\cos^2{\theta}-\ell^2\cot^2{\theta}$ under the identification
\begin{align}
	\label{eq:AngularIdentificationNHEK}
	Q=\mathcal{C}+\frac{3}{4}L^2-\mu^2M^2,\qquad
	P=\frac{L^2}{4}-\mu^2M^2,\qquad
	\ell=L.
\end{align}
Therefore, the NHEK angular integral $\tilde{G}_\theta\pa{\mu^2,L,\mathcal{C}}$ takes the same form as the Kerr angular integral $G_\theta(Q,P,\ell)$ with
\begin{align}
	\tilde{G}_\theta\pa{\mu^2,L,\mathcal{C}}=G_\theta\pa{\mathcal{C}+\frac{3}{4}L^2-\mu^2M^2,\frac{L^2}{4}-\mu^2M^2,L}.
\end{align}
From Eq.~\eqref{eq:AngularIdentificationNHEK}, we see that NHEK geodesics (unlike their Kerr counterparts) are necessarily Type A (non-vortical) because $Q$ cannot be negative, since positivity of the angular potential requires that
\begin{align}
	Q=\mathcal{C}+\frac{3}{4}L^2-\mu^2M^2
	\ge\mathcal{C}+\pa{1-\frac{1}{\Lambda^2}}L^2-\mu^2M^2
	\ge\mathcal{C}+\pa{1-\frac{1}{\Lambda^2}}L^2-\mu^2M^2\pa{1+\cos^2{\theta}}
	=\Theta_n
	\ge 0.
\end{align}
Moreover, it is straightforward to check that $Q=0$ is only allowed for purely equatorial geodesics, a special case requiring a separate (and simpler) treatment. Generically, we therefore have the strict inequality $Q>0$, and the angular motion is necessarily of Type A with bounds given by Eq.~\eqref{eq:Q>0,P=0} (when $P=0$) or Eq.~\eqref{eq:Q>0,P!=0} (when $P\neq0$).

Having completed the analysis of the polar motion, we now turn to the radial motion. Its allowed range is heavily constrained by the requirement that the radial potential $\mathcal{Y}(y)$ remain positive at every point along the trajectory. As in Kerr, real zeroes of the radial potential $\mathcal{Y}(y)$ correspond to turning points $y_\pm$ in the radial motion. In global NHEK, the radial motion can be of two qualitatively different types (and is therefore simpler than in Kerr \cite{Bicak1989}):
\begin{itemize}\setlength{\itemindent}{.5in}
\item[\underline{Type I:}\ ]
Oscillatory motion between turning points $y_\pm$, corresponding to bound particles that are confined to NHEK and traverse the totality of the global strip, $\tau\in\pa{-\infty,+\infty}$. The global-time-lapse between successive turning points is $\Delta\tau=\pi$: each period of motion lies in a single Poincar\'e patch of the global strip.
\item[\underline{Type II:}]
Single-bounce motion (from a boundary at $y=\pm\infty$ to a turning point $y_t$ and back), corresponding to unbound particles that probe a single Poincar\'e patch of the global strip, $\tau\in\br{\tau_0,\tau_0+2\pi}$. The continuation of the geodesic beyond the intersection with the boundary depends on boundary conditions.\footnote{In Anti-de Sitter space, one typically imposes reflective boundary conditions in order to describe a closed system. In the present case, such a choice of boundary conditions would lead to periodic multiple-bounce trajectories. We will not explicitly consider such solutions as they are unlikely to be relevant to astrophysical black holes, but it is trivial to construct them by gluing together a sequence of single bounces.}
\end{itemize}
The type of motion is determined by the properties of the roots of the radial potential $\mathcal{Y}(y)$.
For generic values of the geodesic parameters, the (possibly complex) roots of $\mathcal{Y}(y)$ are given by 
\begin{align}
	\label{eq:GlobalTurningPoints}
	y_\pm=\frac{\triangle L\pm\sqrt{\pa{\triangle^2-\mathcal{C}}\pa{\mathcal{C}+L^2}}}{\mathcal{C}}.
\end{align}
As we vary the geodesic parameters, these roots move around in the complex $y$-plane. When one or more of the roots approaches or pinches the contour of integration, the radial motion of the allowed geodesics is then constrained to lie exclusively on one side of the root.

Positivity of energy in the local frame of the particle, $P^\tau\ge0$, imposes the additional constraint
\begin{align}
	\label{eq:PositiveEnergyCondition}
	\triangle+Ly\ge0.
\end{align}
In terms of the critical radius
\begin{align}
	y_c=-\frac{\triangle}{L},
\end{align}
this constraint becomes
\begin{align}
	y\gtrless y_c,\qquad
	L\gtrless 0.
\end{align}
The roots $y_\pm$ can never equal $y_c$ for real values of the geodesic parameter $\triangle$, since
\begin{align}
	\label{eq:RootSeparation}
	y_c=y_\pm\qquad\Longleftrightarrow\qquad
	\triangle^2=-L^2.
\end{align}
The roots $y_\pm$ will be real when the quantity $\pa{\triangle^2-\mathcal{C}}\pa{\mathcal{C}+L^2}$ is positive. Positivity of $Q$ provides the bound
\begin{align}
	\label{eq:ThetaBound}
	\mathcal{C}+\frac{3}{4}L^2-\mu^2M^2=Q
	>0\qquad\Longrightarrow\qquad
	\mathcal{C}>-\frac{3}{4}L^2+\mu^2M^2,
\end{align}
from which it follows that $\mathcal{C}+L^2>0$. Therefore the polynomial $\mathcal{Y}(y)$ has a pair of real roots whenever $\triangle^2-\mathcal{C}\ge0$:
\begin{align}
	y_\pm\in\mathbb{R}\qquad\Longleftrightarrow\qquad
	\triangle^2-\mathcal{C}\ge0.
\end{align}
Generic geodesics probe both positive and negative values of $y$, so the sign of $y_\pm$ is less important than its relation to $y_c$. Since $\mathcal{Y}(y)=-\mathcal{C}y^2+\O{y}$, there are now three cases to examine separately:
\begin{enumerate}
\item
\underline{$\mathcal{C}>0$:} In this case, we must demand that $y_\pm\in\mathbb{R}$, as $\mathcal{Y}(y)$ is negative outside the interval between its roots. That is, we must require that $0<\mathcal{C}<\triangle^2$ and $y\in\mathcal{I}=\br{y_-,y_+}$. Moreover, one can show that $y_c<y_-<y_+$ whenever $\triangle L>0$, and $y_-<y_+<y_c$ whenever $\triangle L<0$. For all values of $L$, the condition \eqref{eq:PositiveEnergyCondition} then requires that $\triangle>0$ and the entire interval $\mathcal{I}$ is allowed. These bounded geodesics are not permitted to escape NHEK.\footnote{This bound agrees with the bound derived for equatorial geodesics in Section 2 of Ref.~\cite{Bardeen1999}, whose authors set $M=2^{-1/2}$.}
\item
\underline{$\mathcal{C}<0$:} In this case, the roots $y_\pm$ are manifestly real because $\triangle^2-\mathcal{C}>0$. Moreover, $y_+<y_c<y_-$.\footnote{By Eq.~\eqref{eq:RootSeparation}, each of the roots $y_\pm$ must lie strictly to one side of $y_c$. When $\triangle=0$, it is evident that $y_+<0=y_c<y_-$, so this ordering must hold in general. One can also prove these inequalities directly with some straightforward algebraic manipulations.} Hence, the radial potential $\mathcal{Y}(y)$ is nonnegative if $y\le y_+$ or $y\ge y_-$. There are two further subcases to consider:\footnote{Note that $L=0$ is forbidden by Eq.~\eqref{eq:ThetaBound}, since $\mathcal{C}<0$ is incompatible with $\mathcal{C}>\mu^2M^2\ge0$.}
\begin{enumerate}
\item
$L>0$, in which case the condition \eqref{eq:PositiveEnergyCondition} requires that $y>y_c$, so only the branch $y\ge y_-$ is allowed.
\item
$L<0$, in which case the condition \eqref{eq:PositiveEnergyCondition} requires that $y<y_c$, so only the branch $y\le y_+$ is allowed.
\end{enumerate}
These unbound geodesics fall in from the boundary $(y=\pm\infty)$ of NHEK,  and eventually reach a turning point before heading back towards the same boundary.
\item
\underline{$\mathcal{C}=0$:} In this case, $\mathcal{Y}(y)$ is linear with a zero at $y_0=\frac{L^2-\triangle^2}{2\triangle L}$ and a slope given by $2\triangle L$. Therefore, for $\triangle L>0$, the allowed range of motion is $y\in[y_0,\infty)$, while for $\triangle L<0$ the allowed range is $y\in(-\infty,y_0]$. The condition \eqref{eq:PositiveEnergyCondition} then requires that $\triangle>0$ and the entire semi-infinite interval is allowed for both signs of $L$. These geodesics are a limiting example of Type II geodesics and are also unbound.
\end{enumerate}
To summarize, the type of radial motion is picked out by the sign of $\mathcal{C}$:\vspace{10pt}

\ovalbox{\parbox{.95\textwidth}{
\begin{enumerate}
\item[I.]
$0<\mathcal{C}\le\triangle^2$ corresponds to Type I oscillations. Geodesic motion is permitted in the range $y\in\br{y_-,y_+}$, provided that $\triangle>0$.
\item[II.]
$-L^2<\mathcal{C}<0$ corresponds to Type II motion. Geodesic motion is allowed either in the range $y\in[y_-,+\infty)$ when $L>0$, or in the range $y\in(-\infty,y_+]$ when $L<0$. (The angular motion imposes a stricter bound $-\frac{3}{4}L^2+\mu^2M^2<\mathcal{C}<0$.)
\item[III.]
$\mathcal{C}=0$ corresponds to a limit of Type II motion in which $\triangle>0$ is required. Geodesic motion is allowed either in the range $[y_0,\infty)$ when $L>0$, or in the range $(-\infty,y_0]$ when $L<0$.
\end{enumerate}
}}
\vspace{10pt}

\begin{figure}[!ht]
	\begin{minipage}[c]{0.45\textwidth}
		\includegraphics[width=.6\textwidth]{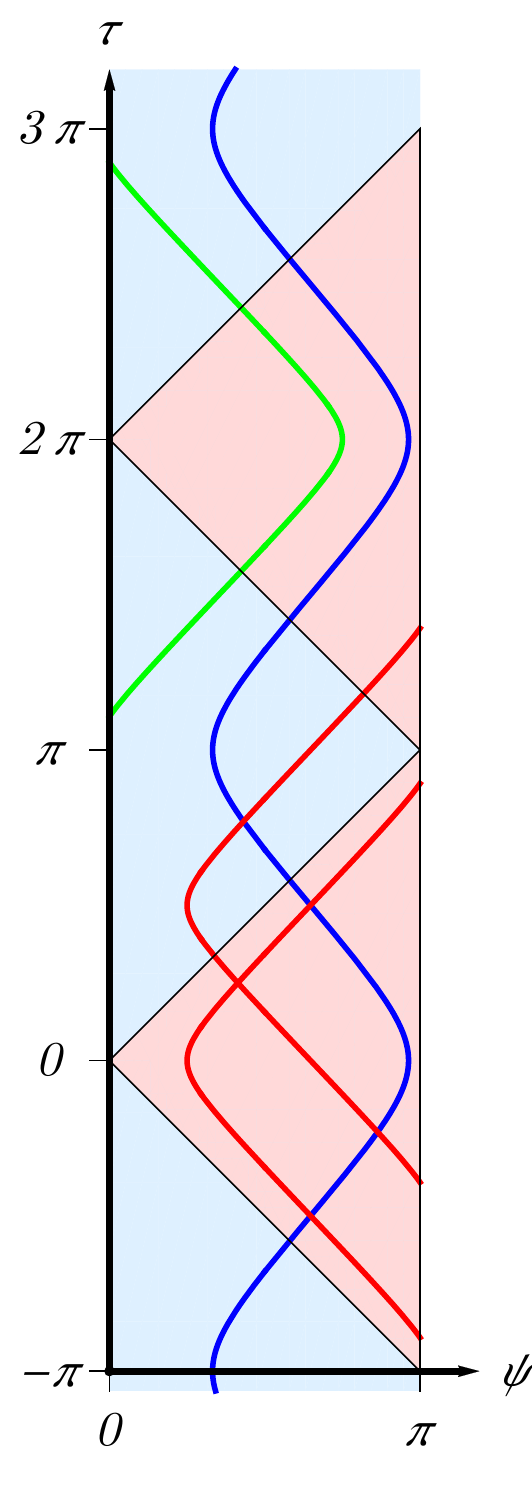}
	\end{minipage}
	\hfill
	\begin{minipage}[c]{0.5\textwidth}
	\caption{Plot of the radial motion $(\tau,\psi(\tau))$ of geodesics in global NHEK, with $\psi=\pi/2+\arctan{y}$ where $y(\tau)$ is given in Eq.~\eqref{eq:GlobalRadialMotion}. The blue geodesic is Type I ($\mathcal{C}>0$) with $\pa{\triangle,L,\mathcal{C}}=(4,2,2)$. The red geodesic is Type II ($\mathcal{C}<0$) and coming in from the right boundary ($L>0$) with $\pa{\triangle,L,\mathcal{C}}=(4,2,-2)$. It appears twice, with a global time shift of $\Delta\tau=\pi/2$. The green geodesic is Type II ($\mathcal{C}<0$) and coming in from the left boundary ($L<0$) with $\pa{\triangle,L,\mathcal{C}}=(4,-2,-2)$. In the Poincar\'e patch, the Type I geodesics come out of the past horizon, travel to increasingly large radius until they encounter a turning point at a maximal allowed radius, and then fall back into the future horizon. The upper red geodesic is Type IIA, with $p^R\gtrless0$ in the upper/lower Poincar\'e patch. The lower red geodesic is Type IIB with $L>0$, while the green geodesic is Type IIB with $L<0$. Type IIA and Type IIB are related by global-time-translation symmetry.}
	\label{fig:GeodesicTypes}
	\end{minipage}
\end{figure}

Note that the sign of the $\mathsf{SL}(2,\mathbb{R})$ Casimir $\mathcal{C}$ determines the eventual fate of a NHEK geodesic: those with $\mathcal{C}>0$ are bound and remain in the throat forever, while those with $\mathcal{C}<0$ can reach the boundary in finite global time and escape. These different types of motion are illustrated in Fig.~\ref{fig:GeodesicTypes}. There is also a special class of timelike bound geodesics of constant global radius arising as a special limit of the Type I ($\mathcal{C}>0$) motion. In the limit $\mathcal{C}\to\triangle^2$, the zeroes of the radial potential coalesce, $y_-\to y_+$, and one has $\mathcal{Y}(y_\pm)=\mathcal{Y}'(y_\pm)=0$. The gravitational pull exerted by the black hole on these particles is exactly balanced by the centrifugal force they experience from its rotation.

After dimensional reduction to two dimensions, NHEK geodesics coincide with Lorentz force trajectories of charged particles evolving in two-dimensional Anti-de Sitter spacetime AdS$_2$ with a background radial electric field (see App.~\ref{app:AdS2}). From that two-dimensional perspective, the geodesics of constant global radius arise when gravity in the AdS$_2$ box is exactly balanced out by the acceleration of the charge by the electric field.

\subsubsection{Geodesic integrals in global NHEK}

We now wish to compute the integrals $\tilde{I}_y$, $\tilde{I}_\tau$, $\tilde{I}_\varphi$, $\tilde{G}_\theta$, and $\tilde{G}_\varphi$ that appear in the NHEK geodesic equation \eqref{eq:GlobalGeodesics}. As previously noted, the NHEK angular integral $\tilde{G}_\theta\pa{\mu^2,L,\mathcal{C}}$ takes the same form as the Kerr angular integral $G_\theta(Q,P,\ell)$ under the indentification
\begin{align}
	\label{eq:AngularIdentificationNHEKbis}
	Q=\mathcal{C}+\frac{3}{4}L^2-\mu^2M^2,\qquad
	P=\frac{L^2}{4}-\mu^2M^2,\qquad
	\ell=L.
\end{align}
Similarly, given that
\begin{align}
	\frac{L}{\Lambda^2}=L\csc^2{\theta}-\frac{L}{4}\cos^2{\theta}-\frac{3L}{4},
\end{align}
the NHEK integral $\tilde{G}_\varphi\pa{\mu^2,L,\mathcal{C}}$ is a linear combination of the Kerr angular integrals $G_\theta$, $G_t$, and $G_\phi$,
\begin{align}
	\tilde{G}_\varphi=\frac{L}{4}\pa{4G_\phi-G_t-3G_\theta},
\end{align}
with the identification of arguments given by \eqref{eq:AngularIdentificationNHEKbis}. Therefore, we can simply apply the results derived in Sec.~\ref{sec:Kerr} for the Kerr angular integrals in order to evaluate their NHEK counterparts.
It remains to evaluate the radial integrals $\tilde{I}_y$, $\tilde{I}_\tau$, and $\tilde{I}_\varphi$. For this purpose, it is useful to define signs
\begin{align}
	\nu_s=\sign\pa{P_s^y}
	=(-1)^w\sign\pa{P_o^y},\qquad
	\nu_o=\sign\pa{P_o^y}
	=(-1)^w\sign\pa{P_s^y},
\end{align}
where $P_s^y$ and $P_o^y$ denote the radial momentum evaluated at the endpoints $X_s^\mu$ and $X_o^\mu$ of the geodesic, respectively, and $w\in\mathbb{N}$ denotes the number of turning points in the radial motion. Then by working through all the possible configurations, one can check that the radial path integral unpacks as follows:
\begin{subequations}
\label{eq:GlobalType}
\begin{align}
	\label{eq:GlobalTypeI}
	\text{Type I:}\qquad&
	\fint_{y_s}^{y_o}=\pa{w\pm\frac{\nu_o-\nu_s}{2}}\int_{y_-}^{y_+}-\nu_s\int_{y_\pm}^{y_s}+\nu_o\int_{y_\pm}^{y_o},\\
	\label{eq:GlobalTypeII}
	\text{Type II, III:}\qquad&
	\fint_{y_s}^{y_o}=-\nu_s\int_{y_\pm}^{y_s}+\nu_o\int_{y_\pm}^{y_o},\qquad
	\pm=\sign\pa{\mathcal{C}L}.
\end{align}
\end{subequations}
For the Type I geodesics, Eq.~\eqref{eq:GlobalTypeI} constitutes two equivalent representations that differ only in the choice of reference turning point for the integrals. For the Type II geodesics, one must select the (unique) turning point that is actually encountered along the trajectory.

We can now explicitly evaluate the radial integrals. For our purposes, the relevant basis of integrals is given by
\begin{subequations}
\begin{align}
	\tilde{\mathcal{I}}_1^\pm&\equiv\int_{y_\pm}^{y_i}\frac{\ed y}{\sqrt{\mathcal{C}\pa{y_+-y}\pa{y-y_-}}},\\
	\tilde{\mathcal{I}}_2^\pm&\equiv\int_{y_\pm}^{y_i}\pa{\frac{1}{1+y^2}}\frac{\ed y}{\sqrt{\mathcal{C}\pa{y_+-y}\pa{y-y_-}}},\\
	\tilde{\mathcal{I}}_3^\pm&\equiv\int_{y_\pm}^{y_i}\pa{\frac{y}{1+y^2}}\frac{\ed y}{\sqrt{\mathcal{C}\pa{y_+-y}\pa{y-y_-}}},
\end{align}
\end{subequations}
in terms of which
\begin{subequations}
\begin{align}
	\int_{y_\pm}^{y_i}\frac{\ed y}{\sqrt{\mathcal{Y}(y)}}&=\tilde{\mathcal{I}}_1^\pm,\\
	\int_{y_\pm}^{y_i}y\pa{\frac{\triangle+Ly}{1+y^2}}\frac{\ed y}{\sqrt{\mathcal{Y}(y)}}&=L\pa{\tilde{\mathcal{I}}_1^\pm-\tilde{\mathcal{I}}_2^\pm}+\triangle\tilde{\mathcal{I}}_3^\pm,\\
	\int_{y_\pm}^{y_i}\pa{\frac{\triangle+Ly}{1+y^2}}\frac{\ed y}{\sqrt{\mathcal{Y}(y)}}&=\triangle\tilde{\mathcal{I}}_2^\pm+L\tilde{\mathcal{I}}_3^\pm.
\end{align}
\end{subequations}
In each case, for the allowed range of $y_i$, the integrals $\tilde{\mathcal{I}}_1^\pm$, $\tilde{\mathcal{I}}_2^\pm$, and $\tilde{\mathcal{I}}_3^\pm$ are all manifestly real. To evaluate them, we will use the identities \eqref{eq:GlobalIntegralIdentities} with the quantities
\begin{align}
	\label{eq:GlobalParameters}
	q_\pm=\pa{\frac{1+y_\pm^2}{1+y_\mp^2}}^{1/4}>0,\qquad
	\cos{\alpha_\pm}=\mp\frac{1+y_-y_+}{\sqrt{\pa{1+y_-^2}\pa{1+y_+^2}}}\in\br{-1,1}.
\end{align}
Observe that $\ab{\cos\alpha_\pm}=1$ if and only if $y_+=y_-$, which only occurs when $\mathcal{C}=\triangle^2$. This corresponds to a degenerate limit of Type I motion in which there is no radial motion and the geodesic remains at fixed constant global radius.

When $\mathcal{C}>0$, the choice of turning point (and thus the sign $\pm$ of $\tilde{\mathcal{I}}_i^\pm$) is arbitrary, so we can use either of the substitutions
\begin{align}
	y=\frac{y_\pm+y_\mp x^2}{1+x^2}\qquad\Longrightarrow\qquad
	x_i^\pm=\ab{\frac{y_i-y_\pm}{y_i-y_\mp}}^{1/2}
	\in[0,+\infty).
\end{align}
In terms of the quantities defined in Eq.~\eqref{eq:GlobalIntegrals}, this results in (assuming $\alpha_-\neq0$)
\begin{subequations}
\begin{align}
	\tilde{\mathcal{I}}_1^\pm&=\mp\frac{2}{\sqrt{\mathcal{C}}}\int_0^{x_i^\pm}\frac{\ed x}{1+x^2}
	=\mp\frac{2}{\sqrt{\mathcal{C}}}\arctan{x_i^\pm},\\
	\tilde{\mathcal{I}}_2^\pm&=\mp\frac{2}{\sqrt{\mathcal{C}}}\int_0^{x_i^\pm}\frac{\pa{1+x^2}\ed x}{\pa{1+y_\pm^2}+2\pa{1+y_-y_+}x^2+\pa{1+y_\mp^2}x^4}
	=\mp\frac{2}{\sqrt{\mathcal{C}}}\br{\frac{\tilde{I}_-\pa{x_i^\pm,q_\pm,\alpha_+}+\tilde{I}_+\pa{x_i^\pm,q_\pm,\alpha_+}}{1+y_\mp^2}},\\
	\tilde{\mathcal{I}}_3^\pm&=\mp\frac{2}{\sqrt{\mathcal{C}}}\int_0^{x_i^\pm}\frac{\pa{y_\pm+y_\mp x^2}\ed x}{\pa{1+y_\pm^2}+2\pa{1+y_-y_+}x^2+\pa{1+y_\mp^2}x^4}
	=\mp\frac{2}{\sqrt{\mathcal{C}}}\br{\frac{y_\pm\tilde{I}_-\pa{x_i^\pm,q_\pm,\alpha_+}+y_\mp\tilde{I}_+\pa{x_i^\pm,q_\pm,\alpha_+}}{1+y_\mp^2}}.
\end{align}
\end{subequations}
On the other hand, when $\mathcal{C}<0$, we must choose the sign $\pm$ according to whether $L\lessgtr0$, and then use the substitution
\begin{align}
	y=\frac{y_\pm-y_\mp x^2}{1-x^2}\qquad\Longrightarrow\qquad
	x_i^\pm=\ab{\frac{y_i-y_\pm}{y_i-y_\mp}}^{1/2}
	\in\br{0,1}.
\end{align}
In terms of the quantities defined in Eq.~\eqref{eq:GlobalIntegrals}, this results in
\begin{subequations}
\begin{align}
	\tilde{\mathcal{I}}_1^\pm&=\mp\frac{2}{\sqrt{\ab{\mathcal{C}}}}\int_0^{x_i^\pm}\frac{\ed x}{1-x^2}
	=\mp\frac{2}{\sqrt{\ab{\mathcal{C}}}}\arctanh{x_i^\pm},\\
	\tilde{\mathcal{I}}_2^\pm&=\mp\frac{2}{\sqrt{\ab{\mathcal{C}}}}\int_0^{x_i^\pm}\frac{\pa{1-x^2}\ed x}{\pa{1+y_\pm^2}-2\pa{1+y_-y_+}x^2+\pa{1+y_\mp^2}x^4}
	=\mp\frac{2}{\sqrt{\ab{\mathcal{C}}}}\br{\frac{\tilde{I}_-\pa{x_i^\pm,q_\pm,\alpha_-}-\tilde{I}_+\pa{x_i^\pm,q_\pm,\alpha_-}}{1+y_\mp^2}},\\
	\tilde{\mathcal{I}}_3^\pm&=\mp\frac{2}{\sqrt{\ab{\mathcal{C}}}}\int_0^{x_i^\pm}\frac{\pa{y_\pm-y_\mp x^2}\ed x}{\pa{1+y_\pm^2}-2\pa{1+y_-y_+}x^2+\pa{1+y_\mp^2}x^4}
	=\mp\frac{2}{\sqrt{\ab{\mathcal{C}}}}\br{\frac{y_\pm\tilde{I}_-\pa{x_i^\pm,q_\pm,\alpha_-}-y_\mp\tilde{I}_+\pa{x_i^\pm,q_\pm,\alpha_-}}{1+y_\mp^2}}.
\end{align}
\end{subequations}
Finally, when $\mathcal{C}=0$, the integrals $\tilde{\mathcal{I}}_1^\pm$, $\tilde{\mathcal{I}}_2^\pm$, and $\tilde{\mathcal{I}}_3^\pm$ degenerate to
\begin{align}
	\tilde{\mathcal{I}}_1^\pm=\int_{y_0}^{y_i}\frac{\ed y}{\sqrt{2\triangle L\pa{y-y_0}}},\qquad
	\tilde{\mathcal{I}}_2^\pm=\int_{y_0}^{y_i}\pa{\frac{1}{1+y^2}}\frac{\ed y}{\sqrt{2\triangle L\pa{y-y_0}}},\qquad
	\tilde{\mathcal{I}}_3^\pm=\int_{y_0}^{y_i}\pa{\frac{y}{1+y^2}}\frac{\ed y}{\sqrt{2\triangle L\pa{y-y_0}}}.
\end{align}
To evaluate these simpler integrals, we use the substitution
\begin{align}
	y=y_0\mp x^2\qquad\Longrightarrow\qquad
	x_i^0=\sqrt{\ab{y_i-y_0}}\in[0,+\infty),
\end{align}
where we must still choose the sign $\pm=\sign\pa{-L}$ since  $L\lessgtr0$ corresponds to $y_i\lessgtr y_0$. This results in
\begin{subequations}
\begin{align}
	\tilde{\mathcal{I}}_1^\pm&=\mp\frac{2}{\sqrt{2\triangle\ab{L}}}\int_0^{x_i^0}\ed x
	=\mp\frac{2x_i^0}{\sqrt{2\triangle\ab{L}}},\\
	\tilde{\mathcal{I}}_2^\pm&=\mp\frac{2}{\sqrt{2\triangle\ab{L}}}\int_0^{x_i^0}\frac{\ed x}{\pa{1+y_0^2}\mp2y_0x^2+x^4}
	=\mp\frac{2}{\sqrt{2\triangle\ab{L}}}\tilde{I}_-\pa{x_i^0,q_0,\alpha_0^\pm},\\
	\tilde{\mathcal{I}}_3^\pm&=\mp\frac{2}{\sqrt{2\triangle\ab{L}}}\int_0^{x_i^0}\frac{\pa{y_0\mp x^2}\ed x}{\pa{1+y_0^2}\mp2y_0x^2+x^4}
	=\mp\frac{2}{\sqrt{2\triangle\ab{L}}}\br{y_0\tilde{I}_-\pa{x_i^0,q_0,\alpha_0^\pm}\mp\tilde{I}_+\pa{x_i^0,q_0,\alpha_0^\pm}},\\
	\label{eq:GlobalParameters0}
	&q_0=\pa{1+y_0^2}^{1/4}>0,\qquad
	\alpha_0^\pm=\frac{\pi}{2}\mp\arctan{y_0}\in\br{0,\pi}.
\end{align}
\end{subequations}

To put everything together, it is useful to define
\begin{subequations}
\begin{align}
	\tilde{\mathcal{I}}_\varphi^\pm(x,\alpha,s)&=\frac{\pa{\triangle y_\pm-L}\tilde{I}_-\pa{x,q_\pm,\alpha}+\pa{s\triangle y_\mp-L}\tilde{I}_+\pa{x,q_\pm,\alpha}}{1+y_\mp^2},\\
	\tilde{\mathcal{I}}_\varphi^0(x,\alpha,s)&=\pa{\triangle y_0-L}\tilde{I}_-\pa{x,q_0,\alpha}-s\triangle\tilde{I}_+\pa{x,q_0,\alpha},\\
	\tilde{\mathcal{I}}_\tau^\pm(x,\alpha,s)&=\frac{\pa{\triangle+Ly_\pm}\tilde{I}_-\pa{x,q_\pm,\alpha}+\pa{\triangle+sLy_\mp}\tilde{I}_+\pa{x,q_\pm,\alpha}}{1+y_\mp^2},\\
	\tilde{\mathcal{I}}_\tau^0(x,\alpha,s)&=\pa{\triangle+Ly_0}\tilde{I}_-\pa{x,q_0,\alpha}-sL\tilde{I}_+\pa{x,q_0,\alpha},
\end{align}
\end{subequations}
where we remind the reader that the quantities $y_\pm$, $q_\pm$, $q_0$, and $\tilde{I}_\pm$ are defined in Eqs.~\eqref{eq:GlobalTurningPoints}, \eqref{eq:GlobalParameters}, \eqref{eq:GlobalParameters0}, and \eqref{eq:GlobalIntegrals}, respectively. Then, we see that when $\mathcal{C}>0$,
\begin{subequations}
\label{eq:GlobalIntegralsTypeI}
\begin{align}
	\int_{y_\pm}^{y_i}\frac{\ed y}{\sqrt{\mathcal{Y}(y)}}&=\mp\frac{2}{\sqrt{\mathcal{C}}}\arctan{x_i^\pm},\\
	\int_{y_\pm}^{y_i}y\pa{\frac{\triangle+Ly}{1+y^2}}\frac{\ed y}{\sqrt{\mathcal{Y}(y)}}&=\mp\frac{2}{\sqrt{\mathcal{C}}}\br{L\arctan{x_i^\pm}+\tilde{\mathcal{I}}_\varphi^\pm\pa{x_i^\pm,\alpha_+,+}},\\
	\int_{y_\pm}^{y_i}\pa{\frac{\triangle+Ly}{1+y^2}}\frac{\ed y}{\sqrt{\mathcal{Y}(y)}}&=\mp\frac{2}{\sqrt{\mathcal{C}}}\tilde{\mathcal{I}}_\tau^\pm\pa{x_i^\pm,\alpha_+,+},
\end{align}
\end{subequations}
where the choice of sign $\pm$ is arbitrary. In particular, for Type I geodesics (with $\mathcal{C}>0$), the half-period integrals are
\begin{align}
	\label{eq:GlobalTimeLapse}
	\int_{y_-}^{y_+}\frac{\ed y}{\sqrt{\mathcal{Y}(y)}}=\frac{\pi}{\sqrt{\mathcal{C}}},\qquad
	\int_{y_-}^{y_+}y\pa{\frac{\triangle+Ly}{1+y^2}}\frac{\ed y}{\sqrt{\mathcal{Y}(y)}}=\frac{\pi L}{\sqrt{\mathcal{C}}},\qquad
	\int_{y_-}^{y_+}\pa{\frac{\triangle+Ly}{1+y^2}}\frac{\ed y}{\sqrt{\mathcal{Y}(y)}}=\pi.
\end{align}
On the other hand, when $\mathcal{C}<0$, we must choose $\pm=\sign\pa{-L}$ and
\begin{subequations}
\label{eq:GlobalIntegralsTypeII}
\begin{align}
	\int_{y_\pm}^{y_i}\frac{\ed y}{\sqrt{\mathcal{Y}(y)}}&=\mp\frac{2}{\sqrt{\ab{\mathcal{C}}}}\arctanh{x_i^\pm},\\
	\int_{y_\pm}^{y_i}y\pa{\frac{\triangle+Ly}{1+y^2}}\frac{\ed y}{\sqrt{\mathcal{Y}(y)}}&=\mp\frac{2}{\sqrt{\ab{\mathcal{C}}}}\br{L\arctanh{x_i^\pm}+\tilde{\mathcal{I}}_\varphi^\pm\pa{x_i^\pm,\alpha_-,-}},\\
	\int_{y_\pm}^{y_i}\pa{\frac{\triangle+Ly}{1+y^2}}\frac{\ed y}{\sqrt{\mathcal{Y}(y)}}&=\mp\frac{2}{\sqrt{\ab{\mathcal{C}}}}\tilde{\mathcal{I}}_\tau^\pm\pa{x_i^\pm,\alpha_-,-}.
\end{align}
\end{subequations}
Likewise, when $\mathcal{C}=0$ and $\pm=\sign\pa{-L}$,
\begin{subequations}
\label{eq:GlobalIntegralsTypeIII}
\begin{align}
	\int_{y_0}^{y_i}\frac{\ed y}{\sqrt{\mathcal{Y}(y)}}&=\mp\frac{2x_i^0}{\sqrt{2\triangle\ab{L}}},\\
	\int_{y_\pm}^{y_i}y\pa{\frac{\triangle+Ly}{1+y^2}}\frac{\ed y}{\sqrt{\mathcal{Y}(y)}}&=\mp\frac{2}{\sqrt{2\triangle\ab{L}}}\br{Lx_i^0+\tilde{\mathcal{I}}_\varphi^0\pa{x_i^0,\alpha_0^\pm,\pm}},\\
	\int_{y_\pm}^{y_i}\pa{\frac{\triangle+Ly}{1+y^2}}\frac{\ed y}{\sqrt{\mathcal{Y}(y)}}&=\mp\frac{2}{\sqrt{2\triangle\ab{L}}}\tilde{\mathcal{I}}_\tau^0\pa{x_i^0,\alpha_0^\pm,\pm}.
\end{align}
\end{subequations}

Substituting Eqs.~\eqref{eq:GlobalIntegralsTypeI}--\eqref{eq:GlobalTimeLapse} into Eq.~\eqref{eq:GlobalTypeI} yields the geodesic integrals for Type I motion:
\begin{subequations}
\begin{empheq}[box=\ovalbox]{align}
	\shortintertext{\centering\underline{Type I:}\qquad$0<\mathcal{C}\le\triangle^2$ with $\pm$ arbitrary}
	\tilde{I}_y&=\frac{1}{\sqrt{\mathcal{C}}}\br{w\pi\mp\nu_s\pa{\frac{\pi}{2}-2\arctan{x_s^\pm}}\pm\nu_o\pa{\frac{\pi}{2}-2\arctan{x_o^\pm}}},\\
	\tilde{I}_\varphi&=L\tilde{I}_y\pm\frac{2}{\sqrt{\mathcal{C}}}\br{\nu_s\tilde{\mathcal{I}}_\varphi^\pm\pa{x_s^\pm,\alpha_+,+}-\nu_o\tilde{\mathcal{I}}_\varphi^\pm\pa{x_o^\pm,\alpha_+,+}},\\
	\label{eq:GlobalTI}
	\tilde{I}_\tau&=w\pi\mp\nu_s\br{\frac{\pi}{2}-\frac{2}{\sqrt{\mathcal{C}}}\tilde{\mathcal{I}}_\tau^\pm\pa{x_s^\pm,\alpha_+,+}}\pm\nu_o\br{\frac{\pi}{2}-\frac{2}{\sqrt{\mathcal{C}}}\tilde{\mathcal{I}}_\tau^\pm\pa{x_o^\pm,\alpha_+,+}}.
\end{empheq}
\end{subequations}

The geodesic integrals for Type II and Type III motion are likewise obtained by substituting Eqs.~\eqref{eq:GlobalIntegralsTypeII}--\eqref{eq:GlobalIntegralsTypeIII} into Eq.~\eqref{eq:GlobalTypeII}, resulting in
\begin{subequations}
\begin{empheq}[box=\ovalbox]{align}
	\shortintertext{\centering\underline{Type II:}\qquad$-L^2<\mathcal{C}<0$ with $\pm=\sign\pa{-L}$}
	\tilde{I}_y&=\pm\frac{2}{\sqrt{\ab{\mathcal{C}}}}\pa{\nu_s\arctanh{x_s^\pm}-\nu_o\arctanh{x_o^\pm}},\\
	\tilde{I}_\varphi&=L\tilde{I}_y\pm\frac{2}{\sqrt{\ab{\mathcal{C}}}}\br{\nu_s\tilde{\mathcal{I}}_\varphi^\pm\pa{x_s^\pm,\alpha_-,-}-\nu_o\tilde{\mathcal{I}}_\varphi^\pm\pa{x_o^\pm,\alpha_-,-}},\\
	\label{eq:GlobalTII}
	\tilde{I}_\tau&=\pm\frac{2}{\sqrt{\ab{\mathcal{C}}}}\br{\nu_s\tilde{\mathcal{I}}_\tau^\pm\pa{x_s^\pm,\alpha_-,-}-\nu_o\tilde{\mathcal{I}}_\tau^\pm\pa{x_o^\pm,\alpha_-,-}}.
\end{empheq}
\end{subequations}
\begin{subequations}
\begin{empheq}[box=\ovalbox]{align}
	\shortintertext{\centering\underline{Type III:}\qquad$\mathcal{C}=0$ with $\pm=\sign\pa{-L}$}
	\tilde{I}_y&=\pm\frac{2}{\sqrt{2\triangle\ab{L}}}\pa{\nu_sx_s^0-\nu_ox_o^0},\\
	\tilde{I}_\varphi&=L\tilde{I}_y\pm\frac{2}{\sqrt{2\triangle\ab{L}}}\br{\nu_s\tilde{\mathcal{I}}_\varphi^0\pa{x_s^0,\alpha_0^\pm,\pm}-\nu_o\tilde{\mathcal{I}}_\varphi^0\pa{x_o^0,\alpha_0^\pm,\pm}},\\
	\label{eq:GlobalTIII}
	\tilde{I}_\tau&=\pm\frac{2}{\sqrt{2\triangle\ab{L}}}\br{\nu_s\tilde{\mathcal{I}}_\tau^0\pa{x_s^0,\alpha_0^\pm,\pm}-\nu_o\tilde{\mathcal{I}}_\tau^0\pa{x_o^0,\alpha_0^\pm,\pm}}.\end{empheq}
\end{subequations}
For the Type II and Type III geodesics (with $\mathcal{C}<0$ and $\mathcal{C}=0$, respectively), the radial integral $\tilde{I}_y$ is in general divergent, indicating simply that the typical geodesic undergoes an infinite number of polar librations as it makes its way towards the boundary. In physical applications, this divergence is of course cut off by the finite distance to the NHEK boundary. Note however that the integral $\tilde{I}_\tau$ is well behaved as $y\to\pm\infty$, reflecting the fact that geodesics can reach the boundary of NHEK in finite global (but infinite proper) time. 

\subsubsection{Explicit solution of the geodesic equation}

Up to this point, we have simply repeated the analysis performed for the Kerr integrals in Sec.~\ref{sec:Kerr}, taking advantage of the fact that the Kerr elliptic integrals reduce to trigonometric integrals in the near-horizon scaling limit. Fortuitously, the simplicity of the global NHEK geodesic equation relative to its Kerr counterpart allows one to go further and obtain an explicit time-parameterization for the global geodesics. Comparing Eqs.~\eqref{eq:RadialGeodesicEquation} and \eqref{eq:TimeGeodesicEquation} to Eqs.~\eqref{eq:GlobalRadialEquation} and \eqref{eq:GlobalTimeEquation}, one sees that the disappearance of $\theta$-dependence on the right-hand-side of Eq.~\eqref{eq:GlobalTimeEquation} allows one to reduce the $(\tau,y)$ motion in global NHEK to a single ordinary differential equation:
\begin{align}
	\label{eq:GlobalODE}
	\frac{dy}{d\tau}=\pm_y\frac{\pa{1+y^2}}{\triangle+Ly}\sqrt{\mathcal{Y}(y)}.
\end{align}
This is known as a first-order, non-linear autonomous system. Such equations are capable of displaying a number of interesting properties worth discussing. For instance, while continuity of the right-hand-side of Eq.~\eqref{eq:GlobalODE} guarantees the existence of a solution locally, a continuous derivative is required to guarantee its uniqueness. Since this condition is violated at zeroes of the radial potential, multiple solutions may be possible for a given set of initial conditions. Indeed, these extra equilibrium solutions are the constant-radius trajectories discussed previously: a geodesic $y(\tau)=y_t$ obeys the same initial conditions as a geodesic with turning point $y_t$ whenever they meet. As we will see, although it is rarely the case for a generic non-linear autonomous system, the general solution to Eq.~\eqref{eq:GlobalODE} includes these special equilibrium states as a subset.

More interestingly, solutions to systems like Eq.~\eqref{eq:GlobalODE} are known to be capable of exhibiting blowup in finite time. In ordinary applications, this signals an instability of the non-linear system, but in our case it represents a reasonable physical effect: geodesics in NHEK can reach the conformal boundary in finite coordinate time. It is often impossible to obtain explicit solutions to systems like Eq.~\eqref{eq:GlobalODE} in terms of elementary functions, so typical methods focus on describing the qualitative behavior of the solutions using fixed-point theory. The equilibrium configurations (both stable and unstable), which correspond to the zeroes of the radial potential, control the late-time behavior of the geodesics. As one varies the parameters $(\triangle,L,\mathcal{C})$ of this dynamical system, the various equilibrium points move around, merge and annihilate, or emerge and separate. This qualitative analysis was performed in Sec.~\ref{subsec:GlobalQualitative}.

When an analytic solution is possible, it is obtainable through the following procedure. Using separation of variables, one integrates Eq.~\eqref{eq:GlobalODE} to determine $\tau(y)$, up to a constant of integration. This was done in Eqs.~\eqref{eq:GlobalTI}, \eqref{eq:GlobalTII}, and \eqref{eq:GlobalTIII}. This procedure defines $y(\tau)$ implicitly. In global NHEK, $\tau(y)$ is in general multivalued because $y(\tau)$ is periodic. If one manages to invert $\tau(y)$ in a small neighborhood, one can then extend the solution $y(\tau)$ to the entire global strip. Indeed, because the system has time-translation symmetry (is autonomous), the general solution to the initial value problem is obtained from $y(\tau)$ by a simple shift of the argument $\tau\to\tau-\tau_\star$. 

Although generic autonomous systems can be difficult to integrate, Eq.~\eqref{eq:GlobalODE} is very special. It arises from a system with extra symmetries generated by the Killing vector fields \eqref{eq:KillingFieldsGlobalNHEK}, and these symmetries can be used to reduce the differential equation to an algebraic one. While we chose to label geodesics in global NHEK by the conserved quantities associated to $iL_0^\mu$, $W_0^\mu$, and $\mathcal{C}^{\mu\nu}$, the quantities $H_+=H_+^\mu p_\mu$, $H_-=H_-^\mu p_\mu$, and $H_0=H_0^\mu p_\mu$ are all conserved along particle trajectories as well. A particularly simple combination to work with is
\begin{align}
	\label{eq:ExtraConservation}
	\frac{H_+-H_-}{2H_0}=\frac{\pa{L-\triangle y}\cos{\tau}\pm_y\sqrt{\mathcal{Y}(y)}\sin{\tau}}{\pa{L-\triangle y}\sin{\tau}\mp_y\sqrt{\mathcal{Y}(y)}\cos{\tau}}.
\end{align}
Differentiating this relation with respect to $\tau$ and solving for $y'(\tau)$, one easily recovers \eqref{eq:GlobalODE}. Therefore, every solution to Eq.~\eqref{eq:GlobalODE} obeys this algebraic relation for some value of the constant $\pa{H_+-H_-}/2H_0$, which depends on initial conditions. In fact, one can check that different choices of this constant amount to a shift $\tau \to \tau-\tau_\star$ in \eqref{eq:ExtraConservation}, in line with our expectations for an autonomous system.  For the particular choice $-L/\triangle$, further manipulation puts this equation in the form
\begin{align}
	\br{\frac{\triangle\cos{\tau}+L\sin{\tau}}{L\cos{\tau}-\triangle\sin{\tau}}}^2=\frac{\mathcal{Y}(y)}{\pa{L-\triangle y}^2},
\end{align}
and it is straightforward to verify that a solution to either algebraic equation satisfies Eq.~\eqref{eq:GlobalODE}. In terms of
\begin{align}
	S(\tau)=\mathcal{C}+\frac{\triangle^2-\mathcal{C}}{\triangle^2+L^2}\pa{\triangle\cos{\tau}+L\sin{\tau}}^2,
\end{align}
these expressions can be inverted to give
\begin{align}
	y(\tau)=\frac{\triangle L}{S(\tau)}\pa{1+\frac{\triangle\sin{\tau}-L\cos{\tau}}{\triangle L}\sqrt{\frac{\triangle^2-\mathcal{C}}{\triangle^2+L^2}\br{S(\tau)+L^2}}}.
\end{align}
(The quadratic equation has a second root with a $-$ sign, which is related to this root by a shift of $\tau\to\tau+\pi$.) Using global-time-translation symmetry, we can shift the argument $\tau\to\tau-\tau_\star$ to ensure that the geodesics pass through any point $(\tau_s,y_s)$ where geodesic motion is allowed:
\begin{subequations}
\label{eq:GlobalRadialMotion}
\begin{align}
	y_o(\tau_o)&=\frac{\triangle L}{S(\tau_o-\tau_\star)}\pa{1+\frac{\triangle\sin\pa{\tau_o-\tau_\star}-L\cos\pa{\tau_o-\tau_\star}}{\triangle L}\sqrt{\frac{\triangle^2-\mathcal{C}}{\triangle^2+L^2}\br{S(\tau_o-\tau_\star)+L^2}}},\\
	\tau_\star&=\tau_s-\arctan\pa{y_s}+\arctan\br{\frac{\pa{L-\triangle y_s}\pa{\mathcal{C}+L^2}}{\pa{\triangle^2-\mathcal{C}}\pa{\triangle+Ly_s}+\pa{\triangle^2+L^2}\sqrt{\mathcal{Y}(y_s)}}}.
\end{align}
\end{subequations}
(In so doing, the values of the constants in Eq.~\eqref{eq:ExtraConservation} change, since they arise from symmetries that do not commute with global-time-translations.) This function describes the radial motion of a geodesic along the entire global strip. In the Type II case, geodesics hit the boundary at global time
\begin{align}
	\tau_b^\pm=\tau_\star\pm\pi+\arctan\br{\frac{\pa{\mathcal{C}-\triangle^2}L\mp\pa{\triangle^2+L^2}\sqrt{\ab{\mathcal{C}}}}{\pa{\mathcal{C}+L^2}\triangle}}\mod_{2\pi}.
\end{align}

Provided that one keeps track of radial turning points encountered along the way (if any), the expression \eqref{eq:GlobalRadialMotion} allows one to obtain $\tilde{I}_y$ as a function of global time by plugging in $y_o(\tau_o)$. In turn, substitution of $\tilde{I}_y(\tau_o)$ into the inversion formulas \eqref{eq:Q>0,P=0}--\eqref{eq:Q>0,P!=0} derived in Sec.~\ref{sec:Kerr} allows one to obtain the polar angle of the particle as a function of global time. For instance, in the generic case $P\neq0$ ($L\neq\pm2\mu M$),
\begin{align}
	\cos{\theta_o(\tau_o)}=\sqrt{u_\pm}\sn\pa{X^\pm(\tau_o)\left|\frac{u_\pm}{u_\mp}\right.},\qquad
	X^\pm(\tau_o)=F\pa{\arcsin\pa{\frac{\cos{\theta_s}}{\sqrt{u_\pm}}}\left|\frac{u_\pm}{u_\mp}\right.}-\sign\pa{p_s^\theta}\sqrt{-u_\mp P}\tilde{I}_y(\tau_o).
\end{align}
Finally, given both $y_o(\tau_o)$ and $\theta_o(\tau_o)$, one can plug them into the expressions for $\tilde{I}_\varphi$ and $\tilde{G}_\varphi$ to obtain the azimuthal angle as a function of global time,
\begin{align}
	\varphi_o(\tau_o)=\varphi_s+\tilde{G}_\varphi(\tau_o)-\tilde{I}_\varphi(\tau_o).
\end{align}
This completes the explicit parameterization of geodesics by the global time elapsed along their trajectory.

\subsection{Geodesics in Poincar\'e NHEK}
\label{subsec:PoincareGeodesics}

Having solved for the most general geodesic motion in global NHEK, we now turn to the description of geodesics in the Poincar\'e patch. Of course, each class of geodesics in Poincar\'e NHEK is in principle obtainable from some geodesic in global NHEK, which is simply the geodesic completion of the Poincar\'e patch. However, since the Poincar\'e coordinates naturally arise in the near-horizon limit of the extreme Kerr black hole, the Poincar\'e patch description is more useful for connecting the discussion to the far region, and many expressions take a much simpler form in these coordinates. Recall from Sec.~\ref{subsec:EmergentThroat} that in Poincar\'e coordinates, the NHEK line element is
\begin{align}
\label{eq:PoincareMetric}
	d\hat{s}^2=2M^2\Gamma\br{-R^2\ed T^2+\frac{\ed R^2}{R^2}+\ed\theta^2+\Lambda^2\pa{\ed\Phi+R\ed T}^2},
\end{align}
and the generators of $\mathsf{SL}(2,\mathbb{R})\times\mathsf{U}(1)$ are
\begin{align}
	H_0=T\pd_T-R\pd_R,\qquad
	H_+=\pd_T,\qquad
	H_-=\pa{T^2+\frac{1}{R^2}}\pd_T-2TR\pd_R-\frac{2}{R}\pd_\Phi,\qquad
	W_0=\pd_\Phi.
\end{align}
The Casimir of $\mathsf{SL}(2,\mathbb{R})$ is the (manifestly reducible) symmetric Killing tensor
\begin{align}
	\mathcal{C}^{\mu\nu}=-H_0^\mu H_0^\nu+\frac{1}{2}\pa{H_+^\mu H_-^\nu+H_-^\mu H_+^\nu}.
\end{align}

The motion of a free particle of mass $\mu$ and four-momentum $P^\mu$ is described by the geodesic equation
\begin{align}
	P^\mu\hat{\nabla}P^\nu=0,\qquad
	\hat{g}^{\mu \nu}P_\mu P_\nu=-\mu^2.
\end{align}
Geodesic motion in Poincar\'e NHEK is completely characterized by the three conserved quantities
\begin{subequations}
\begin{gather}
	E=-H_+^\mu P_\mu
	=-P_T,\qquad
	L=W_0^\mu P_\mu
	=P_\Phi,\\
	\mathcal{C}=\mathcal{C}^{\mu\nu}P_\mu P_\nu
	=P_\theta^2-\pa{1-\frac{1}{\Lambda^2}}P_\Phi^2+\pa{2M^2\Gamma}\mu^2
	=\frac{\pa{P_T-P_\Phi R}^2}{R^2}-R^2P_R^2-P_\Phi^2,
\end{gather}
\end{subequations}
denoting the NHEK energy, angular momentum parallel to the axis of symmetry, and $\mathsf{SL}(2,\mathbb{R})$ Casimir, respectively.

By inverting the above relations for $\pa{\mu^2,E,L,\mathcal{C}}$, we find that a particle following a geodesic in the Poincar\'e patch of the NHEK geometry \eqref{eq:PoincareMetric} has an instantaneous four-momentum $P=P_\mu\ed X^\mu$ of the form
\begin{align}
	\label{eq:GeodesicNHEK}
	P(X^\mu,E,L,\mathcal{C})=-E\ed T\pm_R\frac{\sqrt{\mathcal{R}_n(R)}}{R^2}\ed R\pm_\theta\sqrt{\Theta_n(\theta)}\ed\theta+L\ed\Phi,
\end{align}
where the two choices of sign $\pm_R$ and $\pm_\theta$ depend on the radial and polar directions of travel, respectively. Here, we used the same polar potential $\Theta_n(\theta)$ introduced in Eq.~\eqref{eq:PolarPotentialNHEK}, and additionally defined a new radial potential
\begin{align}
	\label{eq:PoincareRadialPotential}
	\mathcal{R}_n(R)=\pa{E+LR}^2-\pa{\mathcal{C}+L^2}R^2.
\end{align}
One can then raise $P_\mu$ to obtain the equations for the geodesic trajectory,
\begin{subequations}
\label{eq:PoincareMomentum}
\begin{align}
	2M^2\Gamma\frac{dR}{d\sigma}&=\pm_R\sqrt{\mathcal{R}_n(R)},\\
	2M^2\Gamma\frac{d\theta}{d\sigma}&=\pm_\theta\sqrt{\Theta_n(\theta)},\\
	2M^2\Gamma\frac{d\Phi}{d\sigma}&=-\frac{E+LR}{R}+\frac{L}{\Lambda^2},\\
	2M^2\Gamma\frac{dT}{d\sigma}&=\frac{E+LR}{R^2}.
\end{align}
\end{subequations}
The parameter $\sigma$ is the affine parameter for massless particles ($\mu=0$), and is related to the proper time $\delta$ by $\delta=\mu\sigma$ for massive particles.

Following the same procedure as in Kerr and global NHEK, we find from Eqs.~\eqref{eq:PoincareMomentum} that a geodesic labeled by $\pa{\mu^2,E,L,\mathcal{C}}$ connects spacetime points $X_s^\mu=\pa{T_s,R_s,\theta_s,\Phi_s}$ and $X_o^\mu=\pa{T_o,R_o,\theta_o,\Phi_o}$ in Poincar\'e NHEK if
\begin{subequations}
\begin{align}
	&\fint_{R_s}^{R_o}\frac{\ed R}{\pm_R\sqrt{\mathcal{R}_n(R)}}=\fint_{\theta_s}^{\theta_o}\frac{\ed\theta}{\pm_\theta\sqrt{\Theta_n(\theta)}},\\
	\Phi_o-\Phi_s&=\fint_{R_s}^{R_o}\pa{-\frac{E+LR}{R}}\frac{\ed R}{\pm_R\sqrt{\mathcal{R}_n(R)}}+\fint_{\theta_s}^{\theta_o}\frac{L\Lambda^{-2}(\theta)}{\pm_\theta\sqrt{\Theta_n(\theta)}}\ed\theta,\\
	T_o-T_s&=\fint_{R_s}^{R_o}\pa{\frac{E+LR}{R^2}}\frac{\ed R}{\pm_R\sqrt{\mathcal{R}_n(R)}}.
\end{align}
\end{subequations}
We may rewrite these conditions as
\begin{align}
	\label{eq:PoincareGeodesics}
	\hat{I}_R=\hat{G}_\theta,\qquad
	\Phi_o-\Phi_s=-\hat{I}_\Phi+\hat{G}_\Phi,\qquad
	T_o-T_s=\hat{I}_T,
\end{align}
where we have introduced new integrals
\begin{subequations}
\begin{gather}
	\hat{I}_R=\fint_{R_s}^{R_o}\frac{\ed R}{\pm_R\sqrt{\mathcal{R}_n(R)}},\qquad
	\hat{I}_\Phi=\fint_{R_s}^{R_o}\pa{\frac{E+LR}{R}}\frac{\ed R}{\pm_R\sqrt{\mathcal{R}_n(R)}},\qquad
	\hat{I}_T=\fint_{R_s}^{R_o}\pa{\frac{E+LR}{R^2}}\frac{\ed R}{\pm_R\sqrt{\mathcal{R}_n(R)}},\\
	\hat{G}_\theta=\fint_{\theta_s}^{\theta_o}\frac{\ed\theta}{\pm_\theta\sqrt{\Theta_n(\theta)}},\qquad
	\hat{G}_\Phi=\fint_{\theta_s}^{\theta_o}\frac{L\Lambda^{-2}(\theta)}{\pm_\theta\sqrt{\Theta_n(\theta)}}\ed\theta.
\end{gather}
\end{subequations}

\subsubsection{Qualitative description of geodesic motion}
\label{subsec:PoincareQualitative}

As always, before evaluating the geodesic integrals, it is useful to determine the possible classes of geodesic motion. The angular motion is qualitatively the same as in Kerr and global NHEK, with the polar integral given by
\begin{align}
	\label{eq:AngularIntegralsNHEK}
	\hat{G}_\theta=\tilde{G}_\theta
	=G_\theta,
\end{align}
under the identifications \eqref{eq:AngularIdentificationNHEK}. In particular, the angular motion is non-vortical. We now turn to the classification of the radial motion. The analysis is similar to that for global NHEK, but is slightly complicated by the presence of the Poincar\'e horizon. The radial potential $\mathcal{R}_n(R)=E^2+2ELR-\mathcal{C}R^2$ is a quadratic polynomial in $R$ with manifestly positive discriminant $4E^2\pa{\mathcal{C}+L^2}>0$. As such, it always has two real roots
\begin{align}
	R_\pm=\frac{EL}{\mathcal{C}}\pa{1\pm\sqrt{1+\frac{\mathcal{C}}{L^2}}}\in\mathbb{R}.
\end{align}
Positivity of energy in the local frame of the particle, $P^T\geq0$, requires that
\begin{align}
	\label{eq:PoincarePositivity}
	E+LR\geq0.
\end{align}
In terms of the critical radius
\begin{align}
	R_c=-\frac{E}{L},
\end{align}
this requirement becomes
\begin{align}
	R\gtrless R_c,\qquad
	L\gtrless0.
\end{align}
When $EL>0$, $R_c$ lies behind the Poincar\'e horizon and is irrelevant for motion in the Poincar\'e patch. For allowed values of the geodesic parameters, the roots can only coincide with $R_c$ if $E=0$, in which case $R_c=0$. Since this does not affect motion in the Poincar\'e patch, for all practical purposes, the roots $R_\pm$ can never equal $R_c$. Similarly, while both roots $R_\pm$ are real for all values of the geodesic parameters, it is only the positive ones that affect motion in the Poincar\'e patch: a Poincar\'e observer will never know if the geodesic reaches a turning point behind the horizon.

As in Sec.~\ref{subsec:GlobalGeodesics}, the qualitative behavior of the radial motion is determined by the sign of $\mathcal{C}$:
\begin{enumerate}
\item
\underline{$0<\mathcal{C}<\infty$} corresponds to Type I motion.\footnote{The fact that $\mathcal{C}$ is unbounded in Poincar\'e NHEK does not contradict the fact that $\mathcal{C}$ is bounded above by $\triangle^2$ in global NHEK. Taking $\mathcal{C}\to\infty$ at fixed Poincar\'e energy $E$ sends $\triangle\to\infty$ as well, so that the global NHEK bound $\mathcal{C}<\triangle^2$ is still satisfied.} In this case, one root is positive while the other is negative: if $EL>0$, then $R_-<0<R_+$, whereas if $EL<0$, then $R_+<0<R_-<R_c$. The condition \eqref{eq:PoincarePositivity} requires $E>0$ in both cases. In Poincar\'e NHEK, particles on these geodesics come out of the past horizon at $R=0$, travel to increasingly large radius until they encounter a turning point at a maximal allowed radius ($R_\pm$ according to whether $L\gtrless0$), and then fall back into the future horizon at $R=0$.\footnote{Recall that the turning points are separated by $\Delta\tau=\pi$, so there is precisely one per Poincar\'e patch.}
\item
\underline{$-L^2<\mathcal{C}<0$} corresponds to Type II motion. In this case, the roots $R_\pm$ are both simultaneously positive or negative depending on the sign of $EL$. Motion with both $E,L<0$ is not allowed, because it violates  condition \eqref{eq:PoincarePositivity}. This leaves us with two subtypes of geodesics:
\begin{itemize}\setlength{\itemindent}{.5in}
\item[\underline{Type IIA:}]
These geodesics have both $E,L>0$. In this case, the roots are both negative, $R_+<R_-<0$, and energy positivity imposes no constraint on the radius. These geodesics all have one endpoint on the NHEK boundary and another on the future or past horizon: they explore all radii $R\in[0,\infty)$ without encountering any turning point. Those with $P^R<0$ represent particles plunging into the black hole that eventually pass through the future horizon, while those with $P^R>0$ emerge from the past horizon and appear as particles ejected by the white hole. The latter are unlikely to be of astrophysical interest since the past horizon does not exist for black holes formed by collapse.
\item[\underline{Type IIB:}]
These geodesics have $EL<0$. In this case, the roots are both positive and $0<R_-<R_c<R_+$. Hence, the radial potential $\mathcal{R}_n(R)$ is nonnegative if $R\leq R_-$ or $R\geq R_+$. There are two further subcases to consider:
\begin{itemize}\setlength{\itemindent}{.45in}
\item[\underline{$L>0$},] in which case condition \eqref{eq:PoincarePositivity}  requires that $R>R_c$, so that only the branch $R_+<R$ is allowed. These geodesics fall in from the boundary of NHEK, bounce off of the radial potential, and return to the boundary. In global NHEK, these geodesics correspond to Type II motion that begins and ends on the right boundary.
\item[\underline{$L<0$},] in which case condition \eqref{eq:PoincarePositivity} requires that $R<R_c$, so that only the branch $R<R_-$ is allowed. These geodesics enter the Poincar\'e patch of NHEK through the past horizon, bounce off of the radial potential and fall back into the future horizon. From the perspective of a Poincar\'e observer, they are qualitatively similar to the Type I Poincar\'e geodesics. In global NHEK, these geodesics correspond to Type II motion that begins and ends on the left boundary, passing through a single Poincar\'e patch.
\end{itemize}
\end{itemize}
\item
\underline{$\mathcal{C}=0$} corresponds to Type III motion, which is a limit of Type II. The radial potential $\mathcal{R}_n(R)=2EL\pa{R-R_0}$ is linear with a zero at $R_0=-E/\pa{2L}$. Condition \eqref{eq:PoincarePositivity} then requires $E>0$. When $L>0$ (Type IIIA), the only constraint on the radial motion is that $R>0$. When $L<0$ (Type IIIB), the radial motion is constrained to the range $0\leq R\leq R_0$.
\end{enumerate}
To summarize, the type of radial motion is picked out by the sign of $\mathcal{C}$ and $L$:\vspace{10pt}

\ovalbox{\parbox{.95\textwidth}{
\begin{enumerate}
\item[I.]
$\mathcal{C}>0$ corresponds to Type I motion in which $E>0$ is required. Geodesic motion within the Poincar\'e patch is permitted in the range $R\in\br{0,R_\pm}$, with $\pm=\sign\pa{L}$.
\item[II.]
$-L^2<\mathcal{C}<0$ corresponds to Type II motion. There are two subcases:
\begin{itemize}\setlength{\itemindent}{.5in}
\item[Type IIA] motion with $E,L>0$. Geodesic motion within the Poincar\'e patch is allowed in the range $R\in [0,\infty)$ and the geodesics encounter no turning points.
\item[Type IIB] motion with $EL<0$. Geodesic motion within the Poincar\'e patch is allowed in the range $R\in[0,R_-]$ when $L<0$ and $R\in[R_+,\infty)$ when $L>0$.
\end{itemize}
\item[III.]
$\mathcal{C}=0$ corresponds to Type III motion, a limit of Type II motion in which $E>0$ is required. Geodesic motion within the Poincar\'e patch is allowed in the range $R\in[0,\infty)$ when $L>0$ (Type IIIA motion), and in the range $R\in[0,R_0]$ when $L<0$ (Type IIIB motion).
\end{enumerate}
}}
\vspace{10pt}

Note that since global NHEK and the Poincar\'e patch share a spacelike surface that connects the two boundaries of the global strip, every Type I geodesic in global NHEK also appears as a Type I geodesic in Poincar\'e NHEK. However, neither global NHEK nor the Poincar\'e patch are globally hyperbolic, and this spacelike surface is not a Cauchy surface. As a result, not all Type II geodesics in global NHEK appear as Type II geodesics in Poincar\'e NHEK: many simply enter and exit the strip without penetrating the Poincar\'e patch. Indeed, even those that do appear in the patch are not necessarily complete geodesics: Type IIA geodesics look like plunges in the Poincar\'e patch because they cross the horizon and only return to the boundary farther up along the global strip, while Type IIB geodesics with $L<0$ are extendible through both the past and future horizons. Only the Type IIB geodesics with $L>0$ are complete.

\subsubsection{Geodesic integrals in Poincar\'e NHEK}

We now wish to compute the integrals $\hat{I}_R$, $\hat{I}_T$, $\hat{I}_\Phi$, $\hat{G}_\theta$, and $\hat{G}_\Phi$ that appear in the NHEK geodesic equation \eqref{eq:PoincareGeodesics}. As previously noted, the Poincar\'e NHEK angular integral $\hat{G}_\theta\pa{\mu^2,L,\mathcal{C}}$ takes the same form as the Kerr angular integral $G_\theta(Q,P,\ell)$ under the identifications \eqref{eq:AngularIdentificationNHEK}. Similarly,
\begin{align}
	\hat{G}_\Phi\pa{\mu^2,L,\mathcal{C}}=\frac{L}{4}\pa{4G_\phi-G_t-3G_\theta},
\end{align}
under the same identifications. It remains to evaluate the radial integrals $\hat{I}_R$, $\hat{I}_T$, and $\hat{I}_\Phi$. It is useful to define signs
\begin{align}
	\nu_s=\sign\pa{P_s^R}
	=(-1)^w\sign\pa{P_o^R},\qquad
	\nu_o=\sign\pa{P_o^R}
	=(-1)^w\sign\pa{P_s^R},
\end{align}
where $P_s^R$ and $P_o^R$ denote the radial momentum evaluated at the endpoints $X_s^\mu$ and $X_o^\mu$ of the geodesic, respectively, and $w\in\mathbb{N}$ denotes the number of turning points in the radial motion. As we saw in the previous subsection, radial motion in the Poincar\'e patch has either $w=0$ or $w=1$ turning points. For those geodesics capable of encountering a turning point (those for which at least one of $R_{\pm}>0$ or $R_0>0$), it easily follows that the radial path integral unpacks as follows:\footnote{This is essentially the same formula as for Type II motion in global NHEK, since there can never be more than one turning point.}
\begin{align}
	\label{eq:RadialIntegralsNHEK}
	\text{Type I, IIB, IIIB:}\qquad
	\fint_{R_s}^{R_o}=-\nu_s\int_{R_\pm}^{R_s}+\nu_o\int_{R_\pm}^{R_o},\qquad
	\pm=
	\begin{cases}
		\sign\pa{L}&\quad\text{Type I, IIB},\\
		0&\quad\text{Type IIIB}.
	\end{cases}
\end{align}
Since only the Type I, Type IIB, and Type IIIB geodesics encounter a (positive-radius) turning point, we need to specify a different reference radius in order to compute the radial integrals for Type IIA and Type IIIA motion: integrating to $R_-$ and $R_0$ in these cases produces real, finite expressions for $\hat{I}_R$, but requires one to integrate past the non-integrable singularity at $R=0$ when computing $\hat{I}_{\Phi}$ and $\hat{I}_T$. Since $\hat{I}_{\Phi}$ and $\hat{I}_T$ are convergent at large $R$, a convenient reference radius is instead $R\to\infty$. Therefore, the radial path integral for Type IIA and Type IIIA geodesics, which always have $\nu_s=\nu_o$, unpacks as follows:
\begin{align}
	\label{eq:RadialIntegralsNHEK2}
	\text{Type IIA, IIIA:}\qquad
	\fint_{R_s}^{R_o}=-\nu_s\int_{\infty}^{R_s}+\nu_s\int_{\infty}^{R_o}.
\end{align}
To recapitulate: for Type IIA and Type IIIA motion, $\hat{I}_{\Phi}$ and $\hat{I}_T$ converge at large $R$, but are singular as the limit of integration approaches the Poincar\'e horizon. For these integrals, one should use the representation \eqref{eq:RadialIntegralsNHEK2}. Conversely, $\hat{I}_R$ is divergent for large $R$, but converges near the horizon. For this integral, one should instead resort to the representation \eqref{eq:RadialIntegralsNHEK} with $R_-$ for Type IIA and $R_0$ for Type IIIA.  

We can now explicitly evaluate the radial integrals. For our purposes, the relevant basis of integrals is given by
\begin{subequations}
\begin{align}
	\hat{\mathcal{I}}_1^\pm&\equiv\int_{R_\pm}^{R_i}\frac{\ed R}{\sqrt{\mathcal{C}\pa{R_+-R}\pa{R-R_-}}},\\
	\hat{\mathcal{I}}_2^\pm&\equiv\int_{R_\pm}^{R_i}\pa{\frac{E}{R}}\frac{\ed R}{\sqrt{\mathcal{C}\pa{R_+-R}\pa{R-R_-}}},\\
	\hat{\mathcal{I}}_3^\pm&\equiv\int_{R_\pm}^{R_i}\pa{\frac{E+LR}{R^2}}\frac{\ed R}{\sqrt{\mathcal{C}\pa{R_+-R}\pa{R-R_-}}},
\end{align}
\end{subequations}
in terms of which
\begin{subequations}
\label{eq:PoincareIntegrals}
\begin{align}
	\int_{R_\pm}^{R_i}\frac{\ed R}{\sqrt{\mathcal{R}_n(R)}}&=\hat{\mathcal{I}}_1^\pm,\\
	\int_{R_\pm}^{R_i}\pa{\frac{E+LR}{R}}\frac{\ed R}{\sqrt{\mathcal{R}_n(R)}}&=L\hat{\mathcal{I}}_1^\pm+\hat{\mathcal{I}}_2^\pm,\\
	\int_{R_\pm}^{R_i}\pa{\frac{E+LR}{R^2}}\frac{\ed R}{\sqrt{\mathcal{R}_n(R)}}&=\hat{\mathcal{I}}_3^\pm.
\end{align}
\end{subequations}
For Type IIA and Type IIIA geodesics, the radial turning point $R_\pm$ in the limit of integration should be replaced by $R\to\infty$ for $\hat{\mathcal{I}}_2$ and $\hat{\mathcal{I}}_3$.

We define the positive quantity
\begin{align}
	z_i^\pm=\ab{\frac{R_i-R_\pm}{R_i-R_\mp}}^{1/2}.
\end{align}
When $\mathcal{C}>0$, motion is allowed in the range $R\in\br{0,R_\pm}$, with $R_\mp<0<R_\pm$ and $\pm=\sign\pa{L}$. Using the substitution
\begin{align}
	R=\frac{R_\pm+R_\mp z^2}{1+z^2},\qquad
	z^2=\frac{R_\pm-R}{R-R_\mp}>0,
\end{align}
together with the fact that $\mathcal{C}R_+R_-=-E^2$, one finds that
\begin{subequations}
\label{eq:PoincareTypeI}
\begin{align}
	\hat{\mathcal{I}}_1^\pm&=-\frac{2}{\sqrt{\mathcal{C}}}\int_0^{z_i^\pm}\frac{\ed z}{1+z^2}
	=-\frac{2}{\sqrt{\mathcal{C}}}\arctan{z_i^\pm},\\
	\hat{\mathcal{I}}_2^\pm&=-\frac{2E}{\sqrt{\mathcal{C}}}\int_0^{z_i^\pm}\frac{\ed z}{R_\pm+R_\mp z^2}
	=-2\arctanh\pa{\sqrt{-\frac{R_\mp}{R_\pm}}z_i^\pm},\\
	\hat{\mathcal{I}}_3^\pm&=\left.-\frac{\sqrt{\mathcal{R}_n(R)}}{ER}\right|_{R=R_\pm}^{R=R_i}
	=-\frac{\sqrt{\mathcal{R}_n(R_i)}}{ER_i}.
\end{align}
\end{subequations}
When $\mathcal{C}<0$, we choose $\pm=\sign\pa{L}$ for Type IIB geodesics. The appropriate substitution is
\begin{align}
    \label{eq:PoincareSubstitution}
	R=\frac{R_\pm-R_\mp z^2}{1-z^2},\qquad
	z^2=\frac{R-R_\pm}{R-R_\mp}\in\br{0,1}.
\end{align}
The integrals are then given by
\begin{subequations}
\label{eq:PoincareTypeIIB}
\begin{align}
	\hat{\mathcal{I}}_1^\pm&=\pm\frac{2}{\sqrt{\ab{\mathcal{C}}}}\int_0^{z_i^\pm}\frac{\ed z}{1-z^2}
	=\pm\frac{2}{\sqrt{\ab{\mathcal{C}}}}\arctanh{z_i^\pm},\\
	\hat{\mathcal{I}}_2^\pm&=\pm \frac{2E}{\sqrt{\ab{\mathcal{C}}}}\int_0^{z_i^\pm}\frac{\ed z}{R_\pm-R_\mp z^2}
	=-2\arctanh\pa{\sqrt\frac{R_\mp}{R_\pm}z_i^\pm},\\
	\hat{\mathcal{I}}_3^\pm&=\left.-\frac{\sqrt{\mathcal{R}_n(R)}}{ER}\right|_{R=R_\pm}^{R=R_i}
	=-\frac{\sqrt{\mathcal{R}_n(R_i)}}{ER_i}.
\end{align}
\end{subequations}
On the other hand, for the Type IIA geodesics, we have
\begin{align}
	E>0,\qquad
	L>0,\qquad
	R_+<R_-<0.
\end{align}
The appropriate substitution is the same as in Eq.~\eqref{eq:PoincareSubstitution}, with the lower sign for $\hat{\mathcal{I}}_1^-$ and the upper sign for $\hat{\mathcal{I}}_2^\infty$ and $\hat{\mathcal{I}}_3^\infty$. The integrals are given by
\begin{subequations}
\label{eq:PoincareTypeIIA}
\begin{align}
	\hat{\mathcal{I}}_1^-&=\frac{2}{\sqrt{\ab{\mathcal{C}}}}\int_0^{z_i^-}\frac{\ed z}{1-z^2}
	=\frac{2}{\sqrt{\ab{\mathcal{C}}}}\arctanh{z_i^-},\\
	\hat{\mathcal{I}}_2^\infty&=\frac{2E}{\sqrt{\ab{\mathcal{C}}}}\int_1^{z_i^+}\frac{\ed z}{R_+-R_-z^2}
	=2\arctanh\sqrt{\frac{R_-}{R_+}}-2\arctanh\pa{\sqrt{\frac{R_-}{R_+}}z_i^+},\\
	\hat{\mathcal{I}}_3^\infty&=\left.-\frac{\sqrt{\mathcal{R}_n(R)}}{ER}\right|_{R\to\infty}^{R=R_i}
	=\frac{\sqrt{\ab{\mathcal{C}}}}{E}-\frac{\sqrt{\mathcal{R}_n(R_i)}}{ER_i}.
\end{align}
\end{subequations}

Finally, when $\mathcal{C}=0$, the quantities $E$ and $L\pa{R-R_0}$ are separately positive, and the appropriate substitution for evaluating the integrals is $z^2=2EL\pa{R-R_0}$. For Type IIIB motion with $L<0$, they take the simple form
\begin{subequations}
\label{eq:PoincareTypeIIIB}
\begin{align}
	\hat{\mathcal{I}}_1^0&=\int_{R_0}^{R_i}\frac{\ed R}{\sqrt{2EL\pa{R-R_0}}}
	=\frac{\sqrt{\mathcal{R}_n(R_i)}}{EL},\\
	\hat{\mathcal{I}}_2^0&=\int_{R_0}^{R_i}\pa{\frac{E}{R}}\frac{\ed R}{\sqrt{2EL\pa{R-R_0}}}
	=-2\arctanh\pa{\frac{\sqrt{\mathcal{R}_n(R_i)}}{E}},\\
	\hat{\mathcal{I}}_3^0&=\int_{R_0}^{R_i}\pa{\frac{E+LR}{R^2}}\frac{\ed R}{\sqrt{2EL\pa{R-R_0}}}
	=-\frac{\sqrt{\mathcal{R}_n(R_i)}}{ER_i}.
\end{align}
\end{subequations}
For Type IIIA motion with $L>0$, one instead obtains
\begin{subequations}
\label{eq:PoincareTypeIIIA}
\begin{align}
	\hat{\mathcal{I}}_1^0&=\int_{R_0}^{R_i}\frac{\ed R}{\sqrt{2EL\pa{R-R_0}}}
	=\frac{\sqrt{\mathcal{R}_n(R_i)}}{EL},\\
	\hat{\mathcal{I}}_2^\infty&=\int_\infty^{R_i}\pa{\frac{E}{R}}\frac{\ed R}{\sqrt{2EL\pa{R-R_0}}}
	=-2\arcsinh{\sqrt{-\frac{R_0}{R_i}}},\\
	\hat{\mathcal{I}}_3^\infty&=\int_{\infty}^{R_i}\pa{\frac{E+LR}{R^2}}\frac{\ed R}{\sqrt{2EL\pa{R-R_0}}}
	=-\frac{\sqrt{\mathcal{R}_n(R_i)}}{ER_i}.
\end{align}
\end{subequations}

Substituting Eqs.~\eqref{eq:PoincareIntegrals} and \eqref{eq:PoincareTypeI} into Eq.~\eqref{eq:RadialIntegralsNHEK} yields the geodesic integrals for Type I motion:
\begin{subequations}
\begin{empheq}[box=\ovalbox]{align}
	\shortintertext{\centering\underline{Type I:}\qquad$\mathcal{C}>0$ with $\pm=\sign\pa{L}$}
	\hat{I}_R&=\frac{2}{\sqrt{\mathcal{C}}}\br{\nu_s\arctan{z_s^\pm}-\nu_o\arctan{z_o^\pm}},\\
	\hat{I}_\Phi&=L\hat{I}_R+2\br{\nu_s\arctanh\pa{\sqrt{-\frac{R_\mp}{R_\pm}}z_s^\pm}-\nu_o\arctanh\pa{\sqrt{-\frac{R_\mp}{R_\pm}}z_o^\pm}},\\
	\hat{I}_T&=\nu_s\frac{\sqrt{\mathcal{R}_n(R_s)}}{ER_s}-\nu_o\frac{\sqrt{\mathcal{R}_n(R_o)}}{ER_o}.
\end{empheq}
\end{subequations}

The geodesic integrals for Type IIA motion are likewise obtained by substituting Eqs.~\eqref{eq:PoincareIntegrals} and \eqref{eq:PoincareTypeIIA} into Eqs.~\eqref{eq:RadialIntegralsNHEK} and \eqref{eq:RadialIntegralsNHEK2}, resulting in: 
\begin{subequations}
\begin{empheq}[box=\ovalbox]{align}
	\shortintertext{\centering\underline{Type IIA:}\qquad$-L^2<\mathcal{C}<0$ with $E,L>0$}
	\hat{I}_R&=\frac{2\nu_s}{\sqrt{\ab{\mathcal{C}}}}\br{\arctanh{z_o^-}-\arctanh{z_s^-}},\\
	\hat{I}_\Phi&=L\hat{I}_R+2\nu_s\br{\arctanh\pa{\sqrt{\frac{R_-}{R_+}}z_s^+}-\arctanh\pa{\sqrt{\frac{R_-}{R_+}}z_o^+}},\\
	\hat{I}_T&=\nu_s\br{\frac{\sqrt{\mathcal{R}_n(R_s)}}{ER_s}-\frac{\sqrt{\mathcal{R}_n(R_o)}}{ER_o}}.
\end{empheq}
\end{subequations}

The integrals for Type IIB motion are obtained by substituting Eqs.~\eqref{eq:PoincareIntegrals} and \eqref{eq:PoincareTypeIIB} into Eq.~\eqref{eq:RadialIntegralsNHEK}, resulting in:
\begin{subequations}
\begin{empheq}[box=\ovalbox]{align}
	\shortintertext{\centering\underline{Type IIB:}\qquad$-L^2<\mathcal{C}<0$ with $EL<0$ and $\pm=\sign\pa{L}$ }
	\hat{I}_R&=\mp\frac{2}{\sqrt{\ab{\mathcal{C}}}}\br{\nu_s\arctanh{z_s^\pm}-\nu_o\arctanh{z_o^\pm}},\\
	\hat{I}_\Phi&=L\hat{I}_R+2\br{\nu_s\arctanh\pa{\sqrt\frac{R_\mp}{R_\pm}z_i^\pm}-\nu_o\arctanh\pa{\sqrt\frac{R_\mp}{R_\pm}z_o^\pm}},\\
	\hat{I}_T&=\nu_s\frac{\sqrt{\mathcal{R}_n(R_s)}}{ER_s}-\nu_o\frac{\sqrt{\mathcal{R}_n(R_o)}}{ER_o}.
\end{empheq}
\end{subequations}

Substituting Eqs.~\eqref{eq:PoincareIntegrals} and \eqref{eq:PoincareTypeIIIA} into Eqs.~\eqref{eq:RadialIntegralsNHEK} and \eqref{eq:RadialIntegralsNHEK2}, we obtain the Type IIIA integrals:
\begin{subequations}
\begin{empheq}[box=\ovalbox]{align}
	\shortintertext{\centering\underline{Type IIIA:}\qquad$\mathcal{C}=0$ with $L>0$}
	\hat{I}_R&=\nu_s\br{\frac{\sqrt{\mathcal{R}_n(R_o)}}{EL}-\frac{\sqrt{\mathcal{R}_n(R_s)}}{EL}},\\
	\hat{I}_\Phi&=L\hat{I}_R-2\nu_s\br{\arcsinh{\sqrt{-\frac{R_0}{R_o}}}-\arcsinh{\sqrt{-\frac{R_0}{R_s}}}},\\
	\hat{I}_T&=\nu_s\br{\frac{\sqrt{\mathcal{R}_n(R_s)}}{ER_s}-\frac{\sqrt{\mathcal{R}_n(R_o)}}{ER_o}}.
\end{empheq}
\end{subequations}

Finally, substituting Eqs.~\eqref{eq:PoincareIntegrals} and \eqref{eq:PoincareTypeIIIB} into Eq.~\eqref{eq:RadialIntegralsNHEK}, we obtain the Type IIIB integrals:
\begin{subequations}
\begin{empheq}[box=\ovalbox]{align}
	\shortintertext{\centering\underline{Type IIIB:}\qquad$\mathcal{C}=0$ with $L<0$}
	\hat{I}_R&=\nu_o\frac{\sqrt{\mathcal{R}_n(R_o)}}{EL}-\nu_s \frac{\sqrt{\mathcal{R}_n(R_s)}}{EL},\\
	\hat{I}_\Phi&=L\hat{I}_R+2\br{\nu_s\arctanh\pa{\frac{\sqrt{\mathcal{R}_n(R_s)}}{E}}-\nu_o\arctanh\pa{\frac{\sqrt{\mathcal{R}_n(R_o)}}{E}}},\\
	\hat{I}_T&=\nu_s\frac{\sqrt{\mathcal{R}_n(R_s)}}{ER_s}-\nu_o\frac{\sqrt{\mathcal{R}_n(R_o)}}{ER_o}.
\end{empheq}
\end{subequations}

\subsubsection{Explicit solution of the geodesic equation}

As in global NHEK, there are several complementary approaches to the solution of the geodesic equation in Poincar\'e NHEK. First, using Eqs.~\eqref{eq:PoincareMomentum}, one can recast the $(T,R)$ motion as a first-order non-linear autonomous system:
\begin{align}
	\label{eq:PoincareODE}
	\frac{dR}{dT}=\pm_R\frac{R^2}{E+LR}\sqrt{\mathcal{R}_n(R)}.
\end{align}
This formulation emphasizes several important properties of the problem. As in global NHEK, local existence of solutions is guaranteed, but uniqueness is not: the derivative of the right-hand side of Eq.~\eqref{eq:PoincareODE} diverges at zeroes of the radial potential, so multiple solutions (including the physically important constant-radius trajectories) are possible for a given set of initial conditions. Similarly, the nonlinearities of the equation allow for blowup in finite coordinate time, reflecting the fact that it is possible for particles to reach the boundary of NHEK in finite time. Finally, the autonomous character of the equation allows us to obtain the general solution from a particular solution through a time translation. The qualitative fixed-point analysis of this equation was performed in Sec.~\ref{subsec:PoincareQualitative}.

The relative simplicity of the Poincar\'e geodesic integrals allows for an explicit parametrization of the trajectories in terms of Poincar\'e time. According to the previous subsection, the time lapse along any geodesic segment is
\begin{align}
	\label{eq:T(R)NHEK}
	T_o(R_o)-T_s(R_s)=\nu_s\frac{\sqrt{\mathcal{R}_n(R_s)}}{ER_s}-\nu_o\frac{\sqrt{\mathcal{R}_n(R_o)}}{ER_o}.
\end{align}
However, note that it is also possible to obtain this algebraic equation for the radial motion more directly using only the symmetries of the system. Indeed, the dilation charge 
\begin{align}
	H_0\equiv H_0^\mu p_\mu
	=-\pa{ET\pm_R\frac{\sqrt{\mathcal{R}_n(R)}}{R}}
\end{align}
is conserved along the trajectory of each particle. Evaluating this constant at $(T_s,R_s)$ and $(T_o,R_o)$, and then equating the results, immediately yields Eq.~\eqref{eq:T(R)NHEK}. In order to invert Eq.~\eqref{eq:T(R)NHEK}, it is useful to introduce a closely related (but not conserved) quantity
\begin{align}
	S_n(T_o)\equiv\mathcal{C}+\pa{H_0+ET_o}^2
	=\mathcal{C}+\br{\nu_s\frac{\sqrt{\mathcal{R}_n(R_s)}}{R_s}-E\pa{T_o-T_s}}^2.
\end{align}
Rewriting Eq.~\eqref{eq:T(R)NHEK} as 
\begin{align}
	\nu_o\frac{\sqrt{\mathcal{R}_n(R_o)}}{R_o}=\nu_s\frac{\sqrt{\mathcal{R}_n(R_s)}}{R_s}-E\pa{T_o-T_s}
	=-H_0-ET_o,
\end{align}
and then squaring both sides, one finds
\begin{align}
	\frac{E^2+2ELR_o-\mathcal{C}R_o^2}{R_o^2}=\pa{H_0+ET_o}^2
	=S_n(T_o)-\mathcal{C}.
\end{align}
The solution to this quadratic equation is given by
\begin{align}
	\label{eq:R(T)NHEK}
	R_o(T_o)=\frac{EL}{S_n(T_o)}\pa{1\pm\sqrt{1+\frac{S_n(T_o)}{L^2}}}.
\end{align}
The choice of sign in Eq.~\eqref{eq:R(T)NHEK} is fixed by the type of geodesic under consideration. Before we can specify it, note that whenever the time is
\begin{align}
	T_t=T_s+\nu_s\frac{\sqrt{\mathcal{R}_n(R_s)}}{ER_s},
\end{align}
we have $S_n(T_t)=\mathcal{C}$, and therefore $R_o(T_t)=R_\pm$. Hence, $T_t$ is the turning time of the radial motion (provided there is one). Also, note that $S_n(T)$ is strictly positive for Type I geodesics, while for Type II geodesics it has a pair of zeroes
\begin{align}
	T^\pm=T_t\pm\frac{\sqrt{\ab{\mathcal{C}}}}{E}.
\end{align}
We can now explicitly describe the radial motion depending on the sign of $\mathcal{C}$:
\begin{enumerate}
\item[I.]
$0<\mathcal{C}<\infty$ corresponds to Type I motion. In this case, $R(T)$ is given for all times $T\in\mathbb{R}$ by Eq.~\eqref{eq:R(T)NHEK} with the choice of sign $\pm=\sign\pa{L}$. The particle reaches the horizon as $T\to\pm\infty$, and encounters a turning point in between at $(T,R)=(T_t,R_\pm)$.
\item[II.]
$-L^2<\mathcal{C}<0$ corresponds to Type II motion, and there are two further subcases to consider:
\begin{enumerate}
\item[A.]
$E,L>0$ corresponds to Type IIA motion. In this case, $R(T)$ is given by Eq.~\eqref{eq:R(T)NHEK} with the upper choice of sign and is defined on the domain $T\gtrless T^\pm$ according to whether $p^R\lessgtr0$. The particle reaches the boundary at $T=T^\pm$ and the horizon as $T\to\pm\infty$, without ever encountering a turning point.
\item[B.]
$EL<0$ corresponds to Type IIB motion. If $L>0$, then $R(T)$ is given by Eq.~\eqref{eq:R(T)NHEK} with the upper choice of sign and is defined on the domain $T\in\br{T^+,T^-}$. The particle reaches the boundary at $T=T^\pm$, and encounters a turning point in between at $(T,R)=(T_t,R_+)$. If $L<0$, then $R(T)$ is given by Eq.~\eqref{eq:R(T)NHEK} with the lower choice of sign and is defined for all times $T\in\mathbb{R}$. The particle reaches the horizon as $T\to\pm\infty$, and encounters a turning point in between at $(T,R)=(T_t,R_-)$.
\end{enumerate}
\item[III.]
$\mathcal{C}=0$ corresponds to Type III motion, a limit of Type II motion. If $L>0$ (Type IIIA), then $R(T)$ is given by Eq.~\eqref{eq:R(T)NHEK} with the upper choice of sign and is defined on the domain $T\gtrless T_t$ according to whether $p^R\lessgtr0$. The particle reaches the boundary at $T=T_t$ and the horizon as $T\to\pm\infty$, without ever encountering a turning point. If $L<0$ (Type IIIB), then $R(T)$ is given by Eq.~\eqref{eq:R(T)NHEK} with the lower choice of sign and is defined for all times $T\in\mathbb{R}$. The particle reaches the horizon as $T\to\pm\infty$, and encounters a turning point in between at $(T,R)=(T_t,R_0)$.
\end{enumerate}

Provided that one keeps track of radial turning points encountered along the way (if any), expression \eqref{eq:R(T)NHEK} allows one to obtain $\hat{I}_R$ as a function of Poincar\'e time by plugging in $R_o(T_o)$. In turn, substitution of $\hat{I}_R(T_o)$ into the inversion formulas \eqref{eq:Q>0,P=0}--\eqref{eq:Q>0,P!=0} derived in Sec.~\ref{sec:Kerr} allows one to obtain the polar angle of the particle as a function of time. For instance, in the generic case $P\neq0$ ($L\neq\pm2\mu M$),
\begin{align}
	\cos{\theta_o(T_o)}=\sqrt{u_\pm}\sn\pa{X^\pm(T_o)\left|\frac{u_\pm}{u_\mp}\right.},\qquad
	X^\pm(T_o)=F\pa{\arcsin\pa{\frac{\cos{\theta_s}}{\sqrt{u_\pm}}}\left|\frac{u_\pm}{u_\mp}\right.}-\sign\pa{p_s^\theta}\sqrt{-u_\mp P}\hat{I}_R(T_o),
\end{align}
where $P$ and $u_\pm$ are to be evaluated according to the identifications \eqref{eq:AngularIdentificationNHEK}, and $\pm=\sign\pa{P}$. Finally, given both $R_o(T_o)$ and $\theta_o(T_o)$, one can plug them into the expressions for $\hat{I}_\Phi$ and $\hat{G}_\Phi$ to obtain the azimuthal angle,
\begin{align}
	\Phi_o(T_o)=\Phi_s-\hat{I}_\Phi(T_o)+\hat{G}_\Phi(T_o).
\end{align}
This completes the explicit parameterization of Poincar\'e NHEK geodesics by the time elapsed along their trajectory.

\subsubsection{Isometry group orbits of geodesics}

We can now explain how NHEK geodesics transform into one another under the action of the isometry group $\mathsf{SL}(2,\mathbb{R})\times\mathsf{U}(1)$. First, note that none of the isometries act on the polar angle $\theta$. Therefore, two NHEK geodesics can only possibly be mapped into each other if they already share the same polar motion. This requires that they share the same $\mathsf{SL}(2,\mathbb{R})$ Casimir $\mathcal{C}$ and angular momentum $L$. If that is the case, then the geodesics could potentially be related by an isometry, and indeed, this must necessarily be the case.

To see why, first focus on their motion in the $(T,R)$ plane. The path $R_o(T_o)$ traced by one geodesic can always be mapped into that traced by the other using an $\mathsf{SL}(2,\mathbb{R})$ transformation, whose explicit form is given at the end of App.~\ref{app:AdS2}. Once their radial motion $R_o(T_o)$ and (by assumption) their polar motion $\theta_o(T_o)$ both match, it is guaranteed that their azimuthal motion $\Phi_o(T_o)$ differs by at most a constant. This last discrepancy can be eliminated by acting with the $\mathsf{U}(1)$ isometry to produce a countering constant shift. This completes the argument.

\subsection{Geodesics in near-NHEK}
\label{subsec:NearGeodesics}

Having solved for the most general geodesic motion in global and Poincar\'e NHEK, we now turn to the description of geodesics in the smaller patch of near-NHEK. Of course, each class of geodesics in near-NHEK is in principle obtainable from some class of geodesics in global NHEK, its geodesic completion. However, since the Poincar\'e coordinates naturally arise in the near-horizon limit of the near-extreme Kerr black hole, the near-NHEK description is more useful for connecting the discussion to the far region, and many expressions take a simpler form in these coordinates. Recall from Sec.~\ref{subsec:ScalingLimits} that the near-NHEK line element is
\begin{align}
	\label{eq:NearMetric}
	d\bar{s}^2=2M^2\Gamma\br{-\pa{R^2-\kappa^2}\ed T^2+\frac{\ed R^2}{R^2-\kappa^2}+\ed\theta^2+\Lambda^2\pa{\ed\Phi+R\ed T}^2},
\end{align}
and the generators of $\mathsf{SL}(2,\mathbb{R})\times\mathsf{U}(1)$ are
\begin{align}
	H_0=\frac{1}{\kappa}\pd_T,\qquad
	H_\pm=\frac{e^{\mp\kappa T}}{\sqrt{R^2-\kappa^2}}\br{\frac{R}{\kappa}\pd_T\pm\pa{R^2-\kappa^2}\pd_R-\kappa\pd_\Phi},\qquad
	W_0=\pd_\Phi.
\end{align}
The Casimir of $\mathsf{SL}(2,\mathbb{R})$ is the symmetric Killing tensor
\begin{align}
	\mathcal{C}^{\mu\nu}=-H_0^\mu H_0^\nu+\frac{1}{2}\pa{H_+^\mu H_-^\nu+H_-^\mu H_+^\nu}.
\end{align}
The motion of a free particle of mass $\mu$ and four-momentum $P^\mu$ is described by the geodesic equation
\begin{align}
	P^\mu\bar{\nabla}_\mu P^\nu=0,\qquad
	\bar{g}^{\mu\nu}P_\mu P_\nu=-\mu^2.
\end{align}
Geodesic motion in near-NHEK is completely characterized by three conserved quantities
\begin{subequations}
\begin{gather}
	E=-\kappa H_0^\mu P_\mu
	=-P_T,\qquad
	L=W_0^\mu P_\mu
	=P_\Phi,\\
	\mathcal{C}=\mathcal{C}^{\mu\nu}P_\mu P_\nu
	=P_\theta^2-\pa{1-\frac{1}{\Lambda^2}}P_\Phi^2+\pa{2M^2\Gamma}\mu^2
	=\frac{\pa{P_T-P_\Phi R}^2}{R^2-\kappa^2}-P_R^2\pa{R^2-\kappa^2}-P_\Phi^2,
\end{gather}
\end{subequations}
denoting the dilation weight, angular momentum parallel to the axis of symmetry, and $\mathsf{SL}(2,\mathbb{R})$ Casimir, respectively.

By inverting the above relations for $\pa{\mu^2,E,L,\mathcal{C}}$, we find that a particle following a geodesic in the near-NHEK geometry \eqref{eq:NearMetric} has an instantaneous four-momentum $P=P_\mu\ed X^\mu$ of the form
\begin{align}
	\label{eq:GeodesicNearNHEK}
	P(X^\mu,E,L,\mathcal{C})=-E\ed T\pm_R\frac{\sqrt{\mathcal{R}_\kappa(R)}}{R^2-\kappa^2}\ed R\pm_\theta\sqrt{\Theta_n(\theta)}\ed\theta+L\ed\Phi,
\end{align}
where the two choices of sign $\pm_R$ and $\pm_\theta$ depend on the radial and polar directions of travel, respectively. Here, we used the same polar potential $\Theta_n(\theta)$ introduced in Eq.~\eqref{eq:PolarPotentialNHEK}, and additionally defined a new radial potential
\begin{align}
	\label{eq:RadialPotentialNearNHEK}
	\mathcal{R}_\kappa(R)=\pa{E+LR}^2-\pa{\mathcal{C}+L^2}\pa{R^2-\kappa^2}.
\end{align}
One can then raise $P_\mu$ to obtain the equations for the geodesic trajectory,
\begin{subequations}
\label{eq:NearMomentum}
\begin{align}
	2M^2\Gamma\frac{dR}{d\sigma}&=\pm_R\sqrt{\mathcal{R}_\kappa(R)},\\
	2M^2\Gamma\frac{d\theta}{d\sigma}&=\pm_\theta\sqrt{\Theta_n(\theta)},\\
	2M^2\Gamma\frac{d\Phi}{d\sigma}&=-\frac{R\pa{E+LR}}{R^2-\kappa^2}+\frac{L}{\Lambda^2},\\
	2M^2\Gamma\frac{dT}{d\sigma}&=\frac{E+LR}{R^2-\kappa^2}.
\end{align}
\end{subequations}
The parameter $\sigma$ is the affine parameter for massless particles ($\mu=0$), and is related to the proper time $\delta$ by $\delta=\mu\sigma$ for massive particles. Note that
\begin{align}
	\lim_{\kappa\to0}\mathcal{R}_\kappa(R)=\mathcal{R}_n(R).
\end{align}
In fact, by comparing with Eq.~\eqref{eq:PoincareMomentum}, we see that geodesics in near-NHEK with temperature $\kappa$ smoothly approach geodesics in NHEK as $\kappa\to0$, even though the mapping between the geometries becomes singular.

Following the same procedure as always, we find from Eqs.~\eqref{eq:NearMomentum} that a geodesic labeled by $\pa{\mu^2,E,L,\mathcal{C}}$ connects spacetime points $X_s^\mu=\pa{T_s,R_s,\theta_s,\Phi_s}$ and $X_o^\mu=\pa{T_o,R_o,\theta_o,\Phi_o}$ in near-NHEK if
\begin{subequations}
\begin{align}
	&\fint_{R_s}^{R_o}\frac{\ed R}{\pm_R\sqrt{\mathcal{R}_\kappa(R)}}=\fint_{\theta_s}^{\theta_o}\frac{\ed\theta}{\pm_\theta\sqrt{\Theta_n(\theta)}},\\
	\Phi_o-\Phi_s&=\fint_{R_s}^{R_o}\pa{-\frac{R\pa{E+LR}}{R^2-\kappa^2}}\frac{\ed R}{\pm_R\sqrt{\mathcal{R}_\kappa(R)}}+\fint_{\theta_s}^{\theta_o}\frac{L\Lambda^{-2}(\theta)}{\pm_\theta\sqrt{\Theta_n(\theta)}}\ed\theta,\\
	T_o-T_s&=\fint_{R_s}^{R_o}\pa{\frac{E+LR}{R^2-\kappa^2}}\frac{\ed R}{\pm_R\sqrt{\mathcal{R}_\kappa(R)}}.
\end{align}
\end{subequations}
We may rewrite these conditions as
\begin{align}
	\label{eq:NearGeodesics}
	\bar{I}_R=\bar{G}_\theta,\qquad
	\Phi_o-\Phi_s=-\bar{I}_\Phi+\bar{G}_\Phi,\qquad
	T_o-T_s=\bar{I}_T,
\end{align}
where we have introduced the integrals
\begin{subequations}
\begin{gather}
	\bar{I}_R=\fint_{R_s}^{R_o}\frac{\ed R}{\pm_R\sqrt{\mathcal{R}_\kappa(R)}},\qquad
	\bar{I}_\Phi=\fint_{R_s}^{R_o}R\pa{\frac{E+LR}{R^2-\kappa^2}}\frac{\ed R}{\pm_R\sqrt{\mathcal{R}_\kappa(R)}},\qquad
	\bar{I}_T=\fint_{R_s}^{R_o}\pa{\frac{E+LR}{R^2-\kappa^2}}\frac{\ed R}{\pm_R\sqrt{\mathcal{R}_\kappa(R)}},\\
	\bar{G}_\theta=\fint_{\theta_s}^{\theta_o}\frac{\ed\theta}{\pm_\theta\sqrt{\Theta_n(\theta)}},\qquad
	\bar{G}_\Phi=\fint_{\theta_s}^{\theta_o}\frac{L\Lambda^{-2}(\theta)}{\pm_\theta\sqrt{\Theta_n(\theta)}}\ed\theta.
\end{gather}
\end{subequations}

\subsubsection{Qualitative description of geodesic motion}
\label{subsec:NearQualitative}

The qualitative behavior of the geodesics in near-NHEK closely resembles that of geodesics in the Poincar\'e patch, with small adjustments due to the nonzero temperature $\kappa>0$. The angular motion is the same as in global and Poincar\'e NHEK, with the angular integrals given by
\begin{align}
	\bar{G}_\theta=\tilde{G}_\theta
	=G_\theta,\qquad
	\bar{G}_\Phi=\tilde{G}_\varphi
	=\frac{L}{4}\pa{4G_\phi-G_t-3G_\theta},
\end{align}
under the identifications \eqref{eq:AngularIdentificationNHEK}. The angular motion is non-vortical. Properties of the radial motion are determined by the properties of the zeroes of the radial potential $\mathcal{R}_\kappa(R)=E^2+\pa{\mathcal{C}+L^2}\kappa^2+2ELR-\mathcal{C}R^2$, which is a quadratic polynomial in $R$ with discriminant $4\pa{E^2+\mathcal{C}\kappa^2}\pa{\mathcal{C}+L^2}$. Since $\mathcal{C}+L^2>0$, its two roots
\begin{align}
	R_\pm=\frac{EL}{\mathcal{C}}\br{1\pm\sqrt{\pa{1+\frac{\mathcal{C}}{L^2}}\pa{1+\frac{\mathcal{C}\kappa^2}{E^2}}}}
\end{align}
are real whenever $\mathcal{C}>-\pa{E/\kappa}^2$. According to Eq.~\eqref{eq:NearMomentum}, positivity of near-NHEK energy in a local frame of the particle, $P^T\geq0$, requires that
\begin{align}
	\label{eq:NearPositivity}
	E+LR>0.
\end{align}
Note that although the form of the radial potential changes, the Poincar\'e NHEK constraint \eqref{eq:PoincarePositivity} arising from positivity of energy in a local frame remains the same. In terms of the critical radius
\begin{align}
	R_c=-\frac{E}{L},
\end{align}
positivity of energy requires
\begin{align}
	R\gtrless R_c,\qquad
	L\gtrless0.
\end{align}
In near-NHEK, it is possible for the roots $R_\pm$ to coincide with the critical radius $R_c$ if
\begin{align}\label{eq:specialcond}
	L^2=\pa{\frac{E}{\kappa}}^2.
\end{align}
This set of parameters corresponds to $R_c^2=\kappa^2$.

The remainder of the analysis is more intricate in the presence of a nonzero temperature $\kappa>0$ because, unlike NHEK, near-NHEK does not contain a complete spacelike surface connecting the two boundaries of the global strip. This means that, in contrast with the Poincar\'e patch, there exist Type I geodesics in global NHEK which never enter the near-NHEK patch. Similar issues arise in the analysis of the Type II motion, and extra care is needed in delineating all of the separate cases. Despite these complications, we still find it useful to describe the qualitative behavior of the radial motion in terms of the sign of $\mathcal{C}$, as in Sec.~\ref{subsec:GlobalGeodesics}:
\begin{enumerate}
\item
\underline{$0<\mathcal{C}<\infty$} still corresponds to Type I motion. In this case, one root is positive while the other is negative. The energy positivity condition \eqref{eq:NearPositivity} rules out motion with both $E,L<0$. On the other hand:
\begin{itemize}
\item
If both $E,L>0$, then one has both $R_-<0<R_+$ and $R_c<0<\kappa<R_+$,\footnote{To see this, note that $R_+^2-\kappa^2$ is a sum of positive terms when $\mathcal{C}>0$. } so that geodesic motion is allowed in the range $R\in\br{\kappa,R_+}$. These geodesics correspond to Type I trajectories in global NHEK that enter the near-NHEK patch through the past horizon at $R=\kappa$ and continue on to a maximum radius before plunging back into the future horizon at $R=\kappa$.
\item
If $EL<0$, then one has $R_+<0<R_-$ and motion is tentatively allowed in the range $R\in\br{\kappa,R_-}$, provided that $\kappa<R_-<R_c$ when $L<0$ and $R_c<\kappa<R_-$ when $L>0$. The condition for $R_-=\kappa$ is the same as the condition for $R_-=R_c$, namely $L=-E/\kappa$. If $L<0$, then one has $\kappa<R_-<R_c$ whenever $-E/\kappa<L<0$, whereas if $L>0$, one has $R_c<\kappa<R_-$ whenever $0<-E/\kappa<L$. Geodesics outside of this range of parameters do not enter the near-NHEK patch.
\end{itemize}
In conclusion, Type I motion occurs in the range $R\in\br{\kappa,R_\pm}$ with $\pm=\sign\pa{EL}$. When $EL<0$, one additionally requires $-E/\kappa<L$. Particles on these geodesics correspond to Type I geodesics in global NHEK that come out of the past horizon at $R=\kappa$, travel to increasingly large radius until they encounter a turning point at a maximal allowed radius ($R_\pm$ according to whether $EL\gtrless0$), and then fall back into the future horizon at $R=\kappa$.
\item
\underline{$-L^2<\mathcal{C}<0$} still corresponds to Type II motion. Condition \eqref{eq:NearPositivity} again rules out the case $E,L<0$. However, there are additional cases to consider now, since the qualitative behavior depends on whether $-(E/\kappa)^2\gtrless\mathcal{C}$. 
\begin{itemize}
\item[] 
\underline{Case 1: $-(E/\kappa)^2<\mathcal{C}$.} In this case, the radial potential has two real roots, and the analysis resembles that of the radial geodesics in the Poincar\'e patch, with two subtypes of geodesics:
\begin{itemize}
\item
\underline{Type IIA:} These geodesics have both $E,L>0$. In this case, both roots are negative, $R_+<R_-<0<\kappa$, and $R_c<0$. Geodesic motion is permitted in the full range $R\in[\kappa,\infty)$ and the geodesics encounter no turning points. Those with $P^R>0$ emerge from the past horizon and reach the boundary of near-NHEK, while those with $P^R<0$ plunge in from the near-NHEK boundary into the future horizon.
\item
\underline{Type IIB:} These geodesics have $EL<0$. In this case, both roots are positive, $0<R_-<R_+$, and $0<R_c$. When $L>0$, geodesic motion is allowed in the region $R\in[R_+,\infty)$, with $0<R_c<\kappa<R_+$ if $-E/\kappa<L$ and $0<\kappa<R_c<R_+$ if $-E/\kappa>L$. When $L<0$, geodesic motion is allowed in the region $R\in\br{\kappa,R_-}$, provided that $\kappa<R_-<R_c$, which requires that $-E/\kappa<L$. The geodesics with $L>0$ fall in from the boundary of near-NHEK, bounce off of the radial potential, and return to the boundary. The geodesics with $L<0$ enter near-NHEK through the past horizon, bounce off of the radial potential, and fall back into the future horizon.
\end{itemize}
\item[] 
\underline{Case 2: $-L^2<\mathcal{C}<-(E/\kappa)^2$.} The roots of the radial potential are complex and the geodesic encounters no turning points. In this case, geodesics with $E,L>0$ necessarily have $E/\kappa<L$ and are free to explore the entire near-NHEK patch $R\in[\kappa,\infty)$. These are also Type IIA geodesics. When $E<0$ but $L>0$, geodesic motion in the near-NHEK patch is only possible if $0<R_c<\kappa<R$. Therefore, geodesics with $0<-E/\kappa<L$ explore the range $R\in[\kappa,\infty)$. This is a new class of Type IIA geodesic with $EL<0$. Condition \eqref{eq:NearPositivity} rules out the case $E>0$ with $L<0$.
\end{itemize}
\item
\underline{$\mathcal{C}=0$} corresponds to Type III motion, a limit of Type II motion. The radial potential $\mathcal{R}_\kappa(R)=2EL\pa{R-R_0}$ is linear with a zero at $R_0=-\pa{E^2+L^2\kappa^2}/\pa{2EL}$. Condition \eqref{eq:NearPositivity} eliminates the case $E,L<0$. When $E,L>0$ (Type IIIA), the only constraint on the radial motion is that $R>\kappa$. When $EL<0$ (Type IIIB), the radial motion is allowed within the range $R\in\br{\kappa,R_0}$, provided that $R_c<\kappa$ when $L>0$ or that $R_c>R_0$ when $L<0$. For both signs of $L$, this happens if and only if $-E/\kappa<L$.
\end{enumerate}
In summary, the types of radial motion in near-NHEK are:\vspace{10pt}

\ovalbox{\parbox{.95\textwidth}{
\begin{enumerate}
\item[I.]
$\mathcal{C}>0$ corresponds to Type I motion. Unlike in Poincar\'e NHEK, $E>0$ is not required, but $L+E/\kappa>0$ is. Geodesic motion within near-NHEK is permitted in the range $R\in\br{\kappa,R_\pm}$, with $\pm=\sign\pa{EL}$.
\item[II.]
$-L^2<\mathcal{C}<0$ corresponds to Type II motion. There are two subcases:
\begin{itemize}\setlength{\itemindent}{.5in}
\item[Type IIA] motion with $E,L>0$, or $E<0<-E/\kappa<L$ and $\mathcal{C}<-\pa{E/\kappa}^2$. Geodesic motion within near-NHEK is allowed in the range $R\in[\kappa,\infty)$ and the geodesics encounter no turning points.
\item[Type IIB] motion with $EL<0$ and $-(E/\kappa)^2<\mathcal{C}$. Geodesic motion within near-NHEK is allowed in the range $R\in[R_+,\infty)$ when $L>0$, and in the range $R\in\br{\kappa,R_-}$ when $-E/\kappa<L<0$.
\end{itemize}
\item[III.]
$\mathcal{C}=0$ corresponds to Type III motion, a limit of Type II motion. Unlike in Poincar\'e NHEK, $E>0$ is not required, but $L+E/\kappa>0$ is. Geodesic motion within near-NHEK is allowed in the range $R\in[\kappa,\infty)$ when $E,L>0$ (Type IIIA motion), and in the range $R\in\br{\kappa,R_0}$ when $EL<0$ (Type IIIB motion).
\end{enumerate}
}}
\vspace{10pt}

\subsubsection{Geodesic integrals in near-NHEK}

We now wish to compute the integrals $\bar{I}_R$, $\bar{I}_T$, $\bar{I}_\Phi$, $\bar{G}_\theta$, and $\bar{G}_\Phi$ that appear in the near-NHEK geodesic equation \eqref{eq:NearGeodesics}. The near-NHEK angular integral $\bar{G}_\theta\pa{\mu^2,L,\mathcal{C}}$ matches the Kerr angular integral $G_\theta(Q,P,\ell)$ under the identifications \eqref{eq:AngularIdentificationNHEK}. With the same identifications,
\begin{align}
	\bar{G}_\Phi\pa{\mu^2,L,\mathcal{C}}=\frac{L}{4}\pa{4G_\phi-G_t-3G_\theta}.
\end{align}
In order to evaluate the radial integrals $\bar{I}_R$, $\bar{I}_T$, and $\bar{I}_\Phi$, we define the signs
\begin{align}
	\nu_s=\sign\pa{P_s^R}
	=(-1)^w\sign\pa{P_o^R},\qquad
	\nu_o=\sign\pa{P_o^R}
	=(-1)^w\sign\pa{P_s^R},
\end{align}
where $P_s^R$ and $P_o^R$ denote the radial momentum evaluated at the endpoints $X_s^\mu$ and $X_o^\mu$ of the geodesic, respectively, and $w\in\mathbb{N}$ denotes the number of turning points in the radial motion. Radial motion in near-NHEK has either $w=0$ or $w=1$ turning points. Therefore, for geodesics that can encounter a turning point (those for which at least one of $R_\pm$ or $R_0$ lies outside the horizon at $R=\kappa$), the radial path integral unpacks as follows:
\begin{align}
	\label{eq:RadialIntegralsNearNHEK}
	\text{Type I, IIB, IIIB:}\qquad
	\fint_{R_s}^{R_o}=-\nu_s\int_{R_\pm}^{R_s}+\nu_o\int_{R_\pm}^{R_o},\qquad
	\pm=\begin{cases}
		\sign\pa{EL}&\quad\text{Type I},\\
		\sign\pa{L}&\quad\text{Type IIB},\\
		0&\quad\text{Type IIIB}.
	\end{cases}
\end{align}
Since only the Type I, Type IIB and Type IIIB geodesics encounter a turning point outside the horizon, we need to specify an alternate reference radius for the Type IIA geodesics and Type IIIA geodesics. As in Poincar\'e NHEK, $\bar{I}_R$ is divergent at large $R$, but converges when integrating across the horizon at $R=\kappa$: to evaluate this integral one uses the representation \eqref{eq:RadialIntegralsNearNHEK}, with the lower endpoint of integration given by $R_-$ for Type IIA and $R_0$ for Type IIIA. The integrals $\bar{I}_{\Phi}$ and $\bar{I}_T$ are convergent at large $R$, but have a non-integrable singularity at the horizon of near-NHEK: to evaluate these integrals, one should use the alternate representation 
\begin{align}
	\label{eq:RadialIntegralsNearNHEK2}
	\text{Type IIA, IIIA:}\qquad
	\fint_{R_s}^{R_o}=-\nu_s\int_{\infty}^{R_s}+\nu_s\int_{\infty}^{R_o}.
\end{align}

We can now explicitly evaluate the radial integrals. For our purposes, the relevant basis of integrals is given by 
\begin{subequations}
\begin{align}
	\bar{\mathcal{I}}_1^\pm&\equiv\int_{R_\pm}^{R_i}\frac{\ed R}{\sqrt{\mathcal{C}\pa{R_+-R}\pa{R-R_-}}},\\
	\bar{\mathcal{I}}_2^\pm&\equiv\int_{R_\pm}^{R_i}\pa{\frac{L\kappa^2+ER}{R^2-\kappa^2}}\frac{\ed R}{\sqrt{\mathcal{C}\pa{R_+-R}\pa{R-R_-}}}
	=\int_{R_\pm}^{R_i}\frac{1}{2}\pa{\frac{E+L\kappa}{R-\kappa}+\frac{E-L\kappa}{R+\kappa}}\frac{\ed R}{\sqrt{\mathcal{C}\pa{R_+-R}\pa{R-R_-}}},\\
	\bar{\mathcal{I}}_3^\pm&\equiv\int_{R_\pm}^{R_i}\pa{\frac{E+LR}{R^2-\kappa^2}}\frac{\ed R}{\sqrt{\mathcal{C}\pa{R_+-R}\pa{R-R_-}}}
	=\int_{R_\pm}^{R_i}\frac{1}{2\kappa}\pa{\frac{E+L\kappa}{R-\kappa}-\frac{E-L\kappa}{R+\kappa}}\frac{\ed R}{\sqrt{\mathcal{C}\pa{R_+-R}\pa{R-R_-}}},
\end{align}
\end{subequations}
in terms of which
\begin{subequations}
\label{eq:NearIntegrals}
\begin{align}
	\int^{R_i}_{R_\pm}\frac{\ed R}{\sqrt{\mathcal{R}_\kappa(R)}}&=\bar{\mathcal{I}}_1^\pm,\\
	\int^{R_i}_{R_\pm}R\pa{\frac{E+LR}{R^2-\kappa^2}}\frac{\ed R}{\sqrt{\mathcal{R}_\kappa(R)}}&=L\bar{\mathcal{I}}_1^\pm+\bar{\mathcal{I}}_2^\pm,\\
	\int^{R_i}_{R_\pm}\pa{\frac{E+LR}{R^2-\kappa^2}}\frac{\ed R}{\sqrt{\mathcal{R}_\kappa(R)}}&=\bar{\mathcal{I}}_3^\pm.
\end{align}
\end{subequations}
For Type IIA and Type IIIA geodesics, the radial turning point $R_\pm$ in the limit of integration should be replaced by $R\to\infty$ for $\bar{\mathcal{I}}_2$ and $\bar{\mathcal{I}}_3$.

We define the positive quantity and signs
\begin{align}
	w_i^\pm=\ab{\frac{R_i-R_\pm}{R_i-R_\mp}}^{1/2},\qquad
	\rho^\pm=\sign\pa{L\pm\frac{E}{\kappa}}.
\end{align}
When $\mathcal{C}>0$, we have the following set of inequalities for the allowed geodesic motion:
\begin{align}
	R\in\br{\kappa,R_\pm},\qquad
	R_\mp<0<R_\pm,\qquad
	R_\pm-\kappa\geq0,\qquad
	R_\mp+\kappa<0,\qquad
	L+\frac{E}{\kappa}>0,\qquad
	\pm=\sign\pa{EL}.
\end{align}
Using the substitution
\begin{align}
	R=\frac{R_\pm+R_\mp w^2}{1+w^2},\qquad
	w^2=\frac{R_\pm-R}{R-R_\mp}>0,
\end{align}
together with the fact that
\begin{align}
	\mathcal{C}\pa{R_+\pm\kappa}\pa{R_-\pm\kappa}=-\pa{E\mp L\kappa}^2,
\end{align}
one finds that
\begin{subequations}
\label{eq:NearTypeI}
\begin{align}
	\bar{\mathcal{I}}_1^\pm&=-\frac{2}{\sqrt{\mathcal{C}}}\int_0^{w_i^\pm}\frac{\ed w}{1+w^2}
	=-\frac{2}{\sqrt{\mathcal{C}}}\arctan{w_i^\pm},\\
	\bar{\mathcal{I}}_2^\pm&=-\frac{\kappa}{\sqrt{\mathcal{C}}}\br{-\int_0^{w_i^\pm}\frac{\pa{L-E/\kappa}\ed w}{\pa{R_\pm+\kappa}+\pa{R_\mp+\kappa}w^2}+ \int_0^{w_i^\pm}\frac{\pa{L+E/\kappa}\ed w}{\pa{R_\pm-\kappa}+\pa{R_\mp-\kappa}w^2}}\\
	&=-\frac{\kappa}{\sqrt{\mathcal{C}}}\br{-\frac{\pa{L-E/\kappa}}{\sqrt{-\pa{R_\pm+\kappa}\pa{R_\mp+\kappa}}}\arctanh\pa{\sqrt{-\frac{R_\mp+\kappa}{R_\pm+\kappa}}w_i^\pm}+\frac{\pa{L+E/\kappa}}{\sqrt{\pa{R_\pm-\kappa}\pa{\kappa-R_\mp}}}\arctanh\pa{\sqrt{\frac{\kappa-R_\mp}{R_\pm-\kappa}}w_i^\pm}}\notag\\
	&=\rho^-\arctanh\pa{\sqrt{-\frac{R_\mp+\kappa}{R_\pm+\kappa}}w_i^\pm}-\arctanh\pa{\sqrt{\frac{\kappa-R_\mp}{R_\pm-\kappa}}w_i^\pm},\notag\\
	\bar{\mathcal{I}}_3^\pm&=-\frac{1}{\sqrt{\mathcal{C}}}\br{\int_0^{w_i^\pm}\frac{\pa{L-E/\kappa}\ed w}{\pa{R_\pm+\kappa}+\pa{R_\mp+\kappa}w^2}+\int_0^{w_i^\pm}\frac{\pa{L+E/\kappa}\ed w}{\pa{R_\pm-\kappa}+\pa{R_\mp-\kappa}w^2}}\\
	&=-\frac{1}{\sqrt{\mathcal{C}}}\br{\frac{\pa{L-E/\kappa}}{\sqrt{-\pa{R_\pm+\kappa}\pa{R_\mp+\kappa}}}\arctanh\pa{\sqrt{-\frac{R_\mp+\kappa}{R_\pm+\kappa}}w_i^\pm}+\frac{\pa{L+E/\kappa}}{\sqrt{\pa{R_\pm-\kappa}\pa{\kappa-R_\mp}}}\arctanh\pa{\sqrt{\frac{\kappa-R_\mp}{R_\pm-\kappa}}w_i^\pm}}\notag\\
	&=-\frac{1}{\kappa}\br{\rho^-\arctanh\pa{\sqrt{-\frac{R_\mp+\kappa}{R_\pm+\kappa}}w_i^\pm}+\arctanh\pa{\sqrt{\frac{\kappa-R_\mp}{R_\pm-\kappa}}w_i^\pm}}.\notag
\end{align}
\end{subequations}
When $\mathcal{C}<0$, we have the following set of inequalities for the allowed Type IIB motion:
\begin{align}
	R\gtrless R_\pm,\qquad
	0<\kappa<R_-<R_+,\qquad
	\pm=\sign\pa{L}.
\end{align}
Using the substitution
\begin{align}
	\label{eq:SubstitutionTypeII}
	R=\frac{R_\pm-R_\mp w^2}{1-w^2},\qquad
	w^2=\frac{R_\pm-R}{R_\mp-R}\in[0,1],
\end{align}
one finds that
\begin{subequations}
\label{eq:NearTypeIIB}
\begin{align}
	\bar{\mathcal{I}}_1^\pm&=\pm\frac{2}{\sqrt{\ab{\mathcal{C}}}}\int_0^{w_i^\pm}\frac{\ed w}{1-w^2}
	=\pm\frac{2}{\sqrt{\ab{\mathcal{C}}}}\arctanh{w_i^\pm},\\
	\bar{\mathcal{I}}_2^\pm&=\pm\frac{\kappa}{\sqrt{\ab{\mathcal{C}}}}\br{-\int_0^{w_i^\pm}\frac{\pa{L-E/\kappa}\ed w}{\pa{R_\pm+\kappa}-\pa{R_\mp+\kappa}w^2}+\int_0^{w_i^\pm}\frac{\pa{L+E/\kappa}\ed w}{\pa{R_\pm-\kappa}-\pa{R_\mp-\kappa}w^2}}\\
	&=\pm\frac{\kappa}{\sqrt{\mathcal{\ab{C}}}}\br{-\frac{\pa{L-E/\kappa}}{\sqrt{\pa{R_\pm+\kappa}\pa{R_\mp+\kappa}}}\arctanh\pa{\sqrt{\frac{R_\mp+\kappa}{R_\pm+\kappa}}w_i^\pm}+\frac{\pa{L+E/\kappa}}{\sqrt{\pa{R_\pm-\kappa}\pa{R_\mp-\kappa}}}\arctanh\pa{\sqrt{\frac{R_\mp-\kappa}{R_\pm-\kappa}}w_i^\pm}}\notag\\
	&=\mp\br{\rho^-\arctanh\pa{\sqrt{\frac{R_\mp+\kappa}{R_\pm+\kappa}}w_i^\pm}-\rho^+\arctanh\pa{\sqrt{\frac{R_\mp-\kappa}{R_\pm-\kappa}}w_i^\pm}},\notag\\
	\bar{\mathcal{I}}_3^\pm&=\pm\frac{1}{\sqrt{\ab{\mathcal{C}}}}\br{\int_0^{w_i^\pm}\frac{\pa{L-E/\kappa}\ed w}{\pa{R_\pm+\kappa}-\pa{R_\mp+\kappa}w^2}+\int_0^{w_i^\pm}\frac{\pa{L+E/\kappa}\ed w}{\pa{R_\pm-\kappa}-\pa{R_\mp-\kappa}w^2}}\\
	&=\pm\frac{1}{\sqrt{\mathcal{\ab{C}}}}\br{\frac{\pa{L-E/\kappa}}{\sqrt{\pa{R_\pm+\kappa}\pa{R_\mp+\kappa}}}\arctanh\pa{\sqrt{\frac{R_\mp+\kappa}{R_\pm+\kappa}}w_i^\pm}+\frac{\pa{L+E/\kappa}}{\sqrt{\pa{R_\pm-\kappa}\pa{R_\mp-\kappa}}}\arctanh\pa{\sqrt{\frac{R_\mp-\kappa}{R_\pm-\kappa}}w_i^\pm}}\notag\\
	&=\pm\frac{1}{\kappa}\br{\rho^-\arctanh\pa{\sqrt{\frac{R_\mp+\kappa}{R_\pm+\kappa}}w_i^\pm}+\rho^+\arctanh\pa{\sqrt{\frac{R_\mp-\kappa}{R_\pm-\kappa}}w_i^\pm}}.\notag
\end{align}
\end{subequations}

For the Type IIA geodesics, we have two further subcases, according to whether the roots $R_\pm$ are real or complex. In the former subcase,
\begin{align}
	-(E/\kappa)^2<\mathcal{C}<0:\qquad
	E>0,\qquad
	L>0,\qquad
	R_+<R_-<0,\qquad
	R_\pm\pm\kappa<0.
\end{align}
The appropriate substitution is given in Eq.~\eqref{eq:SubstitutionTypeII}, with the lower choice of sign for $\bar{\mathcal{I}}_1$ and the upper choice of sign for $\bar{\mathcal{I}}_2$ and $\bar{\mathcal{I}}_3$:
\begin{subequations}
\label{eq:NearTypeIIARealRoots}
\begin{align}
	\bar{\mathcal{I}}_1^{-}&=\frac{2}{\sqrt{\ab{\mathcal{C}}}}\int_0^{w_i^-}\frac{\ed w}{1-w^2}
	=\frac{2}{\sqrt{\ab{\mathcal{C}}}}\arctanh{w_i^-},\\
	\bar{\mathcal{I}}_2^\infty&=\frac{\kappa}{\sqrt{\ab{\mathcal{C}}}}\br{-\int_1^{w_i^+}\frac{\pa{L-E/\kappa}\ed w}{\pa{R_++\kappa}-\pa{R_-+\kappa}w^2}+\int_1^{w_i^+}\frac{\pa{L+E/\kappa}\ed w}{\pa{R_+-\kappa}-\pa{R_--\kappa}w^2}}\\
	&=\frac{\kappa}{\sqrt{\mathcal{\ab{C}}}}\Bigg\{-\frac{\pa{L-E/\kappa}}{\sqrt{\pa{R_++\kappa}\pa{R_-+\kappa}}}\br{\arctanh\sqrt{\frac{R_-+\kappa}{R_++\kappa}}-\arctanh\pa{\sqrt{\frac{R_-+\kappa}{R_++\kappa}}w_i^+}}\notag\\
	&\qquad\qquad\quad+\frac{\pa{L+E/\kappa}}{\sqrt{\pa{R_+-\kappa}\pa{R_--\kappa}}}\br{\arctanh\sqrt{\frac{R_--\kappa}{R_+-\kappa}}-\arctanh\pa{\sqrt{\frac{R_--\kappa}{R_+-\kappa}}w_i^+}}\Bigg\}\notag\\
	&=\rho^-\br{\arctanh\pa{\sqrt{\frac{R_-+\kappa}{R_++\kappa}}w_i^+}-\arctanh\sqrt{\frac{R_-+\kappa}{R_++\kappa}}}-\br{\arctanh\pa{\sqrt{\frac{R_--\kappa}{R_+-\kappa}}w_i^+}-\arctanh\sqrt{\frac{R_--\kappa}{R_+-\kappa}}},\notag\\
	\bar{\mathcal{I}}_3^\infty&=\frac{1}{\sqrt{\ab{\mathcal{C}}}}\br{\int_1^{w_i^+}\frac{\pa{L-E/\kappa}\ed w}{\pa{R_++\kappa}-\pa{R_-+\kappa}w^2}+\int_1^{w_i^+}\frac{\pa{L+E/\kappa}\ed w}{\pa{R_+-\kappa}-\pa{R_--\kappa}w^2}}\\
	&=\frac{1}{\sqrt{\ab{\mathcal{C}}}}\Bigg\{\frac{\pa{L-E/\kappa}}{\sqrt{\pa{R_++\kappa}\pa{R_-+\kappa}}}\br{\arctanh\sqrt{\frac{R_-+\kappa}{R_++\kappa}}-\arctanh\pa{\sqrt{\frac{R_-+\kappa}{R_++\kappa}}w_i^+}}\notag\\
	&\qquad\qquad\quad+\frac{\pa{L+E/\kappa}}{\sqrt{\pa{R_+-\kappa}\pa{R_--\kappa}}}\br{\arctanh\sqrt{\frac{R_--\kappa}{R_+-\kappa}}-\arctanh\pa{\sqrt{\frac{R_--\kappa}{R_+-\kappa}}w_i^+}}\Bigg\}\notag\\
	&=-\frac{\rho^-}{\kappa}\br{\arctanh\pa{\sqrt{\frac{R_-+\kappa}{R_++\kappa}}w_i^+}-\arctanh\sqrt{\frac{R_-+\kappa}{R_++\kappa}}}-\frac{1}{\kappa}\br{\arctanh\pa{\sqrt{\frac{R_--\kappa}{R_+-\kappa}}w_i^+}-\arctanh\sqrt{\frac{R_--\kappa}{R_+-\kappa}}}.\notag
\end{align}
\end{subequations}
The second subcase is
\begin{align}
	-L^2<\mathcal{C}<-(E/\kappa)^2<0:\qquad
	L>0,\qquad
	L\pm\frac{E}{\kappa}>0,\qquad
	R_+=\overline{R_-}.
\end{align}
Note that since $R_-$ is complex, the integral $\bar{\mathcal{I}}_1^-$ is in general also complex. However, the imaginary part cancels out in the sum \eqref{eq:RadialIntegralsNearNHEK2}, since the full contour is  deformable to the real axis. Similarly, since $R_+$ is complex for this case, the integrals $\bar{\mathcal{I}}_2$ and $\bar{\mathcal{I}}_3$ can be integrated from $R_+$ to $R_i$: one avoids the non-integrable singularity at $R=\kappa$ when $R_+$ moves into the complex plane. In terms of
\begin{align}
	W_i^\pm=\sqrt{\frac{R_i-R_\pm}{R_i-R_\mp}},\qquad
	\xi^\pm=\sign\pa{\frac{\kappa}{\sqrt{\ab{\mathcal{C}}}}\frac{\pa{L\pm E/\kappa}}{\pa{R_+\mp\kappa}W_{\pm\kappa}^-}}
	=\sign\pa{\frac{\sqrt{\pa{R_+\pm\kappa}\pa{R_-\pm\kappa}}}{\pa{R_+\mp\kappa}W_{\pm\kappa}^-}},
\end{align}
one finds
\begin{subequations}
\label{eq:NearTypeIIAComplexRoots}
\begin{align}
	\bar{\mathcal{I}}_1^{-}&=\frac{2}{\sqrt{\ab{\mathcal{C}}}}\int_0^{W_i^-}\frac{\ed w}{1-w^2}
	=\frac{2}{\sqrt{\ab{\mathcal{C}}}}\arctanh{W_i^-},\\
	\bar{\mathcal{I}}_2^+&=\frac{\kappa}{\sqrt{\ab{\mathcal{C}}}}\br{-\int_0^{W_i^+}\frac{\pa{L-E/\kappa}\ed w}{\pa{R_++\kappa}-\pa{R_-+\kappa}w^2}+\int_0^{W_i^+}\frac{\pa{L+E/\kappa}\ed w}{\pa{R_+-\kappa}-\pa{R_--\kappa}w^2}}\\
	&=\frac{\kappa}{\sqrt{\mathcal{\ab{C}}}}\br{-\frac{\pa{L-E/\kappa}}{(R_++\kappa)W^-_{-\kappa}}\arctanh\pa{\sqrt{\frac{R_-+\kappa}{R_++\kappa}}W_i^+}+\frac{\pa{L+E/\kappa}}{(R_+-\kappa)W^-_\kappa}\arctanh\pa{\sqrt{\frac{R_--\kappa}{R_+-\kappa}}W_i^+}}\notag\\
	&=-\xi^-\arctanh\pa{\sqrt{\frac{R_-+\kappa}{R_++\kappa}}W_i^+}+\xi^+\arctanh\pa{\sqrt{\frac{R_--\kappa}{R_+-\kappa}}W_i^+},\notag\\
	\bar{\mathcal{I}}_3^+&=\frac{1}{\sqrt{\ab{\mathcal{C}}}}\br{\int_0^{W_i^+}\frac{\pa{L-E/\kappa}\ed w}{\pa{R_++\kappa}-\pa{R_-+\kappa}w^2}+\int_0^{W_i^+}\frac{\pa{L+E/\kappa}\ed w}{\pa{R_+-\kappa}-\pa{R_--\kappa}w^2}}\\
	&=\frac{1}{\sqrt{\ab{\mathcal{C}}}}\br{\frac{\pa{L-E/\kappa}}{(R_++\kappa)W^-_{-\kappa}}\arctanh\pa{\sqrt{\frac{R_-+\kappa}{R_++\kappa}}W_i^+}+\frac{\pa{L+E/\kappa}}{(R_+-\kappa)W^-_\kappa}\arctanh\pa{\sqrt{\frac{R_--\kappa}{R_+-\kappa}}W_i^+}}\notag\\
	&=\frac{1}{\kappa}\br{\xi^-\arctanh\pa{\sqrt{\frac{R_-+\kappa}{R_++\kappa}}W_i^+}+\xi^+\arctanh\pa{\sqrt{\frac{R_--\kappa}{R_+-\kappa}}W_i^+}}.\notag
\end{align}
\end{subequations}

Finally, when $\mathcal{C}=0$,
\begin{align}
	\mathcal{R}_\kappa(R)=2EL\pa{R-R_0}>0,\qquad
	R_0=-\frac{E^2+L^2\kappa^2}{2EL},\qquad
	\mathcal{R}_\kappa(\pm\kappa)=\kappa^2\pa{L\pm E/\kappa}^2,\qquad
	L+\frac{E}{\kappa}>0,
\end{align}
and the integrals degenerate. We use the substitution $w^2=2EL\pa{R-R_0}$. For Type IIIB motion with $EL<0$ and $R\in\br{\kappa,R_0}$, the integrals evaluate to
\begin{subequations}
\label{eq:NearTypeIIIB}
\begin{align}
	\bar{\mathcal{I}}_1^0&=\int_{R_0}^{R_i}\frac{\ed R}{\sqrt{2EL\pa{R-R_0}}}
	=\frac{\sqrt{\mathcal{R}_\kappa(R_i)}}{EL},\\
	\bar{\mathcal{I}}_2^0&=\int_{R_0}^{R_i}\pa{\frac{L\kappa^2+ER}{R^2-\kappa^2}}\frac{\ed R}{\sqrt{2EL\pa{R-R_0}}}
	=\kappa\br{\frac{\pa{L-E/\kappa}}{\sqrt{\mathcal{R}_\kappa(-\kappa)}}\arctanh{\sqrt{\frac{R_0-R_i}{R_0+\kappa}}}-\frac{\pa{L+E/\kappa}}{\sqrt{\mathcal{R}_\kappa(\kappa)}}\arctanh{\sqrt{\frac{R_0-R_i}{R_0-\kappa}}}}\notag\\
	&=\rho_-\arctanh\sqrt{\frac{R_0-R_i}{R_0+\kappa}}-\arctanh\sqrt{\frac{R_0-R_i}{R_0-\kappa}},\\
	\bar{\mathcal{I}}_3^0&=\int_{R_0}^{R_i}\pa{\frac{E+LR}{R^2-\kappa^2}}\frac{\ed R}{\sqrt{2EL\pa{R-R_0}}}
	=-\br{\frac{\pa{L-E/\kappa}}{\sqrt{\mathcal{R}_\kappa(-\kappa)}}\arctanh{\sqrt{\frac{R_0-R_i}{R_0+\kappa}}}+\frac{\pa{L+E/\kappa}}{\sqrt{\mathcal{R}_\kappa(\kappa)}}\arctanh{\sqrt{\frac{R_0-R_i}{R_0-\kappa}}}}\notag\\
	&=-\frac{1}{\kappa}\br{\rho_-\arctanh\sqrt{\frac{R_0-R_i}{R_0+\kappa}}+\arctanh\sqrt{\frac{R_0-R_i}{R_0-\kappa}}}.
\end{align}
\end{subequations}
For Type IIIA motion with $EL>0$ and $R\in[\kappa,\infty)$, one has the additional relations
\begin{align}
	R_0^2-\kappa^2=\pa{\frac{E^2-L^2\kappa^2}{2EL}}^2>0 \qquad\Longrightarrow\qquad
	R_0+\kappa<0.
\end{align}
Using the substitution $w^{-2}=2EL\pa{R-R_0}$, the integrals take the form
\begin{subequations}
\label{eq:NearTypeIIIA}
\begin{align}
	\bar{\mathcal{I}}_1^0&=\int_{R_0}^{R_i}\frac{\ed R}{\sqrt{2EL\pa{R-R_0}}}
	=\frac{\sqrt{\mathcal{R}_\kappa(R_i)}}{EL},\\
	\bar{\mathcal{I}}_2^\infty&=\int_\infty^{R_i}\pa{\frac{L\kappa^2+ER}{R^2-\kappa^2}}\frac{\ed R}{\sqrt{2EL\pa{R-R_0}}}
	=\kappa\br{\frac{\pa{L-E/\kappa}}{\sqrt{\mathcal{R}_\kappa(-\kappa)}}\arctanh{\sqrt{\frac{R_0+\kappa}{R_0-R_i}}}-\frac{\pa{L+E/\kappa}}{\sqrt{\mathcal{R}_\kappa(\kappa)}}\arctanh{\sqrt{\frac{R_0-\kappa}{R_0-R_i}}}}\notag\\
	&=\rho_-\arctanh\sqrt{\frac{R_0+\kappa}{R_0-R_i}}-\arctanh\sqrt{\frac{R_0-\kappa}{R_0-R_i}},\\
	\bar{\mathcal{I}}_3^\infty&=\int_\infty^{R_i}\pa{\frac{E+LR}{R^2-\kappa^2}}\frac{\ed R}{\sqrt{2EL\pa{R-R_0}}}
	=-\br{\frac{\pa{L-E/\kappa}}{\sqrt{\mathcal{R}_\kappa(-\kappa)}}\arctanh{\sqrt{\frac{R_0+\kappa}{R_0-R_i}}}+\frac{\pa{L+E/\kappa}}{\sqrt{\mathcal{R}_\kappa(\kappa)}}\arctanh{\sqrt{\frac{R_0-\kappa}{R_0-R_i}}}}\notag\\
	&=-\frac{1}{\kappa}\br{\rho_-\arctanh\sqrt{\frac{R_0+\kappa}{R_0-R_i}}+\arctanh\sqrt{\frac{R_0-\kappa}{R_0-R_i}}}.
\end{align}
\end{subequations}

Substituting Eqs.~\eqref{eq:NearIntegrals} and \eqref{eq:NearTypeI} into Eq.~\eqref{eq:RadialIntegralsNearNHEK} yields the geodesic integrals for Type I motion:
\begin{subequations}
\begin{empheq}[box=\ovalbox]{align}
	\shortintertext{\centering\underline{Type I:}\qquad$\mathcal{C}>0$ and $L+E/\kappa>0$ with $\pm=\sign\pa{EL}$}
	\bar{I}_R&=\frac{2}{\sqrt{\mathcal{C}}}\br{\nu_s\arctan{w_s^\pm}-\nu_o\arctan{w_o^\pm}},\\
	\bar{I}_\Phi&=L\bar{I}_R-\nu_s\br{\rho^-\arctanh\pa{\sqrt{-\frac{R_\mp+\kappa}{R_\pm+\kappa}}w_s^\pm}-\arctanh\pa{\sqrt{\frac{\kappa-R_\mp}{R_\pm-\kappa}}w_s^\pm}}\\
	&\quad+\nu_o\br{\rho^-\arctanh\pa{\sqrt{-\frac{R_\mp+\kappa}{R_\pm+\kappa}}w_o^\pm}-\arctanh\pa{\sqrt{\frac{\kappa-R_\mp}{R_\pm-\kappa}}w_o^\pm}},\notag\\
	\bar{I}_T&=\frac{\nu_s}{\kappa}\br{\rho^-\arctanh\pa{\sqrt{-\frac{R_\mp+\kappa}{R_\pm+\kappa}}w_s^\pm}+\arctanh\pa{\sqrt{\frac{\kappa-R_\mp}{R_\pm-\kappa}}w_s^\pm}}\\
	&\quad-\frac{\nu_o}{\kappa}\br{\rho^-\arctanh\pa{\sqrt{-\frac{R_\mp+\kappa}{R_\pm+\kappa}}w_o^\pm}+\arctanh\pa{\sqrt{\frac{\kappa-R_\mp}{R_\pm-\kappa}}w_o^\pm}}.\notag
\end{empheq}
\end{subequations}

Substituting Eqs.~\eqref{eq:NearTypeIIARealRoots} and \eqref{eq:NearTypeIIAComplexRoots} into Eqs.~\eqref{eq:RadialIntegralsNearNHEK} and \eqref{eq:RadialIntegralsNearNHEK2} yields the geodesic integrals for Type IIA motion:
\begin{subequations}
\begin{empheq}[box=\ovalbox]{align}
	\shortintertext{\centering\underline{Type IIA:}\qquad$-L^2<\mathcal{C}<0$, $-\pa{E/\kappa}^2<\mathcal{C}$, and $E,L>0$ ($R_\pm\in\mathbb{R}$)}
	\bar{I}_R&=\frac{2\nu_s}{\sqrt{\ab{\mathcal{C}}}}\br{\arctanh{w_o^-}-\arctanh{w_s^-}},\\
	\bar{I}_\Phi&=L\bar{I}_R-\nu_s\br{\rho^-\arctanh\pa{\sqrt{\frac{R_-+\kappa}{R_++\kappa}}w_s^+}-\arctanh\pa{\sqrt{\frac{R_--\kappa}{R_+-\kappa}}w_s^+}}\\
	&\quad+\nu_s\br{\rho^-\arctanh\pa{\sqrt{\frac{R_-+\kappa}{R_++\kappa}}w_o^+}-\arctanh\pa{\sqrt{\frac{R_--\kappa}{R_+-\kappa}}w_o^+}},\notag\\
	\bar{I}_T&=\frac{\nu_s}{\kappa}\br{\rho^-\arctanh\pa{\sqrt{\frac{R_-+\kappa}{R_++\kappa}}w_s^+}+\arctanh\pa{\sqrt{\frac{R_--\kappa}{R_+-\kappa}}w_s^+}}\\
	&\quad-\frac{\nu_s}{\kappa}\br{\rho^-\arctanh\pa{\sqrt{\frac{R_-+\kappa}{R_++\kappa}}w_o^+}+\arctanh\pa{\sqrt{\frac{R_--\kappa}{R_+-\kappa}}w_o^+}}.\notag
\end{empheq}
\end{subequations}
\begin{subequations}
\begin{empheq}[box=\ovalbox]{align}
	\shortintertext{\centering\underline{Type IIA:}\qquad$-L^2<\mathcal{C}<-\pa{E/\kappa}^2<0$ and $L\pm E/\kappa>0$ ($R_\pm\in\mathbb{C}$)}
	\bar{I}_R&=\frac{2\nu_s}{\sqrt{\ab{\mathcal{C}}}}\br{\arctanh{W_o^-}-\arctanh{W_s^-}},\\
	\bar{I}_\Phi&=L\bar{I}_R+\nu_s\br{\xi^-\arctanh\pa{\sqrt{\frac{R_-+\kappa}{R_++\kappa}}W_s^+}-\xi^+\arctanh\pa{\sqrt{\frac{R_--\kappa}{R_+-\kappa}}W_s^+}}\\
	&\quad-\nu_s\br{\xi^-\arctanh\pa{\sqrt{\frac{R_-+\kappa}{R_++\kappa}}W_o^+}-\xi^+\arctanh\pa{\sqrt{\frac{R_--\kappa}{R_+-\kappa}}W_o^+}},\notag\\
	\bar{I}_T&=-\frac{\nu_s}{\kappa}\br{\xi^-\arctanh\pa{\sqrt{\frac{R_-+\kappa}{R_++\kappa}}W_s^+}+\xi^+\arctanh\pa{\sqrt{\frac{R_--\kappa}{R_+-\kappa}}W_s^+}}\\
	&\quad+\frac{\nu_s}{\kappa}\br{\xi^-\arctanh\pa{\sqrt{\frac{R_-+\kappa}{R_++\kappa}}W_o^+}+\xi^+\arctanh\pa{\sqrt{\frac{R_--\kappa}{R_+-\kappa}}W_o^+}}.\notag
\end{empheq}
\end{subequations}

Substituting Eqs.~\eqref{eq:NearIntegrals} and \eqref{eq:NearTypeIIB} into Eq.~\eqref{eq:RadialIntegralsNearNHEK} yields the geodesic integrals for Type IIB motion:
\begin{subequations}
\begin{empheq}[box=\ovalbox]{align}
	\shortintertext{\centering\underline{Type IIB:}\qquad$-L^2<\mathcal{C}<0$ with $\pm=\sign L$}
	\bar{I}_R&=\mp\frac{2}{\sqrt{\mathcal{\ab{C}}}}\br{\nu_s\arctanh{w_s^\pm}-\nu_o\arctanh{w_o^\pm}},\\
	\bar{I}_\Phi&=L\bar{I}_R\pm\nu_s\br{\rho^-\arctanh\pa{\sqrt{\frac{R_\mp+\kappa}{R_\pm+\kappa}}w_s^\pm}-\rho^+\arctanh\pa{\sqrt{\frac{R_\mp-\kappa}{R_\pm-\kappa}}w_s^\pm}}\\
	&\quad\mp\nu_o\br{\rho^-\arctanh\pa{\sqrt{\frac{R_\mp+\kappa}{R_\pm+\kappa}}w_o^\pm}-\rho^+\arctanh\pa{\sqrt{\frac{R_\mp-\kappa}{R_\pm-\kappa}}w_o^\pm}},\notag\\
	\bar{I}_T&=\mp\frac{\nu_s}{\kappa}\br{\rho^-\arctanh\pa{\sqrt{\frac{R_\mp+\kappa}{R_\pm+\kappa}}w_s^\pm}+\rho^+\arctanh\pa{\sqrt{\frac{R_\mp-\kappa}{R_\pm-\kappa}}w_s^\pm}}\\
	&\quad\pm\frac{\nu_o}{\kappa}\br{\rho^-\arctanh\pa{\sqrt{\frac{R_\mp+\kappa}{R_\pm+\kappa}}w_o^\pm}+\rho^+\arctanh\pa{\sqrt{\frac{R_\mp-\kappa}{R_\pm-\kappa}}w_o^\pm}}.\notag
\end{empheq}
\end{subequations}

Substituting Eqs.~\eqref{eq:NearIntegrals} and \eqref{eq:NearTypeIIIA} into Eq.~\eqref{eq:RadialIntegralsNearNHEK2} yields the geodesic integrals for Type IIIA motion:
\begin{subequations}
\begin{empheq}[box=\ovalbox]{align}
	\shortintertext{\centering\underline{Type IIIA:}\qquad$\mathcal{C}=0$ and $EL>0$}
	\bar{I}_R&=\nu_s\br{\frac{\sqrt{\mathcal{R}_\kappa(R_o)}}{EL}-\frac{\sqrt{\mathcal{R}_\kappa(R_s)}}{EL}},\\
	\bar{I}_\Phi&=L\bar{I}_R-\nu_s\br{\rho_-\arctanh\sqrt{\frac{R_0+\kappa}{R_0-R_s}}-\arctanh\sqrt{\frac{R_0-\kappa}{R_0-R_s}}}\\
	&\quad+\nu_s\br{\rho_-\arctanh\sqrt{\frac{R_0+\kappa}{R_0-R_o}}-\arctanh\sqrt{\frac{R_0-\kappa}{R_0-R_o}}},\notag\\
	\bar{I}_T&=\frac{\nu_s}{\kappa}\br{\rho_-\arctanh\sqrt{\frac{R_0+\kappa}{R_0-R_s}}+\arctanh\sqrt{\frac{R_0-\kappa}{R_0-R_s}}}\\
	&\quad-\frac{\nu_s}{\kappa}\br{\rho_-\arctanh\sqrt{\frac{R_0+\kappa}{R_0-R_o}}+\arctanh\sqrt{\frac{R_0-\kappa}{R_0-R_o}}}.\notag
\end{empheq}
\end{subequations}

Finally, substituting Eqs.~\eqref{eq:NearIntegrals} and \eqref{eq:NearTypeIIIB} into Eq.~\eqref{eq:RadialIntegralsNearNHEK} yields the geodesic integrals for Type IIIB motion:
\begin{subequations}
\begin{empheq}[box=\ovalbox]{align}
	\shortintertext{\centering\underline{Type IIIB:}\qquad$\mathcal{C}=0$, $EL<0$ and $L+E/\kappa>0$}
	\bar{I}_R&=-\nu_s\frac{\sqrt{\mathcal{R}_\kappa(R_s)}}{EL}+\nu_o\frac{\sqrt{\mathcal{R}_\kappa(R_o)}}{EL},\\
	\bar{I}_\Phi&=L\bar{I}_R-\nu_s\br{\rho_-\arctanh\sqrt{\frac{R_0-R_s}{R_0+\kappa}}-\arctanh\sqrt{\frac{R_0-R_s}{R_0-\kappa}}}\\
	&\quad+\nu_o\br{\rho_-\arctanh\sqrt{\frac{R_0-R_o}{R_0+\kappa}}-\arctanh\sqrt{\frac{R_0-R_o}{R_0-\kappa}}},\notag\\
	\bar{I}_T&=\frac{\nu_s}{\kappa}\br{\rho_-\arctanh\sqrt{\frac{R_0-R_s}{R_0+\kappa}}+\arctanh\sqrt{\frac{R_0-R_s}{R_0-\kappa}}}\\
	&\quad-\frac{\nu_o}{\kappa}\br{\rho_-\arctanh\sqrt{\frac{R_0-R_o}{R_0+\kappa}}+\arctanh\sqrt{\frac{R_0-R_o}{R_0-\kappa}}}.\notag
\end{empheq}
\end{subequations}

\subsubsection{Explicit solution of the geodesic equation}

We would now like to explicitly solve for the $(T,R)$ motion in near-NHEK. As always, the problem can be recast as a first-order non-linear autonomous system:
\begin{align}
	\label{eq:NearODE}
	\frac{dR}{dT}=\pm_R\frac{R^2-\kappa^2}{E+LR}\sqrt{\mathcal{R}_\kappa(R)}.
\end{align}
The physically relevant properties of the analogous ODEs in global and Poincar\'e NHEK manifest themselves in near-NHEK as well. The non-linearity of the system allows for blowup of solutions in finite coordinate time, allowing particles to reach the boundary of near-NHEK. Likewise, non-differentiability of the right-hand side of Eq.~\eqref{eq:NearODE} at zeroes of the radial potential allows for multiple solutions with the same initial conditions: constant-radius trajectories also exist in near-NHEK. The qualitative fixed-point analysis of this equation was performed in Sec.~\ref{subsec:NearQualitative}.

To proceed in solving this equation, we make use of the symmetries of near-NHEK. The quantities $H_+=H_+^\mu p_\mu$ and $H_-=H_-^\mu p_\mu$ are conserved along the particle trajectory. We can therefore evaluate the quantity
\begin{align}
	\frac{H_+}{H_-}=e^{-2\kappa T}\frac{ER+L\kappa^2\mp_R\kappa\sqrt{\mathcal{R}_\kappa(R)}}{ER+L\kappa^2\pm_R\kappa\sqrt{\mathcal{R}_\kappa(R)}}
\end{align}
at the separate points $(T_s,R_s)$ and $(T_o,R_o)$, and then set the results equal to each other to obtain
\begin{align}
	e^{2\kappa\pa{T_o-T_s}}=\pa{\frac{ER_o+L\kappa^2-\nu_o\kappa\sqrt{\mathcal{R}_\kappa(R_o)}}{ER_o+L\kappa^2+\nu_o\kappa\sqrt{\mathcal{R}_\kappa(R_o)}}}\pa{\frac{ER_s+L\kappa^2+\nu_s\kappa\sqrt{\mathcal{R}_\kappa(R_s)}}{ER_s+L\kappa^2-\nu_s\kappa\sqrt{\mathcal{R}_\kappa(R_s)}}}.
\end{align}
Taking the logarithm of this expression yields 
\begin{align}
	\label{eq:T(R)NearNHEK}
	T(R_o)-T(R_s)=\left.\frac{1}{2\kappa}\log\br{\frac{ER+L\kappa^2\mp_R\kappa\sqrt{\mathcal{R}_\kappa(R)}}{ER+L\kappa^2\pm_R\kappa\sqrt{\mathcal{R}_\kappa(R)}}}\right|_{R=R_s}^{R=R_o}.
\end{align}
We would now like to invert this relation to obtain $R_o(T_o)$. In terms of the (non-conserved) quantity
\begin{align}
	X(T_o)&\equiv e^{2\kappa T_o}\left.\frac{H_+}{H_-}\right|_{(T,R)=(T_s,R_s)}
	=e^{2\kappa\pa{T_o-T_s}}\frac{ER_s+L\kappa^2-\nu_s\kappa\sqrt{\mathcal{R}_\kappa(R_s)}}{ER_s+L\kappa^2+\nu_s\kappa\sqrt{\mathcal{R}_\kappa(R_s)}}
	=1+\O{\kappa},
\end{align}
Eq.~\eqref{eq:T(R)NearNHEK} takes the form
\begin{align}
	\frac{ER_o+L\kappa^2-\nu_o\kappa\sqrt{\mathcal{R}_\kappa(R_o)}}{ER_o+L\kappa^2+\nu_o\kappa\sqrt{\mathcal{R}_\kappa(R_o)}}=X(T_o).
\end{align}
Isolating the terms with radicals, squaring, and solving the resulting quadratic equation, one finds
\begin{align}
	\label{eq:R(T)NearNHEK}
	R_o(T_o)=\frac{EL}{S_\kappa(T_o)}\br{1\pm\sqrt{\pa{1+\frac{S_\kappa(T_o)}{L^2}}\pa{1+\frac{S_\kappa(T_o)\kappa^2}{E^2}}}},
\end{align}
where
\begin{align}
	S_\kappa(T_o)\equiv\frac{1}{X(T_o)}\br{\mathcal{C}\pa{\frac{1+X(T_o)}{2}}^2+\frac{E^2}{\kappa^2}\pa{\frac{1-X(T_o)}{2}}^2}
	=S_n(T_o)+\O{\kappa^2}.
\end{align}
In the $\kappa\to0$ limit, this expression reduces to Eq.~\eqref{eq:R(T)NHEK}, as expected. The choice of sign in Eq.~\eqref{eq:R(T)NearNHEK} is fixed by the type of geodesic under consideration. Before we can specify it, note that whenever the time is
\begin{align}
	T_t=T_s-\frac{1}{2\kappa}\log\br{\frac{ER_s+L\kappa^2-\nu_s\kappa\sqrt{\mathcal{R}_\kappa(R_s)}}{ER_s+L\kappa^2+\nu_s\kappa\sqrt{\mathcal{R}_\kappa(R_s)}}},
\end{align}
we have $X(T_t)=1$, and therefore $S_\kappa(T_t)=\mathcal{C}$, which in turn implies $R_o(T_t)=R_\pm$. Hence, $T_t$ is the turning time of the radial motion (provided there is one). Also, note that $S_\kappa(T_t)$ is strictly positive for Type I geodesics, while for Type II geodesics, it has a pair of real zeroes
\begin{align}
	T^\pm=T_s-\frac{1}{2\kappa}\log\br{\pa{\frac{ER_s+L\kappa^2-\nu_s\kappa\sqrt{\mathcal{R}_\kappa(R_s)}}{ER_s+L\kappa^2+\nu_s\kappa\sqrt{\mathcal{R}_\kappa(R_s)}}}\frac{\pa{\mathcal{C}+E^2/\kappa^2}^{\pm1}}{\pa{\sqrt{\ab{\mathcal{C}}}+E/\kappa}^{\pm2}}}.
\end{align}
We can now explicitly describe the radial motion depending on the sign of $\mathcal{C}$:
\begin{enumerate}
\item[I.]
$0<\mathcal{C}<\infty$ corresponds to Type I motion. In this case, $R(T)$ is given for all times $T\in\mathbb{R}$ by Eq.~\eqref{eq:R(T)NearNHEK} with the choice of sign $\pm=\sign\pa{EL}$, provided that $R_\pm>\kappa$. The particle reaches the horizon at $T\to\pm\infty$, and encounters a turning point in between at $(T,R)=(T_t,R_\pm)$.
\item[II.]
$-L^2<\mathcal{C}<0$ corresponds to Type II motion, and there are two further subcases to consider:
\begin{enumerate}
\item[A.]
$E,L>0$, or $E<0<-E/\kappa<L$ with $\mathcal{C}<-\pa{E/\kappa}^2$, corresponds to Type IIA motion. If $E,L>0$, then $R(T)$ is given by Eq.~\eqref{eq:R(T)NearNHEK} with the upper choice of sign. If $E<0<-E/\kappa<L$ with $\mathcal{C}<-\pa{E/\kappa}^2$, then $R(T)$ is given by Eq.~\eqref{eq:R(T)NearNHEK} with the choice of sign flipping at $T=\pa{T_++T_-}/2$. One must choose the upper/lower sign for $T<\pa{T_++T_-}/2$ according to whether $p^R\lessgtr0$, and the opposite sign for $T>\pa{T_++T_-}/2$. In both cases, $R(T)$ is defined on the domain $T\gtrless T^\pm$ according to whether $p^R\lessgtr0$. The particle reaches the boundary at $T=T^\pm$ and the horizon as $T\to\pm\infty$, without ever encountering a turning point.
\item[B.]
$EL<0$ with $-\pa{E/\kappa}^2<\mathcal{C}$ corresponds to Type IIB motion. If $L>0$, then $R(T)$ is given by Eq.~\eqref{eq:R(T)NearNHEK} with the upper choice of sign and is defined on the domain $T\in\br{T^+,T^-}$. The particle reaches the boundary at $T=T^\pm$, and encounters a turning point in between at $(T,R)=(T_t,R_+)$. If $L<0$, then $R(T)$ is given by Eq.~\eqref{eq:R(T)NearNHEK} with the lower choice of sign and is defined for all times $T\in\mathbb{R}$. The particle reaches the horizon as $T\to\pm\infty$, and encounters a turning point in between at $(T,R)=(T_t,R_-)$.
\end{enumerate}
\item[III.]
$\mathcal{C}=0$ corresponds to Type III motion, a limit of Type II motion. If $E,L>0$ (Type IIIA), then $R(T)$ is given by Eq.~\eqref{eq:R(T)NearNHEK} with the upper choice of sign and is defined on the domain $T\gtrless T_t$ according to whether $p^R\lessgtr0$. The particle reaches the boundary at $T=T_t$ and the horizon as $T\to\pm\infty$, without ever encountering a turning point. If $EL<0$ (Type IIIB), then $R(T)$ is given by Eq.~\eqref{eq:R(T)NearNHEK} with the lower choice of sign and is defined for all times $T\in\mathbb{R}$. The particle reaches the horizon as $T\to\pm\infty$, and encounters a turning point in between at $(T,R)=(T_t,R_0)$.
\end{enumerate}

Provided that one keeps track of radial turning points encountered along the way (if any), the expression \eqref{eq:R(T)NearNHEK} allows one to obtain $\bar{I}_R$ as a function of time by plugging in $R_o(T_o)$. In turn, substitution of $\bar{I}_R(T_o)$ into the inversion formulas \eqref{eq:Q>0,P=0}--\eqref{eq:Q>0,P!=0} derived in Sec.~\ref{sec:Kerr} allows one to obtain the polar angle of the particle as a function of time. For instance, in the generic case $P\neq0$ ($L\neq\pm2\mu M$),
\begin{align}
	\cos{\theta_o(T_o)}=\sqrt{u_\pm}\sn\pa{X^\pm(T_o)\left|\frac{u_\pm}{u_\mp}\right.},\qquad
	X^\pm(T_o)=F\pa{\arcsin\pa{\frac{\cos{\theta_s}}{\sqrt{u_\pm}}}\left|\frac{u_\pm}{u_\mp}\right.}-\sign\pa{p_s^\theta}\sqrt{-u_\mp P}\bar{I}_R(T_o),
\end{align}
where $P$ and $u_\pm$ are to be evaluated according to the identifications \eqref{eq:AngularIdentificationNHEK}, and $\pm=\sign\pa{P}$. Finally, given both $R_o(T_o)$ and $\theta_o(T_o)$, one can plug them into the expressions for $\bar{I}_\Phi$ and $\bar{G}_\Phi$ to obtain the azimuthal angle,
\begin{align}
	\Phi_o(T_o)=\Phi_s-\bar{I}_\Phi(T_o)+\bar{G}_\Phi(T_o).
\end{align}
This completes the explicit parameterization of near-NHEK geodesics by the time elapsed along their trajectory.

\section*{Acknowledgments}

This work was supported in part by NSF grant 1205550 to Harvard University. DK gratefully acknowledges support from DOE grant DE-SC0009988. The authors thank Geoffrey Comp\`ere, Samuel Gralla, Shahar Hadar, Abhishek Pathak, Achilleas Porfyriadis, Andrew Strominger, and Peter Zimmerman for fruitful conversations and comments on the draft, and Yichen Shi for collaboration during early stages of the project.$\hfill{\tiny\checkmark}$

\appendix

\section{Elliptic and pseudo-elliptic integrals}
\label{app:EllipticIntegrals}

In this appendix, we define our conventions for the elliptic integrals used throughout the text. We also present two integral identities that are needed in Sec.~\ref{subsec:GlobalGeodesics} to compute the radial geodesic integrals in global NHEK.

\subsubsection{Incomplete elliptic integrals}

The incomplete elliptic integral of the first kind $F$ is defined as
\begin{align}
	F(x|k)=\int_0^x\frac{\ed\theta}{\sqrt{1-k\sin^2{\theta}}}
	=\int_0^{\sin{x}}\frac{\ed t}{\sqrt{\pa{1-t^2}\pa{1-kt^2}}}.
\end{align}
The incomplete elliptic integral of the second kind $E$ is defined as
\begin{align}
	E(x|k)=\int_0^x\sqrt{1-k\sin^2{\theta}}\ed\theta
	=\int_0^{\sin{x}}\sqrt{\frac{1-kt^2}{1-t^2}}\ed t.
\end{align}
We use a $\prime$ to denote its derivative with respect to its second argument:
\begin{align}
	E'(x|k)\equiv\frac{d}{dk}E(x|k)
	=-\frac{1}{2}\int_0^{\sin{x}}\frac{t^2\ed t}{\sqrt{\pa{1-t^2}\pa{1-kt^2}}}.
\end{align}
The incomplete elliptic integral of the third kind $\Pi$ is defined as
\begin{align}
	\Pi(n;x|k)=\int_0^x\frac{1}{\pa{1-n\sin^2{\theta}}}\frac{\ed\theta}{\sqrt{1-k\sin^2{\theta}}}
	=\int_0^{\sin{x}}\frac{1}{1-nt^2}\frac{\ed t}{\sqrt{\pa{1-t^2}\pa{1-kt^2}}}.
\end{align}

\subsubsection{Complete elliptic integrals}

Elliptic integrals are said to be ``complete'' when the amplitude $x=\pi/2$. The complete elliptic integral of the first kind $K$ is denoted
\begin{align}
	K(k)=F\pa{\left.\frac{\pi}{2}\right|k}
	=\frac{\pi}{2}{_2F_1}\pa{\frac{1}{2},\frac{1}{2};1;k},
\end{align}
where ${_2F_1}$ is Gauss' hypergeometric function. The complete elliptic integral of the second kind $E$ is defined as
\begin{align}
	E(k)=E\pa{\left.\frac{\pi}{2}\right|k}
	=\frac{\pi}{2}{_2F_1}\pa{\frac{1}{2},-\frac{1}{2};1;k}.
\end{align}
The complete elliptic integral of the third kind $\Pi$ is defined as
\begin{align}
	\Pi(n|k)=\Pi\pa{n;\left.\frac{\pi}{2}\right.|k}
	=\frac{\pi}{2}F_1\pa{\frac{1}{2};\frac{1}{2},1;1;k,n},
\end{align}
where $F_1$ denotes the first Appell series.

\subsubsection{Pseudo-elliptic integrals for global NHEK}

Let $x>0$, $q>0$, and $\alpha\in\pa{0,\pi}$. Define the manifestly real and positive quantities
\begin{align}
	\label{eq:GlobalIntegrals}
	\tilde{I}_\pm(x,q,\alpha)=\frac{q^{\pm1}}{\pa{2q}^2\sin{\frac{\alpha}{2}}}\br{\frac{\pi}{2}+\arctan\pa{\frac{x^2-q^2}{2qx\sin{\frac{\alpha}{2}}}}\mp\frac{1}{2}\tan\pa{\frac{\alpha}{2}}\log\pa{\frac{x^2+2qx\cos{\frac{\alpha}{2}}+q^2}{x^2-2qx\cos{\frac{\alpha}{2}}+q^2}}}.
\end{align}
One can verify by direct calculation that\footnote{The first identity is essentially identical to Eq.~(2.161.1) in Ref.~\cite{Gradshteyn2007}.}
\begin{align}
	\pd_x\tilde{I}_-(x,q,\alpha)=\frac{1}{q^4-2q^2x^2\cos{\alpha}+x^4},\qquad
	\pd_x\tilde{I}_+(x,q,\alpha)=\frac{x^2}{q^4-2q^2x^2\cos{\alpha}+x^4},
\end{align}
and also that
\begin{align}
	\qquad \lim_{x\to0^+}\tilde{I}_\pm(x,q,\alpha)
	=0.
\end{align}
This implies that for all $x_i>0$,
\begin{align}
	\label{eq:GlobalIntegralIdentities}
	\int_0^{x_i}\frac{\ed x}{q^4-2q^2x^2\cos{\alpha}+x^4}=\tilde{I}_-(x_i,q,\alpha),\qquad
	\int_0^{x_i}\frac{x^2\ed x}{q^4-2q^2x^2\cos{\alpha}+x^4}=\tilde{I}_+(x_i,q,\alpha).
\end{align}

\section{Charged geodesics in \texorpdfstring{AdS$_2$}{AdS2}}
\label{app:AdS2}

The Carter-Penrose diagram of NHEK in Fig.~\ref{fig:PenroseDiagrams} obviously resembles that of two-dimensional Anti de-Sitter space (AdS$_2$). This is simply explained by the observation that, for any fixed $\theta$, the NHEK geometry is a particular fibration of $\mathsf{U}(1)$ over an AdS$_2$ base. One would therefore expect the NHEK geodesics with trivial polar motion\footnote{This class of geodesics includes both the principal null congruences as well as all the equatorial geodesics, which were analyzed in Ref.~\cite{Compere2018}.} and fixed angular momentum $L$ to be related to some motion in AdS$_2$. Upon dimensional reduction, we find that this class of NHEK geodesics reduces to the Lorentz-force trajectories of an electrically charged particle moving in AdS$_2$ with a background electric field.\footnote{A special subset of these geodesics was derived in Ref.~\cite{Maldacena1999}.} We then solve the corresponding charged geodesic equation in global AdS$_2$, Poincar\'e AdS$_2$ and near-AdS$_2$, and relate the solutions to the NHEK geodesics of Sec.~\ref{sec:GeodesicsInNHEK}. Finally, we study the action of $\mathsf{SL}(2,\mathbb{R})$ on this space of trajectories, and show that it acts transitively, as expected.

\subsubsection{Dimensional reduction of NHEK to \texorpdfstring{AdS$_2$}{AdS2}}

The four-dimensional metric ansatz appropriate for the NHEK geometry is
\begin{align}
	\label{eq:MetricAnsatz}
	ds_{(4)}^2=e^\varphi\br{ds_{(2)}^2+\ed\theta^2+e^{2\psi}\pa{\ed\phi-A}^2},
\end{align}
where the metric $ds_{(2)}^2\equiv g_{ab}\ed x^a\ed x^b$ and one-form $A\equiv A_a\ed x^a$ depend only on the coordinates $x^a$, while $\varphi(\theta)$ and $\psi(\theta)$ are scalar functions of $\theta$ only. For instance, the NHEK metric in Poincar\'e coordinates \eqref{eq:NHEK} is of this form with
\begin{align}
	\label{eq:MetricAnsatzNHEK}
	ds_{(2)}^2=-R^2\ed T^2+\frac{\ed R^2}{R^2},\qquad
	A=- R\ed T,\qquad
	e^\varphi=2M^2\Gamma(\theta),\qquad
	e^\psi=\Lambda(\theta),
\end{align}
while for near-NHEK one has
\begin{align}
	\label{eq:MetricAnsatznearNHEK}
	ds_{(2)}^2=-\pa{R^2-\kappa^2}\ed T^2+\frac{\ed R^2}{R^2-\kappa^2},\qquad
	A=- R\ed T,\qquad
	e^\varphi=2M^2\Gamma(\theta),\qquad
	e^\psi=\Lambda(\theta).
\end{align}
For global NHEK, the ansatz reads 
\begin{align}
	ds_{(2)}^2=-(1+y^2)\ed\tau^2+\frac{\ed y^2}{1+y^2},\qquad
	A=-y\ed\tau,\qquad
	e^\varphi=2M^2\Gamma(\theta),\qquad
	e^\psi=\Lambda(\theta).
\end{align}

The Christoffel connection for a metric ansatz of this form is given by
\begin{subequations}
\begin{gather}
	\Gamma^a_{bc}=\hat{\Gamma}^a_{bc}+\frac{1}{2}e^{2\psi}g^{ad}\br{A_cF_{bd}+A_bF_{cd}},\qquad
	\Gamma^a_{b\phi}=-\frac{1}{2}e^{2\psi}g^{ac}F_{bc},\qquad
	\Gamma^a_{b\theta}=\frac{1}{2}\dot{\phi}\delta^a_b,\\
	\Gamma^\phi_{ab}=A^cg_{cd}\hat{\Gamma}^d_{ab}-\frac{1}{2}\pa{\pd_aA_b+\pd_bA_a}+\frac{1}{2}e^{2\psi}A^c\pa{A_aF_{bc}+A_bF_{ac}},\\
	\Gamma^\phi_{a\theta}=-\dot{\psi}A_a,\qquad
	\Gamma^\phi_{a\phi}=-\frac{1}{2}e^{2\psi}A^bF_{ab},\qquad
	\Gamma^\phi_{\theta\phi}=\frac{1}{2}\pa{\dot{\varphi}+2\dot{\psi}},\\
	\Gamma^\theta_{ab}=-\frac{1}{2}\br{\dot{\varphi}\pa{g_{ab}+e^{2\psi}A_aA_b}+2\dot{\psi}e^{2\psi}A_aA_b},\qquad
	\Gamma^\theta_{a\phi}=\frac{1}{2}A_a\pa{\dot{\varphi}+2\dot{\psi}}e^{2\psi},\\
	\Gamma^\theta_{\theta\theta}=\frac{1}{2}\dot{\varphi},\qquad
	\Gamma^\theta_{\phi\phi}=-\frac{1}{2}\pa{\dot{\varphi}+2\dot{\psi}}e^{2\psi}.
\end{gather}
\end{subequations}
Here and henceforth, $\hat{\Gamma}^a_{bc}$ and $D_a$ are the two-dimensional Christoffel connection and covariant derivative associated to $g_{ab}$, while $F_{ab}=\pd_aA_b-\pd_bA_a$ is the curvature of $A_a$, and we use an overdot to denote a derivative with respect to $\theta$. By reducing the four-dimensional Einstein-Hilbert action
\begin{align}
	S=\frac{1}{16\pi G_N}\int\ed^4x\sqrt{-g_4}\mathcal{R}_{(4)}
\end{align}
on the metric ansatz \eqref{eq:MetricAnsatz}, one obtains an effective two-dimensional action for the metric $g_{ab}$ and gauge field $A_a$ with AdS$_2$ solutions supported by electric flux. In terms of these variables, the four-dimensional density is given by
\begin{align}
	\sqrt{-g_4}\mathcal{R}_{(4)}=\sqrt{-g_2}e^{\varphi+\psi}\br{\mathcal{R}_{(2)}-\pa{3\Ddot{\varphi}+\frac{3}{2}\dot{\varphi}^2+3\dot{\varphi}\dot{\psi}+2\dot{\psi}^2+2\Ddot{\psi}}-\frac{1}{4}e^{2\psi}F_{ab}F^{ab}}.
\end{align}
Substituting the NHEK expressions \eqref{eq:MetricAnsatzNHEK} for $\varphi(\theta)$ and $\psi(\theta)$ and then performing the angular integrals yields the two-dimensional action
\begin{align}
	S=\frac{M^2}{2G_N}\int\ed^2x\sqrt{-g_2}\pa{\mathcal{R}_{(2)}+1-\frac{1}{2}F^2}.
\end{align}
This action gives rise to the Einstein-Maxwell equations of motion
\begin{align}
	R_{ab}-\frac{1}{2}Rg_{ab}-\frac{1}{2}g_{ab}=T_{ab}^\mathrm{EM},
\end{align}
where $T_{ab}^\mathrm{EM}$ is the electromagnetic stress-energy tensor
\begin{align}
	T_{ab}^\mathrm{EM}=F_{ac}{F_b}^c-\frac{1}{4}F^2.
\end{align}
In two dimensions, the Einstein tensor identically vanishes,
\begin{align}
	G_{ab}\equiv R_{ab}-\frac{1}{2}Rg_{ab}
	=0,
\end{align}
leaving the much simpler equation of motion
\begin{align}
	-\frac{1}{2}g_{ab}=T_{ab}^\mathrm{EM}.
\end{align}
The metric $g_{ab}$ and gauge potential $A_a$ given in Eq.~\eqref{eq:MetricAnsatzNHEK} manifestly satisfy this equation. This solution corresponds to AdS$_2$ with a background electric field strength given by the $\mathsf{SL}(2,\mathbb{R})$-invariant volume form on AdS$_2$:
\begin{align}
	\label{eq:ElectricField}
	F=\ed\tau\wedge\ed y
	=\ed T\wedge\ed R.
\end{align}
Note that while $F$ is $\mathsf{SL}(2,\mathbb{R})$-invariant and satisfies the two-dimensional Maxwell equations,
\begin{align}
	\ed F=\ed \star F
	=0,
\end{align}
any choice of the gauge potential $A$ shifts by a gauge transformation under at least one isometry of AdS$_2$.

This example exhibits the general mechanism that supports the AdS$_2$-like throats of \textit{vacuum} near-extremal black hole solutions and reveals the obstruction to obtaining higher-dimensional analogues of AdS with such a setup. When reducing along the isometries of a spinning black hole, the frame-dragging term in the metric always appears as a gauge potential $A$ in the lower-dimensional spacetime, thereby providing a natural 2-form (namely, the electromagnetic field strength $F=\ed A$). The throat region outside of the black hole needs to inherit an effective cosmological constant from the four-dimensional solution, and a 2-form can only serve as a cosmological constant in two dimensions.

\subsubsection{Dimensional reduction of planar NHEK geodesics}

In this subsection, we will restrict our attention to NHEK geodesics with no polar motion. Since the reduction of NHEK along the angular directions yields AdS$_2$ with a background electric field, one expects to be able to recast the NHEK geodesic equation in terms of an effective Lorentz-force geodesic equation, with the Kaluza-Klein angular momentum $L$ in NHEK playing the role of the $\mathsf{U}(1)$ gauge charge in AdS$_2$. The  NHEK momentum, denoted by $P_\mu$, is given by Eq.~\eqref{eq:KerrGeodesic} in global NHEK, by Eq.~\eqref{eq:GeodesicNHEK} in the Poincar\'e patch, and by Eq.~\eqref{eq:GeodesicNearNHEK} in near-NHEK. In the two-dimensional description, the NHEK momentum plays the role of the generalized momentum
\begin{align}
	P_a=p_a-LA_a,\qquad
	p^a\equiv g^{ab}\pa{P_a+LA_a},
\end{align}
where $p_a$ is the two-dimensional momentum. The NHEK momentum satisfies the four-dimensional geodesic equation
\begin{align}
	P^\mu\nabla_\mu P_\nu
	=0.
\end{align}
The non-angular components of this equation take the form
\begin{align}
	P^\mu\nabla_\mu P_a=e^{-\phi}\pa{p^bD_bp_a+LF_{ab}p^b}
	=0.
\end{align}
This is the Lorentz-force geodesic equation for a charged particle propagating in AdS$_2$ with a background electric field. The four-dimensional mass, when expressed in terms of the metric ansatz \eqref{eq:MetricAnsatz}, takes the form
\begin{align}
	-\mu^2=g^{\mu\nu}P_\mu P_\mu
	=e^{-\phi}\pa{g^{ab}p_ap_b+e^{-2\psi}L^2}.
\end{align}
In terms of the effective two-dimensional mass $m^2=-g^{ab}p_ap_b$, one therefore finds
\begin{align}
	m^2=e^\phi\mu^2+e^{-2\psi}L^2
	=\mathcal{C}+L^2.
\end{align}
The last equality follows from Eq.~\eqref{eq:GlobalConservedQuantities} with $P_\theta=0$, which holds for each choice of NHEK coordinate system.

We have therefore reduced the problem of planar geodesic motion in NHEK to the motion of a particle with charge $q=L$ and mass $m=\sqrt{\mathcal{C}+L^2}$ in AdS$_2$ with a constant background electric field:
\begin{align}
	\label{eq:LorentzForceLaw}
	p^bD_bp_a=Lp^bF_{ba},\qquad
	g^{ab}p_a p_b=-\pa{\mathcal{C}+L^2}.
\end{align}
In the absence of the background field, particles are confined by the gravitational potential of AdS$_2$. The superradiant effects in the four-dimensional geometry are reflected in the electric field, which exerts a force on charged particles and can expel them from the AdS$_2$ throat.

\subsubsection{Charged geodesics on the global strip}

Global AdS$_2$ is covered by the coordinates $(\tau,y)$ with $-\infty<\tau,y<+\infty$. The line element is given by
\begin{align}
	ds^2=-\pa{1+y^2}\ed\tau^2+\frac{\ed y^2}{1+y^2}.
\end{align}
This geometry has constant negative curvature $\mathcal{R}=-2$ and admits three Killing vector fields
\begin{align}
	H_0=\frac{y\sin{\tau}}{\sqrt{1+y^2}}\pd_\tau-\cos{\tau}\sqrt{1+y^2}\pd_y,\qquad
	H_\pm=\pa{1\pm\frac{y\cos{\tau}}{\sqrt{1+y^2}}}\pd_\tau\pm\sin{\tau}\sqrt{1+y^2}\pd_y,
\end{align}
which generate an $\mathsf{SL}(2,\mathbb{R})$ isometry group,
\begin{align}
	\br{H_0,H_\pm}=\mp H_\pm,\qquad
	\br{H_+,H_-}=2H_0.
\end{align}
In global coordinates, it is convenient to complexify this algebra by introducing new (complex) generators
\begin{align}
	L_0=i\pd_\tau,\qquad
	L_\pm=e^{\pm i\tau}\br{\pm\frac{y}{\sqrt{1+y^2}}\pd_\tau-i\sqrt{1+y^2}\pd_y},
\end{align}
which obey the same commutation relations:
\begin{align}
	\br{L_0,L_\pm}=\mp L_\pm,\qquad
	\br{L_+,L_-}=2L_0.
\end{align}
The relation between these two sets of generators is the same as in Eq.~\eqref{eq:Automorphism}. The metric is the Casimir of $\mathsf{SL}(2,\mathbb{R})$:
\begin{align}
	\label{eq:Casimir}
	g^{ab}=H_0^aH_0^b-\frac{1}{2}\pa{H_+^aH_-^b+H_-^aH_+^b}.
\end{align}
In the global strip, the background gauge potential 
\begin{align}
	A=-y\ed\tau
\end{align}
gives rise to the symmetric field strength \eqref{eq:ElectricField} and preserves global-time-translations:
\begin{align}
	\label{eq:GlobalInvariance}
	\L_{L_0}A=0.
\end{align}

To solve Eq.~\eqref{eq:LorentzForceLaw} in this background, it is convenient to introduce the generalized momentum (NHEK momentum)
\begin{align}
	P_a=p_a-LA_a.
\end{align}
By virtue of Eq.~\eqref{eq:GlobalInvariance}, the global energy
\begin{align}
	\triangle=iL_0^a P_a
	=-P_\tau
	=-p_\tau-Ly
\end{align}
is conserved along the trajectory. By inverting the above relations for $\pa{\triangle,\mathcal{C}+L^2}$, we find that a charged particle following a Lorentz-force trajectory in global AdS$_2$ with radial electric field \eqref{eq:ElectricField} has an instantaneous momentum $p=p_a\ed x^a$ of the form
\begin{align}
	p(x^a,\triangle,L,\mathcal{C})=-\pa{\triangle+Ly}\ed\tau\pm_y\frac{\sqrt{\mathcal{Y}(y)}}{1+y^2}\ed y,
\end{align}
where the choice of sign $\pm_y$ depends on the radial direction of travel, and we recovered the radial potential \eqref{eq:GlobalRadialPotential},
\begin{align}
	\mathcal{Y}(y)=\pa{\triangle+Ly}^2-\pa{\mathcal{C}+L^2}\pa{1+y^2}.
\end{align}
One can then raise $p_a$ to obtain the equations for the trajectory,
\begin{align}
	\frac{d\tau}{d\sigma}=\frac{\triangle+Ly}{1+y^2},\qquad
	\frac{dy}{d\sigma}=\pm_y\sqrt{\mathcal{Y}(y)}.
\end{align}
Hence, a charged geodesic with global energy $\triangle$ connects spacetime points $X_s^a=\pa{\tau_s,y_s}$ and $X_o^a=\pa{\tau_o,y_o}$ if
\begin{align}
	\tau_o-\tau_s&=\fint_{y_s}^{y_o}\pa{\frac{\triangle+Ly}{1+y^2}}\frac{\ed y}{\pm_y\sqrt{\mathcal{Y}(y)}}.
\end{align}
This equation is solved in Sec.~\ref{subsec:GlobalGeodesics}.

\subsubsection{Charged geodesics on the Poincar\'e patch of AdS$_2$}

The Poincar\'e patch of AdS$_2$ is covered by coordinates $(T,R)$ obtained from the global coordinates $(\tau,y)$ using the coordinate transformation \eqref{eq:PoincareToGlobal}, which results in the line element
\begin{align}
	ds^2=-R^2\ed T^2+\frac{\ed R^2}{R^2}.
\end{align}
The generators of $\mathsf{SL}(2,\mathbb{R})$ are given by
\begin{align}
	H_0=T\pd_T-R\pd_R,\qquad
	H_+=\pd_T,\qquad
	H_-=\pa{T^2+\frac{1}{R^2}}\pd_T-2TR\pd_R,
\end{align}
and its Casimir reproduces the metric as in Eq.~\eqref{eq:Casimir}. In the Poincar\'e patch of AdS$_2$, the background gauge potential
\begin{align}
	A=-R\ed T
\end{align}
gives rise to the symmetric field strength \eqref{eq:ElectricField} and preserves both time-translations and dilations (though not special conformal transformations):
\begin{align}
	\label{eq:PoincareInvariance}
	\L_{H_+}A=\L_{H_0}A=0.
\end{align}

To solve Eq.~\eqref{eq:LorentzForceLaw} in this background, it is convenient to introduce the generalized momentum (NHEK momentum)
\begin{align}
	P_a=p_a-LA_a.
\end{align}
By virtue of Eq.~\eqref{eq:PoincareInvariance}, the Poincar\'e energy
\begin{align}
	E=-H_+^a P_a
	=-P_T
	=-p_T-LR
\end{align}
is conserved along the trajectory. By inverting the above relations for $\pa{E,\mathcal{C}+L^2}$, we find that a charged particle following a Lorentz-force trajectory in a Poincar\'e patch of AdS$_2$ with radial electric field \eqref{eq:ElectricField} has an instantaneous momentum $p=p_a\ed x^a$ of the form
\begin{align}
	p(x^a,E,L,\mathcal{C})=-\pa{E+LR}\ed T\pm_R\frac{\sqrt{\mathcal{R}_n(R)}}{R^2}\ed R,
\end{align}
where the choice of sign $\pm_R$ depends on the radial direction of travel, and we recovered the radial potential \eqref{eq:PoincareRadialPotential},
\begin{align}
	\mathcal{R}_n(R)=\pa{E+LR}^2-\pa{\mathcal{C}+L^2}R^2.
\end{align}
One can then raise $p_a$ to obtain the equations for the trajectory,
\begin{align}
	\frac{dT}{d\sigma}=\frac{E+LR}{R^2},\qquad
	\frac{dR}{d\sigma}=\pm_R\sqrt{\mathcal{R}_n(R)}.
\end{align}
Hence, a charged geodesic with Poincar\'e energy $E$ connects spacetime points $X_s^a=\pa{T_s,R_s}$ and $X_o^a=\pa{T_o,R_o}$ if
\begin{align}
	T_o-T_s=\fint_{R_s}^{R_o}\pa{\frac{E+LR}{R^2}}\frac{\ed R}{\pm_R\sqrt{\mathcal{R}_n(R)}}.
\end{align}
This equation is solved in Sec.~\ref{subsec:PoincareGeodesics}.

\subsubsection{Charged geodesics on \texorpdfstring{near-AdS$_2$}{near-AdS2}}

The coordinate transformation \eqref{eq:NHEK2NearNHEK} maps Poincar\'e AdS$_2$ to a smaller patch with line element
\begin{align}
	ds^2=-\pa{R^2-\kappa^2}\ed T^2+\frac{\ed R^2}{R^2-\kappa^2}.
\end{align}
As in Sec.~\ref{subsec:ConformalSymmetryInTheSky}, this transformation implements a boundary time reparameterization $T\to e^{\kappa T}$ that introduces a small temperature $\kappa$. This smaller patch can thus be interpreted as a black hole in AdS$_2$ (see, $e.g.$, Ref.~\cite{Maldacena2016b}).

The generators of $\mathsf{SL}(2,\mathbb{R})$ are given by
\begin{align}
	H_0=\frac{1}{\kappa}\pd_T,\qquad
	H_\pm=\frac{e^{\mp\kappa T}}{\sqrt{R^2-\kappa^2}}\br{\frac{R}{\kappa}\pd_T\pm\pa{R^2-\kappa^2}\pd_R},
\end{align}
and its Casimir reproduces the metric as in Eq.~\eqref{eq:Casimir}. In near-AdS$_2$, the gauge  potential
\begin{align}
	A=-R\ed T
\end{align}
gives rise to the symmetric field strength \eqref{eq:ElectricField} and preserves near-AdS$_2$ time-translations:\footnote{Note that the gauge-equivalent connection $A=-R\ed T-\ed\arctanh(R/\kappa)$ preserves both $H_+$ and $H_0$.}
\begin{align}
	\label{eq:NearInvariance}
	\L_{H_0}A=0.
\end{align}

To solve Eq.~\eqref{eq:LorentzForceLaw} in this background, it is convenient to introduce the generalized momentum (NHEK momentum)
\begin{align}
	P_a=p_a-LA_a.
\end{align}
By virtue of Eq.~\eqref{eq:NearInvariance}, the near-AdS$_2$ energy
\begin{align}
	E=-\kappa H_0^a P_a
	=-P_T
	=-p_T-LR
\end{align}
is conserved along the trajectory. By inverting the above relations for $\pa{E,\mathcal{C}+L^2}$, we find that a charged particle following a Lorentz force trajectory in near-AdS$_2$ with radial electric field \eqref{eq:ElectricField} has an instantaneous momentum $p=p_a\ed x^a$ of the form
\begin{align}
	p(x^a,E,L,\mathcal{C})=-\pa{E+LR}\ed T\pm_R\frac{\sqrt{\mathcal{R}_\kappa(R)}}{R^2-\kappa^2}\ed R,
\end{align}
where the choice of sign $\pm_R$ depends on the radial direction of travel, and we recovered the radial potential \eqref{eq:RadialPotentialNearNHEK},
\begin{align}
	\mathcal{R}_\kappa(R)=\pa{E+LR}^2-\pa{\mathcal{C}+L^2}\pa{R^2-\kappa^2}.
\end{align}
One can then raise $p_\mu$ to obtain the equations for the trajectory,
\begin{align}
	\frac{dT}{d\sigma}=\frac{E+LR}{R^2-\kappa^2},\qquad
	\frac{dR}{d\sigma}=\pm_R\sqrt{\mathcal{R}_\kappa(R)}.
\end{align}
Hence, a charged geodesic with near-AdS$_2$ energy $E$ connects spacetime points $X_s^a=\pa{T_s,R_s}$ and $X_o^a=\pa{T_o,R_o}$ if
\begin{align}
	T_o-T_s=\fint_{R_s}^{R_o}\pa{\frac{E+LR}{R^2-\kappa^2}}\frac{\ed R}{\pm_R\sqrt{\mathcal{R}_\kappa(R)}}.
\end{align}
This equation is solved in Sec.~\ref{subsec:NearGeodesics}.

\subsubsection{Isometry group orbits of geodesics}

The NHEK equatorial geodesics were classified and studied in detail in Ref.~\cite{Compere2018}. Remarkably, it was found that all equatorial orbits can be transformed to a circular orbit by a (possibly complex) isometry. In Anti-de Sitter spacetime (which is maximally symmetric), any two geodesics can be mapped into each other by an isometry (up to parity or time-reversal transformation): all geodesics fall into a small number of orbits of the isometry group action on the manifold.

In this subsection, we prove that in AdS$_2$, this statement still holds in the presence of the unique symmetric electromagnetic field \eqref{eq:ElectricField}. In other words, we show that any two charged geodesics with the same charge $L$ and mass $\sqrt{\mathcal{C}+L^2}$ are related by an $\mathsf{SL}(2,\mathbb{R})$ isometry. For convenience, we will only prove this explicity in the Poincar\'e patch of AdS$_2$, but of course it then follows that the result holds in every coordinate system. This fact is crucial for our proof in Sec.~\ref{subsec:PoincareGeodesics} that all NHEK geodesics with the same angular motion (and therefore the same mass $\mu$, angular momentum $L$ about the axis of symmetry, and Casimir $\mathcal{C}$) are related by an isometry of the throat.

To begin with, recall from Sec.~\ref{subsec:PoincareGeodesics} that the general solution to the motion in the Poincar\'e patch is given by Eq.~\eqref{eq:R(T)NHEK}, which can be rearranged to obtain
\begin{align}
	R_o(T_o)-\frac{EL}{S_n(T_o)}=\pm\frac{EL}{S_n(T_o)}\sqrt{1+\frac{S_n(T_o)}{L^2}}.
\end{align}
Squaring both sides and expanding terms out yields
\begin{align}
	R_o^2(T_o)=\frac{E^2}{S_n(T_o)}+\frac{2ELR_o(T_o)}{S_n(T_o)}.
\end{align}
Multiplying by $S_n(T_o)R_o^{-2}(T_o)$ and then completing the square results in
\begin{align}
	S_n(T_o)=\br{\frac{E}{R_o(T_o)}+L}^2-L^2.
\end{align}
Recalling the definition of $S_n(T)$, this is equivalent to
\begin{align}
	\label{eq:PoincareMotion}
	\mathcal{C}+L^2=\br{\frac{E}{R_o(T_o)}+L}^2-(H_0+ET_o)^2=\br{\frac{E}{R_o(T_o)}+L}^2	-\br{E\pa{T_o-T_s}-\nu_s\frac{\sqrt{\mathcal{R}_n(R_s)}}{R_s}}^2,
\end{align}
which is yet another form of the solution to the charged geodesic equation in Poincar\'e coordinates.

Now consider two different geodesics: the first has energy $E_1$ and trajectory $\pa{T_1(\sigma),R_1(\sigma)}$ with starting point $\pa{T_{s,1},R_{s,1}}$, while the second has energy $E_2$ and trajectory $\pa{T_2(\sigma),R_2(\sigma)}$ with starting point $\pa{T_{s,2},R_{s,2}}$. However, we assume they both have the same charge $q=L$ and mass-squared $m^2=\mathcal{C}+L^2$. Then Eq.~\eqref{eq:PoincareMotion} implies that
\begin{align}
	\label{eq:TwoTrajectories}
	-\pa{E_iT_i-c_i}^2+\pa{\frac{E_i}{R_i}+L}^2=\mathcal{C}+L^2,\qquad
	c_i\equiv E_iT_{s,i}+\nu_{s,i}\frac{\sqrt{\mathcal{R}_n(R_{s,i})}}{R_{s,i}},\qquad
	i=1,2.
\end{align}
Notice that the these two trajectories would match if
\begin{align}
	T_2=\frac{E_1}{E_2}T_1+\frac{c_2-c_1}{E_2},\qquad
	R_2=\frac{E_2}{E_1}R_1.
\end{align}
But these relations are exactly the product of two consecutive $\mathsf{SL}(2,\mathbb{R})$ transformations: first, a dilation
\begin{align}
	\pa{T_1,R_1}\to\pa{T_1^\prime,R_1^\prime}
	=\pa{\frac{T_1}{\lambda},\lambda R_1},\qquad
	\lambda=\frac{E_2}{E_1},
\end{align}
followed by a time-translation
\begin{align}
	\pa{T_1^\prime,R_1^\prime}\to\pa{T_1^{\prime\prime},R_1^{\prime\prime}}
	=\pa{T_1^\prime+c,R_1^\prime},\qquad
	c=\frac{c_2-c_1}{E_2}.
\end{align}
Therefore, the second geodesic is an $\mathsf{SL}(2,\mathbb{R})$ image of the first geodesic. To complete the demonstration, we must consider the constant radius trajectories
\begin{align}
	R(T)=R_0,
\end{align}
with $E=\mathcal{C}=0$ and $L$ arbitrary. Under the special conformal transformation generated by $H_-$, which maps
\begin{align}
	\label{eq:SpecialConformalTransformation}
	T\to\frac{T-\lambda\pa{T^2-\frac{1}{R^2}}}{\pa{1-\lambda T}^2-\frac{\lambda^2}{R^2}},\qquad
	R\to R\br{\pa{1-\lambda T}^2-\frac{\lambda^2}{R^2}},
\end{align}
each such curve is mapped into a new curve obeying
\begin{align}
	-\pa{\frac{2\lambda^2L}{R_0}T-\frac{2\lambda L}{R_0}}^2+\pa{\frac{2\lambda^2L/R_0}{R}+L}^2=L^2.
\end{align}
This is of the form \eqref{eq:PoincareMotion} with arbitrary $L$ and
\begin{align}
	E=\frac{2\lambda^2 L}{R_0},\qquad
	\mathcal{C}=0,\qquad
	ET_s+\nu_s\frac{\sqrt{\mathcal{R}_n(R_s)}}{R_s}=\frac{2\lambda L}{R_0},
\end{align}
corresponding to a charged geodesic with the same mass $m^2=L^2$ and charge $q=L$, but different energy and initial position. More generally, two trajectories \eqref{eq:TwoTrajectories} with the same $\mathsf{SL}(2,\mathbb{R})$ Casimir $\mathcal{C}$ and charge $L$ are related by a special conformal transformation \eqref{eq:SpecialConformalTransformation} if
\begin{align}
	E_2=E_1-2\lambda c_1+\frac{c_1^2+\mathcal{C}}{E_1}\lambda^2,\qquad
	c_2=c_1-\frac{c_1^2+\mathcal{C}}{E_1}\lambda.
\end{align}

In conclusion, by a combined dilation, time-translation and special conformal transformation, it is always possible to map any two charged geodesics in AdS$_2$ with the same mass and charge into each other. Note that the required transformations fully exhaust the isometries available.

\bibliography{NHEKgeo}

\providecommand{\href}[2]{#2}\begingroup\raggedright\begin{thebibliography}{10}

\bibitem{Strominger1996}
A.~{Strominger} and C.~{Vafa}, ``{Microscopic origin of the Bekenstein-Hawking
  entropy},'' \href{http://dx.doi.org/10.1016/0370-2693(96)00345-0}{{\em
  Physics Letters B} {\bfseries 379} (Feb., 1996) 99--104},
  \href{http://arxiv.org/abs/hep-th/9601029}{{\ttfamily hep-th/9601029}}.

\bibitem{Maldacena1997}
J.~{Maldacena}, ``{The Large-N Limit of Superconformal Field Theories and
  Supergravity},'' \href{http://dx.doi.org/10.1023/A:1026654312961}{{\em
  International Journal of Theoretical Physics} {\bfseries 38} (1999)
  1113--1133}, \href{http://arxiv.org/abs/hep-th/9711200}{{\ttfamily
  hep-th/9711200}}.

\bibitem{Guica2009}
M.~{Guica}, T.~{Hartman}, W.~{Song}, and A.~{Strominger}, ``{The Kerr/CFT
  correspondence},'' \href{http://dx.doi.org/10.1103/PhysRevD.80.124008}{{\em
  \prd} {\bfseries 80} no.~12, (Dec., 2009) 124008},
  \href{http://arxiv.org/abs/0809.4266}{{\ttfamily arXiv:0809.4266 [hep-th]}}.

\bibitem{Maldacena2016a}
J.~{Maldacena} and D.~{Stanford}, ``{Remarks on the Sachdev-Ye-Kitaev model},''
  \href{http://dx.doi.org/10.1103/PhysRevD.94.106002}{{\em \prd} {\bfseries 94}
  no.~10, (Nov., 2016) 106002},
  \href{http://arxiv.org/abs/1604.07818}{{\ttfamily arXiv:1604.07818
  [hep-th]}}.

\bibitem{Maldacena2016b}
J.~{Maldacena}, D.~{Stanford}, and Z.~{Yang}, ``{Conformal symmetry and its
  breaking in two dimensional Nearly Anti-de-Sitter space},''
  \href{http://dx.doi.org/10.1093/ptep/ptw12}{{\em Progress of Theoretical and
  Experimental Physics} {\bfseries 2016} no.~12, (Nov., 2016) 12C104},
  \href{http://arxiv.org/abs/1606.01857}{{\ttfamily arXiv:1606.01857
  [hep-th]}}.

\bibitem{Abbott2016}
B.~P. {Abbott}, R.~{Abbott}, T.~D. {Abbott}, M.~R. {Abernathy}, F.~{Acernese},
  K.~{Ackley}, C.~{Adams}, T.~{Adams}, P.~{Addesso}, R.~X. {Adhikari}, and
  et~al., ``{Observation of Gravitational Waves from a Binary Black Hole
  Merger},'' \href{http://dx.doi.org/10.1103/PhysRevLett.116.061102}{{\em \prl}
  {\bfseries 116} no.~6, (Feb., 2016) 061102},
  \href{http://arxiv.org/abs/1602.03837}{{\ttfamily arXiv:1602.03837 [gr-qc]}}.

\bibitem{EHT2019}
{Event Horizon Telescope Collaboration}, K.~{Akiyama}, A.~{Alberdi}, W.~{Alef},
  K.~{Asada}, R.~{Azulay}, A.-K. {Baczko}, D.~{Ball}, M.~{Balokovi{\'c}},
  J.~{Barrett}, and et~al., ``{First M87 Event Horizon Telescope Results. I.
  The Shadow of the Supermassive Black Hole},''
  \href{http://dx.doi.org/10.3847/2041-8213/ab0ec7}{{\em \apjl} {\bfseries 875}
  (Apr., 2019) L1}.

\bibitem{Lupsasca2014}
A.~{Lupsasca}, M.~J. {Rodriguez}, and A.~{Strominger}, ``{Force-free
  electrodynamics around extreme Kerr black holes},''
  \href{http://dx.doi.org/10.1007/JHEP12(2014)185}{{\em Journal of High Energy
  Physics} {\bfseries 12} (Dec., 2014) 185},
  \href{http://arxiv.org/abs/1406.4133}{{\ttfamily arXiv:1406.4133 [hep-th]}}.

\bibitem{Zhang2014}
F.~{Zhang}, H.~{Yang}, and L.~{Lehner}, ``{Towards an understanding of the
  force-free magnetosphere of rapidly spinning black holes},''
  \href{http://dx.doi.org/10.1103/PhysRevD.90.124009}{{\em \prd} {\bfseries 90}
  no.~12, (Dec., 2014) 124009},
  \href{http://arxiv.org/abs/1409.0345}{{\ttfamily arXiv:1409.0345
  [astro-ph.HE]}}.

\bibitem{Lupsasca2015}
A.~{Lupsasca} and M.~J. {Rodriguez}, ``{Exact solutions for extreme black hole
  magnetospheres},'' \href{http://dx.doi.org/10.1007/JHEP07(2015)090}{{\em
  Journal of High Energy Physics} {\bfseries 7} (July, 2015) 90},
  \href{http://arxiv.org/abs/1412.4124}{{\ttfamily arXiv:1412.4124 [hep-th]}}.

\bibitem{Compere2015}
G.~{Comp{\`e}re} and R.~{Oliveri}, ``{Near-horizon extreme Kerr
  magnetospheres},'' \href{http://dx.doi.org/10.1103/PhysRevD.93.024035}{{\em
  \prd} {\bfseries 93} no.~2, (Jan., 2016) 024035},
  \href{http://arxiv.org/abs/1509.07637}{{\ttfamily arXiv:1509.07637
  [hep-th]}}.

\bibitem{Gralla2016a}
S.~E. {Gralla}, A.~{Lupsasca}, and A.~{Strominger}, ``{Near-horizon Kerr
  magnetosphere},'' \href{http://dx.doi.org/10.1103/PhysRevD.93.104041}{{\em
  \prd} {\bfseries 93} no.~10, (May, 2016) 104041},
  \href{http://arxiv.org/abs/1602.01833}{{\ttfamily arXiv:1602.01833
  [hep-th]}}.

\bibitem{Chen2017}
B.~{Chen} and L.~C. {Stein}, ``{Separating metric perturbations in near-horizon
  extremal Kerr spacetimes},''
  \href{http://dx.doi.org/10.1103/PhysRevD.96.064017}{{\em \prd} {\bfseries 96}
  no.~6, (Sept., 2017) 064017},
  \href{http://arxiv.org/abs/1707.05319}{{\ttfamily arXiv:1707.05319 [gr-qc]}}.

\bibitem{Gralla2017b}
S.~E. {Gralla} and P.~{Zimmerman}, ``{Critical exponents of extremal Kerr
  perturbations},'' \href{http://dx.doi.org/10.1088/1361-6382/aab140}{{\em
  Classical and Quantum Gravity} {\bfseries 35} no.~9, (May, 2018) 095002},
  \href{http://arxiv.org/abs/1711.00855}{{\ttfamily arXiv:1711.00855 [gr-qc]}}.

\bibitem{Gralla2018}
S.~E. {Gralla} and P.~{Zimmerman}, ``{Scaling and universality in extremal
  black hole perturbations},''
  \href{http://dx.doi.org/10.1007/JHEP06(2018)061}{{\em Journal of High Energy
  Physics} {\bfseries 6} (June, 2018) 61},
  \href{http://arxiv.org/abs/1804.04753}{{\ttfamily arXiv:1804.04753 [gr-qc]}}.

\bibitem{Hadar2019}
S.~{Hadar}, ``{Near-extremal black holes at late times, backreacted},''
  \href{http://dx.doi.org/10.1007/JHEP01(2019)214}{{\em Journal of High Energy
  Physics} {\bfseries 1} (Jan., 2019) 214},
  \href{http://arxiv.org/abs/1811.01022}{{\ttfamily arXiv:1811.01022
  [hep-th]}}.

\bibitem{Porfyriadis2014}
A.~P. {Porfyriadis} and A.~{Strominger}, ``{Gravity waves from the Kerr/CFT
  correspondence},'' \href{http://dx.doi.org/10.1103/PhysRevD.90.044038}{{\em
  \prd} {\bfseries 90} no.~4, (Aug., 2014) 044038},
  \href{http://arxiv.org/abs/1401.3746}{{\ttfamily arXiv:1401.3746 [hep-th]}}.

\bibitem{Hadar2014}
S.~{Hadar}, A.~P. {Porfyriadis}, and A.~{Strominger}, ``{Gravity waves from
  extreme-mass-ratio plunges into Kerr black holes},''
  \href{http://dx.doi.org/10.1103/PhysRevD.90.064045}{{\em \prd} {\bfseries 90}
  no.~6, (Sept., 2014) 064045},
  \href{http://arxiv.org/abs/1403.2797}{{\ttfamily arXiv:1403.2797 [hep-th]}}.

\bibitem{Hadar2015}
S.~{Hadar}, A.~P. {Porfyriadis}, and A.~{Strominger}, ``{Fast plunges into Kerr
  black holes},'' \href{http://dx.doi.org/10.1007/JHEP07(2015)078}{{\em Journal
  of High Energy Physics} {\bfseries 7} (July, 2015) 78},
  \href{http://arxiv.org/abs/1504.07650}{{\ttfamily arXiv:1504.07650
  [hep-th]}}.

\bibitem{Hadar2017}
S.~{Hadar} and A.~P. {Porfyriadis}, ``{Whirling orbits around twirling black
  holes from conformal symmetry},''
  \href{http://dx.doi.org/10.1007/JHEP03(2017)014}{{\em Journal of High Energy
  Physics} {\bfseries 3} (Mar., 2017) 14},
  \href{http://arxiv.org/abs/1611.09834}{{\ttfamily arXiv:1611.09834
  [hep-th]}}.

\bibitem{Compere2018}
G.~{Comp{\`e}re}, K.~{Fransen}, T.~{Hertog}, and J.~{Long}, ``{Gravitational
  waves from plunges into Gargantua},''
  \href{http://dx.doi.org/10.1088/1361-6382/aab99e}{{\em Classical and Quantum
  Gravity} {\bfseries 35} no.~10, (May, 2018) 104002},
  \href{http://arxiv.org/abs/1712.07130}{{\ttfamily arXiv:1712.07130 [gr-qc]}}.

\bibitem{Gralla2015}
S.~E. {Gralla}, A.~P. {Porfyriadis}, and N.~{Warburton}, ``{Particle on the
  innermost stable circular orbit of a rapidly spinning black hole},''
  \href{http://dx.doi.org/10.1103/PhysRevD.92.064029}{{\em \prd} {\bfseries 92}
  no.~6, (Sept., 2015) 064029},
  \href{http://arxiv.org/abs/1506.08496}{{\ttfamily arXiv:1506.08496 [gr-qc]}}.

\bibitem{Gralla2016b}
S.~E. {Gralla}, S.~A. {Hughes}, and N.~{Warburton}, ``{Inspiral into
  Gargantua},'' \href{http://dx.doi.org/10.1088/0264-9381/33/15/155002}{{\em
  Classical and Quantum Gravity} {\bfseries 33} no.~15, (Aug., 2016) 155002},
  \href{http://arxiv.org/abs/1603.01221}{{\ttfamily arXiv:1603.01221 [gr-qc]}}.

\bibitem{Burko2016}
L.~M. {Burko} and G.~{Khanna}, ``{Gravitational waves from a plunge into a
  nearly extremal Kerr black hole},''
  \href{http://dx.doi.org/10.1103/PhysRevD.94.084049}{{\em \prd} {\bfseries 94}
  no.~8, (Oct., 2016) 084049},
  \href{http://arxiv.org/abs/1608.02244}{{\ttfamily arXiv:1608.02244 [gr-qc]}}.

\bibitem{Gralla2016c}
S.~E. {Gralla}, A.~{Zimmerman}, and P.~{Zimmerman}, ``{Transient instability of
  rapidly rotating black holes},''
  \href{http://dx.doi.org/10.1103/PhysRevD.94.084017}{{\em \prd} {\bfseries 94}
  no.~8, (Oct., 2016) 084017},
  \href{http://arxiv.org/abs/1608.04739}{{\ttfamily arXiv:1608.04739 [gr-qc]}}.

\bibitem{Compere2017}
G.~{Comp{\`e}re} and R.~{Oliveri}, ``{Self-similar accretion in thin discs
  around near-extremal black holes},''
  \href{http://dx.doi.org/10.1093/mnras/stx748}{{\em \mnras} {\bfseries 468}
  (July, 2017) 4351--4361}, \href{http://arxiv.org/abs/1703.00022}{{\ttfamily
  arXiv:1703.00022 [astro-ph.HE]}}.

\bibitem{Gralla2017a}
S.~E. {Gralla}, A.~{Lupsasca}, and A.~{Strominger}, ``{Observational signature
  of high spin at the Event Horizon Telescope},''
  \href{http://dx.doi.org/10.1093/mnras/sty039}{{\em \mnras} {\bfseries 475}
  (Apr., 2018) 3829--3853}, \href{http://arxiv.org/abs/1710.11112}{{\ttfamily
  arXiv:1710.11112 [astro-ph.HE]}}.

\bibitem{Porfyriadis2017}
A.~P. {Porfyriadis}, Y.~{Shi}, and A.~{Strominger}, ``{Photon emission near
  extreme Kerr black holes},''
  \href{http://dx.doi.org/10.1103/PhysRevD.95.064009}{{\em \prd} {\bfseries 95}
  no.~6, (Mar., 2017) 064009},
  \href{http://arxiv.org/abs/1607.06028}{{\ttfamily arXiv:1607.06028 [gr-qc]}}.

\bibitem{Lupsasca2018}
A.~{Lupsasca}, A.~P. {Porfyriadis}, and Y.~{Shi}, ``{Critical emission from a
  high-spin black hole},''
  \href{http://dx.doi.org/10.1103/PhysRevD.97.064017}{{\em \prd} {\bfseries 97}
  no.~6, (Mar., 2018) 064017},
  \href{http://arxiv.org/abs/1712.10182}{{\ttfamily arXiv:1712.10182 [gr-qc]}}.

\bibitem{Gates2018}
D.~{Gates}, D.~{Kapec}, A.~{Lupsasca}, Y.~{Shi}, and A.~{Strominger},
  ``{Polarization Whorls from M87 at the Event Horizon Telescope},'' {\em arXiv
  e-prints} (Sept., 2018) , \href{http://arxiv.org/abs/1809.09092}{{\ttfamily
  arXiv:1809.09092 [hep-th]}}.

\bibitem{Carter1968}
B.~{Carter}, ``{Global Structure of the Kerr Family of Gravitational Fields},''
  \href{http://dx.doi.org/10.1103/PhysRev.174.1559}{{\em Physical Review}
  {\bfseries 174} (Oct., 1968) 1559--1571}.

\bibitem{Bardeen1973}
J.~M. {Bardeen}, ``{Timelike and null geodesics in the Kerr metric},'' in {\em
  Black Holes (Les Astres Occlus)}, C.~{Dewitt} and B.~S. {Dewitt}, eds.,
  pp.~215--239.
\newblock Gordon and Breach Science Publishers, New York, 1973.

\bibitem{Chandrasekhar1983}
S.~{Chandrasekhar}, {\em {The mathematical theory of black holes}}.
\newblock Oxford University Press, 1983.

\bibitem{ONeill1995}
B.~{O'Neill}, {\em {The geometry of Kerr black holes}}.
\newblock A~K~Peters, Ltd. Wellesley, Mass., 1995.

\bibitem{Rauch1994}
K.~P. {Rauch} and R.~D. {Blandford}, ``{Optical caustics in a kerr spacetime
  and the origin of rapid X-ray variability in active galactic nuclei},''
  \href{http://dx.doi.org/10.1086/173625}{{\em \apj} {\bfseries 421} (Jan.,
  1994) 46--68}.

\bibitem{Vazquez2004}
S.~E. {V{\'a}zquez} and E.~P. {Esteban}, ``{Strong-field gravitational lensing
  by a Kerr black hole},''
  \href{http://dx.doi.org/10.1393/ncb/i2004-10121-y}{{\em Nuovo Cimento B
  Serie} {\bfseries 119} (May, 2004) 489},
  \href{http://arxiv.org/abs/gr-qc/0308023}{{\ttfamily gr-qc/0308023}}.

\bibitem{Kraniotis2005}
G.~V. {Kraniotis}, ``{Frame dragging and bending of light in Kerr and Kerr
  (anti) de Sitter spacetimes},''
  \href{http://dx.doi.org/10.1088/0264-9381/22/21/001}{{\em Classical and
  Quantum Gravity} {\bfseries 22} (Nov., 2005) 4391--4424},
  \href{http://arxiv.org/abs/gr-qc/0507056}{{\ttfamily gr-qc/0507056}}.

\bibitem{Dexter2009}
J.~{Dexter} and E.~{Agol}, ``{A Fast New Public Code for Computing Photon
  Orbits in a Kerr Spacetime},''
  \href{http://dx.doi.org/10.1088/0004-637X/696/2/1616}{{\em \apj} {\bfseries
  696} (May, 2009) 1616--1629},
  \href{http://arxiv.org/abs/0903.0620}{{\ttfamily arXiv:0903.0620
  [astro-ph.HE]}}.

\bibitem{Fujita2009}
R.~{Fujita} and W.~{Hikida}, ``{Analytical solutions of bound timelike geodesic
  orbits in Kerr spacetime},''
  \href{http://dx.doi.org/10.1088/0264-9381/26/13/135002}{{\em Classical and
  Quantum Gravity} {\bfseries 26} no.~13, (July, 2009) 135002},
  \href{http://arxiv.org/abs/0906.1420}{{\ttfamily arXiv:0906.1420 [gr-qc]}}.

\bibitem{Kraniotis2011}
G.~V. {Kraniotis}, ``{Precise analytic treatment of Kerr and Kerr-(anti) de
  Sitter black holes as gravitational lenses},''
  \href{http://dx.doi.org/10.1088/0264-9381/28/8/085021}{{\em Classical and
  Quantum Gravity} {\bfseries 28} no.~8, (Apr., 2011) 085021},
  \href{http://arxiv.org/abs/1009.5189}{{\ttfamily arXiv:1009.5189 [gr-qc]}}.

\bibitem{Hackmann2010}
E.~{Hackmann}, {\em {Geodesic equations in black hole space-times with
  cosmological constant}}.
\newblock PhD thesis, Bremen University, 2010.

\bibitem{Hackmann2014}
E.~{Hackmann} and C.~{L{\"a}mmerzahl},
  \href{http://dx.doi.org/10.1063/1.4861945}{``{Analytical solution methods for
  geodesic motion},''} in {\em American Institute of Physics Conference
  Series}, vol.~1577 of {\em American Institute of Physics Conference Series},
  pp.~78--88.
\newblock Jan., 2014.
\newblock \href{http://arxiv.org/abs/1506.00807}{{\ttfamily arXiv:1506.00807
  [gr-qc]}}.

\bibitem{Wilkins1972}
D.~C. {Wilkins}, ``{Bound Geodesics in the Kerr Metric},''
  \href{http://dx.doi.org/10.1103/PhysRevD.5.814}{{\em \prd} {\bfseries 5}
  (Feb., 1972) 814--822}.

\bibitem{deFelice1972}
F.~{de Felice} and M.~{Calvani}, ``{Orbital and vortical motion in the Kerr
  metric.},'' {\em Nuovo Cimento B Serie} {\bfseries 10} (1972) 447--458.

\bibitem{Stephani1978}
H.~{Stephani}, ``{A note on Killing tensors},''
  \href{http://dx.doi.org/10.1007/BF00760867}{{\em General Relativity and
  Gravitation} {\bfseries 9} (Sept., 1978) 789--792}.

\bibitem{Fettis1970}
H.~{Fettis}, ``{On the Reciprocal Modulus Relation for Elliptic Integrals},''
  \href{http://dx.doi.org/10.1137/0501045}{{\em SIAM Journal on Mathematical
  Analysis} {\bfseries 470} no.~4, (1970) 524--526}.
  \url{https://doi.org/10.1137/0501045}.

\bibitem{Abramowitz1972}
M.~{Abramowitz} and I.~A. {Stegun}, {\em {Handbook of Mathematical Functions}}.
\newblock United States Department of Commerce, National Bureau of Standards,
  1972.

\bibitem{Sorce2017}
J.~{Sorce} and R.~M. {Wald}, ``{Gedanken experiments to destroy a black hole.
  II. Kerr-Newman black holes cannot be overcharged or overspun},''
  \href{http://dx.doi.org/10.1103/PhysRevD.96.104014}{{\em \prd} {\bfseries 96}
  no.~10, (Nov., 2017) 104014},
  \href{http://arxiv.org/abs/1707.05862}{{\ttfamily arXiv:1707.05862 [gr-qc]}}.

\bibitem{Stuchlik2010}
Z.~{Stuchl{\'{\i}}k} and J.~{Schee}, ``{Appearance of Keplerian discs orbiting
  Kerr superspinars},''
  \href{http://dx.doi.org/10.1088/0264-9381/27/21/215017}{{\em Classical and
  Quantum Gravity} {\bfseries 27} no.~21, (Nov., 2010) 215017},
  \href{http://arxiv.org/abs/1101.3569}{{\ttfamily arXiv:1101.3569 [gr-qc]}}.

\bibitem{Stuchlik2011}
Z.~{Stuchl{\'\i}k}, S.~{Hled{\'\i}k}, and K.~{Truparov{\'a}}, ``{Evolution of
  Kerr superspinars due to accretion counterrotating thin discs},''
  \href{http://dx.doi.org/10.1088/0264-9381/28/15/155017}{{\em Classical and
  Quantum Gravity} {\bfseries 28} no.~15, (Aug., 2011) 155017}.

\bibitem{Brenneman2013}
L.~{Brenneman}, \href{http://dx.doi.org/10.1007/978-1-4614-7771-6}{{\em
  {Measuring the Angular Momentum of Supermassive Black Holes}}}.
\newblock SpringerBriefs in Astronomy, 2013.
\newblock \href{http://arxiv.org/abs/1309.6334}{{\ttfamily arXiv:1309.6334
  [astro-ph.HE]}}.

\bibitem{Reynolds2019}
C.~S. {Reynolds}, ``{Observing black holes spin},''
  \href{http://dx.doi.org/10.1038/s41550-018-0665-z}{{\em Nature Astronomy}
  {\bfseries 3} (Jan., 2019) 41--47}.

\bibitem{Bardeen1972}
J.~M. {Bardeen}, W.~H. {Press}, and S.~A. {Teukolsky}, ``{Rotating Black Holes:
  Locally Nonrotating Frames, Energy Extraction, and Scalar Synchrotron
  Radiation},'' \href{http://dx.doi.org/10.1086/151796}{{\em \apj} {\bfseries
  178} (Dec., 1972) 347--370}.

\bibitem{Novikov1973}
I.~D. {Novikov} and K.~S. {Thorne}, ``{Astrophysics of black holes.},'' in {\em
  Black Holes (Les Astres Occlus)}, C.~{Dewitt} and B.~S. {Dewitt}, eds.,
  pp.~343--450.
\newblock 1973.

\bibitem{Page1974}
D.~N. {Page} and K.~S. {Thorne}, ``{Disk-Accretion onto a Black Hole.
  Time-Averaged Structure of Accretion Disk},''
  \href{http://dx.doi.org/10.1086/152990}{{\em \apj} {\bfseries 191} (July,
  1974) 499--506}.

\bibitem{Cunningham1975}
C.~T. {Cunningham}, ``{The effects of redshifts and focusing on the spectrum of
  an accretion disk around a Kerr black hole},''
  \href{http://dx.doi.org/10.1086/154033}{{\em \apj} {\bfseries 202} (Dec.,
  1975) 788--802}.

\bibitem{Penna2012}
R.~F. {Penna}, A.~{S{\c a}owski}, and J.~C. {McKinney}, ``{Thin-disc theory
  with a non-zero-torque boundary condition and comparisons with
  simulations},''
  \href{http://dx.doi.org/10.1111/j.1365-2966.2011.20084.x}{{\em \mnras}
  {\bfseries 420} (Feb., 2012) 684--698},
  \href{http://arxiv.org/abs/1110.6556}{{\ttfamily arXiv:1110.6556
  [astro-ph.HE]}}.

\bibitem{Jacobson2011}
T.~{Jacobson}, ``{Where is the extremal Kerr ISCO?},''
  \href{http://dx.doi.org/10.1088/0264-9381/28/18/187001}{{\em Classical and
  Quantum Gravity} {\bfseries 28} no.~18, (Sept., 2011) 187001},
  \href{http://arxiv.org/abs/1107.5081}{{\ttfamily arXiv:1107.5081 [gr-qc]}}.

\bibitem{Bardeen1999}
J.~{Bardeen} and G.~T. {Horowitz}, ``{Extreme Kerr throat geometry: A vacuum
  analog of AdS2 x S2},''
  \href{http://dx.doi.org/10.1103/PhysRevD.60.104030}{{\em \prd} {\bfseries 60}
  no.~10, (Nov., 1999) 104030},
  \href{http://arxiv.org/abs/hep-th/9905099}{{\ttfamily hep-th/9905099}}.

\bibitem{Galajinsky2010}
A.~{Galajinsky}, ``{Particle dynamics near extreme Kerr throat and
  supersymmetry},'' \href{http://dx.doi.org/10.1007/JHEP11(2010)126}{{\em
  Journal of High Energy Physics} {\bfseries 11} (Nov., 2010) 126},
  \href{http://arxiv.org/abs/1009.2341}{{\ttfamily arXiv:1009.2341 [hep-th]}}.

\bibitem{AlZahrani2011}
A.~M. {Al Zahrani}, V.~P. {Frolov}, and A.~A. {Shoom}, ``{Particle Dynamics in
  Weakly Charged Extreme Kerr Throat},''
  \href{http://dx.doi.org/10.1142/S0218271811018986}{{\em International Journal
  of Modern Physics D} {\bfseries 20} (2011) 649--660},
  \href{http://arxiv.org/abs/1010.1570}{{\ttfamily arXiv:1010.1570 [gr-qc]}}.

\bibitem{Song2009}
W.~{Song}, D.~{Anninos}, W.~{Li}, M.~{Padi}, and A.~{Strominger}, ``{Warped
  AdS$_{3}$ black holes},''
  \href{http://dx.doi.org/10.1088/1126-6708/2009/03/130}{{\em Journal of High
  Energy Physics} {\bfseries 3} (Mar., 2009) 130},
  \href{http://arxiv.org/abs/0807.3040}{{\ttfamily arXiv:0807.3040 [hep-th]}}.

\bibitem{Aman2012}
J.~E. {Aman}, I.~{Bengtsson}, and H.~F. {Runarsson}, ``{What are extremal Kerr
  Killing vectors up to?},''
  \href{http://dx.doi.org/10.1088/0264-9381/29/21/215017}{{\em Classical and
  Quantum Gravity} {\bfseries 29} no.~21, (Oct., 2012) 215017},
  \href{http://arxiv.org/abs/1206.6306}{{\ttfamily arXiv:1206.6306 [gr-qc]}}.

\bibitem{Israel1986}
W.~{Israel}, ``{Third law of black-hole dynamics - A formulation and proof},''
  \href{http://dx.doi.org/10.1103/PhysRevLett.57.397}{{\em Physical Review
  Letters} {\bfseries 57} (July, 1986) 397--399}.

\bibitem{Bredberg2010}
I.~{Bredberg}, T.~{Hartman}, W.~{Song}, and A.~{Strominger}, ``{Black hole
  superradiance from Kerr/CFT},''
  \href{http://dx.doi.org/10.1007/JHEP04(2010)019}{{\em Journal of High Energy
  Physics} {\bfseries 4} (Apr., 2010) 19},
  \href{http://arxiv.org/abs/0907.3477}{{\ttfamily arXiv:0907.3477 [hep-th]}}.

\bibitem{Dias2009}
{\'O}.~J.~C. {Dias}, H.~S. {Reall}, and J.~E. {Santos}, ``{Kerr-CFT and
  gravitational perturbations},''
  \href{http://dx.doi.org/10.1088/1126-6708/2009/08/101}{{\em Journal of High
  Energy Physics} {\bfseries 8} (Aug., 2009) 101},
  \href{http://arxiv.org/abs/0906.2380}{{\ttfamily arXiv:0906.2380 [hep-th]}}.

\bibitem{Amsel2009}
A.~J. {Amsel}, G.~T. {Horowitz}, D.~{Marolf}, and M.~M. {Roberts}, ``{No
  dynamics in the extremal Kerr throat},''
  \href{http://dx.doi.org/10.1088/1126-6708/2009/09/044}{{\em Journal of High
  Energy Physics} {\bfseries 9} (Sept., 2009) 044},
  \href{http://arxiv.org/abs/0906.2376}{{\ttfamily arXiv:0906.2376 [hep-th]}}.

\bibitem{Compere2012}
G.~{Comp{\`e}re}, ``{The Kerr/CFT Correspondence and its Extensions},''
  \href{http://dx.doi.org/10.12942/lrr-2012-11}{{\em Living Reviews in
  Relativity} {\bfseries 15} (Oct., 2012) 11},
  \href{http://arxiv.org/abs/1203.3561}{{\ttfamily arXiv:1203.3561 [hep-th]}}.

\bibitem{Roberts2012}
M.~M. {Roberts}, ``{Time evolution of entanglement entropy from a pulse},''
  \href{http://dx.doi.org/10.1007/JHEP12(2012)027}{{\em Journal of High Energy
  Physics} {\bfseries 12} (Dec., 2012) 27},
  \href{http://arxiv.org/abs/1204.1982}{{\ttfamily arXiv:1204.1982 [hep-th]}}.

\bibitem{Banados1999}
M.~{Ba{\~n}ados}, \href{http://dx.doi.org/10.1063/1.59661}{``{Three-dimensional
  quantum geometry and black holes},''} in {\em Trends in Theoretical Physics
  II}, H.~{Falomir}, R.~E. {Gamboa Saravi}, and F.~A. {Schaposnik}, eds.,
  vol.~484 of {\em American Institute of Physics Conference Series},
  pp.~147--169.
\newblock July, 1999.
\newblock \href{http://arxiv.org/abs/hep-th/9901148}{{\ttfamily
  hep-th/9901148}}.

\bibitem{Compere2016}
G.~{Comp{\`e}re}, P.~{Mao}, A.~{Seraj}, and M.~M. {Sheikh-Jabbari},
  ``{Symplectic and Killing symmetries of AdS$_{3}$ gravity: holographic vs
  boundary gravitons},'' \href{http://dx.doi.org/10.1007/JHEP01(2016)080}{{\em
  Journal of High Energy Physics} {\bfseries 1} (Jan., 2016) 80},
  \href{http://arxiv.org/abs/1511.06079}{{\ttfamily arXiv:1511.06079
  [hep-th]}}.

\bibitem{Bicak1989}
J.~{Bi{\v{c}}{\'a}k}, Z.~{Stuchl{\'{\i}}k}, and V.~{Balek}, ``{The motion of
  charged particles in the field of rotating charged black holes and naked
  singularities},'' {\em Bulletin of the Astronomical Institutes of
  Czechoslovakia} {\bfseries 40} (Mar., 1989) 65--92.

\bibitem{Gradshteyn2007}
I.~S. {Gradshteyn}, I.~M. {Ryzhik}, A.~{Jeffrey}, and D.~{Zwillinger}, {\em
  {Table of Integrals, Series, and Products}}.
\newblock Elsevier Academic Press, 2007.

\bibitem{Maldacena1999}
J.~{Maldacena}, J.~{Michelson}, and A.~{Strominger}, ``{Anti-de Sitter
  fragmentation},'' \href{http://dx.doi.org/10.1088/1126-6708/1999/02/011}{{\em
  Journal of High Energy Physics} {\bfseries 2} (Feb., 1999) 011},
  \href{http://arxiv.org/abs/hep-th/9812073}{{\ttfamily hep-th/9812073}}.

\end{thebibliography}\endgroup
\bibliographystyle{utphys}

\end{document}